\documentclass[useAMS,usenatbib]{mn2e}
\usepackage{graphicx}
\usepackage[english]{babel}
\usepackage{txfonts}
\usepackage{multirow}
\usepackage{lscape}
\usepackage{longtable}
\usepackage{natbib}
\usepackage{color}
\usepackage{float}
\usepackage{array}
\usepackage{graphicx,amsfonts,psfrag,fancyhdr,layout,appendix}%,subfigure}
\usepackage{makeidx}
\usepackage{fancyvrb}

   \title[Bias of BAO tracers]{Stochastic bias of colour-selected BAO tracers by joint clustering-weak lensing analysis}
   
  \author[Johan Comparat et al.]
  {Johan Comparat$^1$, Eric Jullo$^1$, Jean-Paul Kneib$^{1,2}$, Carlo Schimd,$^1$  HuanYuan Shan,$^{2,3}$ \newauthor
Thomas Erben,$^4$ Olivier Ilbert$^1$ Joel Brownstein,$^5$ Anne Ealet,$^6$  St\'ephanie Escoffier,$^6$\newauthor
Bruno Moraes,$^{7,8}$ Nick Mostek,$^{9}$ Jeffrey A. Newman,$^{10}$ M. E. S. Pereira,$^{7,8}$\newauthor
 Francisco Prada,$^{11,12,13}$ David J. Schlegel,$^{9}$ Donald P. Schneider,$^{14,15}$ Carlos H. Brandt$^{7,16}$\newauthor \\
$^1$ Aix Marseille Universit\'e, CNRS, LAM (Laboratoire d'Astrophysique de Marseille) UMR 7326, 13388, Marseille, France \\
$^2$  Laboratoire d'astrophysique, \'Ecole Polytechnique F\'ed\'erale de Lausanne (EPFL), Observatoire de Sauverny, 1290 Versoix, Switzerland\\
$^3$ Department of Physics and Tsinghua Center for Astrophysics, Tsinghua University, Beijing, 100084, China \\
$^4$ Argelander Institute for Astronomy, University of Bonn, Auf dem H¬ugel 71, 53121 Bonn, Germany.\\
$^5$ Department of Physics and Astronomy, University of Utah, Salt Lake City, UT 84112, USA\\
$^6$Centre de Physique des Particules de Marseille, Universit\'e d'Aix-Marseille, CNRS/IN2P3, 13288 Marseille cedex 09, France\\
$^{7}$Centro Brasileiro de Pesquisas F\'{i}sicas 
Rua Dr. Xavier Sigaud 150, CEP 22290-180, Rio de Janeiro, RJ, Brazil\\
$^{8}$Laborat\'{o}rio Interinstitucional de e-Astronomia --- LIneA, 
Rua Gal. Jos\'{e} Cristino 77, CEP 20921-400, Rio de Janeiro, RJ, Brazil\\
$^{9}$ Lawrence Berkeley National Laboratory, One Cyclotron Road, Berkeley, CA 94720\\
$^{10}$ Department of Physics and Astronomy and PITT-PACC,
University of Pittsburgh, Pittsburgh, PA, 15260\\
$^{11}$ Instituto de F\'{\i}sica Te\'orica, (UAM/CSIC), Universidad Aut\'onoma de Madrid, Cantoblanco, E-28049 Madrid, Spain \\
$^{12}$ Instituto de Astrof\'isica de Andaluc\'ia (CSIC), Glorieta de la Astronom\'ia, E-18080 Granada, Spain\\
$^{13}$ Campus of International Excellence UAM+CSIC, Cantoblanco, E-28049 Madrid, Spain \\
$^{14}$Institute for Gravitation and the Cosmos, The Pennsylvania State University, University Park, PA 16802, USA\\
$^{15}$Department of Astronomy and Astrophysics, The Pennsylvania State University, University Park, PA 16802, USA\\
$^{16}$Laborat\'orio Nacional de Computa\c c\~ao Cient\'ifica, Av. Get\'ulio Vargas, 333 Petr\'opolis, 25651-075 Rio de Janeiro, RJ Brazil\\
}

\begin{document}

%\pagerange{\pageref{firstpage}--\pageref{lastpage}} \pubyear{2012}
\maketitle
%\label{firstpage}
Accepted 2013 May 3 by MNRAS.
%----------------------abstract----------------------------%   
\begin{abstract}
The baryon acoustic oscillation (BAO) feature in the two-point correlation function of galaxies supplies a standard ruler to probe the expansion history of the Universe. 
We study here several galaxy selection schemes, aiming at building an emission-line galaxy (ELG) sample in the redshift range $0.6<z<1.7$, that would be suitable for future BAO studies, providing a highly biased galaxy sample. We analyse the angular galaxy clustering of galaxy selections at the redshifts 0.5, 0.7, 0.8, 1 and 1.2 and we combine this analysis with a halo occupation distribution (HOD) model to derive the properties of the haloes these galaxies inhabit, in particular the galaxy bias on large scales. We also perform a weak lensing analysis (aperture statistics) to extract the galaxy bias and the cross-correlation coefficient and compare to the HOD prediction. 

We apply this analysis on a data set composed of the photometry of the deep co-addition on Sloan Digital Sky Survey (SDSS) Stripe 82 (225 deg$^2$), of Canda-France-Hawai Telescope/Stripe 82 deep \emph{i}-band weak lensing survey and of the {\it Wide-Field Infrared Survey Explorer }infrared photometric band W1.

The analysis on the SDSS-III/constant mass galaxies selection at $z=0.5$ is in agreement with previous studies on the tracer, moreover we measure its cross-correlation coefficient $r=1.16\pm0.35$. 
For the higher redshift bins, we confirm the trends that the brightest galaxy populations selected are strongly biased ($b>1.5$), but we are limited by current data sets depth to derive precise values of the galaxy bias.
A survey using such tracers of the mass field will guarantee a high significance detection of the BAO. 
\end{abstract}
    
%--------------------------------------------------------------------------------------------------------------
%--------------------------------------------------------------------------------------------------------------
%------------------------------			INTRODUCTION       -----------------------------------
%--------------------------------------------------------------------------------------------------------------
%--------------------------------------------------------------------------------------------------------------
%\section{Introduction}
\section{Introduction}
\label{sec:introduction}
Working within the standard cosmological framework described by the Friedmann model, recent analyses of cosmic microwave background (CMB) and Type Ia supernovae (SNIa) data confirm the picture of our Universe as almost spatially flat, with an energy content known at the per cent level and dominated by dark components, and undergoing a phase of accelerated expansion since the last 7 Gyr (e.g. \citealt{Komatsu_2011,2012ApJ...746...85S}). A confirmation of these results is provided by the use of the baryonic acoustic oscillations (BAO) signature in the clustering of galaxies \citep{doi:10.1146/annurev.astro.46.060407.145243}, which offers a standard ruler for the measurement of the angular distance--redshift relation. 
Current measurements of BAO have precision larger than 2 per cent, due to statistical errors. \citet{2011MNRAS.416.3017B} provides the latest measurement at $z=0.106$ on the 6dF Galaxy Survey \citep{2009MNRAS.399..683J} with a precision of 4.5 per cent. \citet{Blake_2011B} and \citet{2012MNRAS.427.3435A} made the latest measurements at $z \simeq 0.6$ using, the WiggleZ Dark Energy Survey \citep{Drinkwater_2010} and the Sloan Digital Sky Survey III Baryonic Oscillation Spectroscopic Survey (SDSS-III/BOSS hereafter \citealt{2011AJ....142...72E,2013AJ....145...10D}), respectively. At larger redshift $z=2.3$, using the BAO from the Ly$\alpha$ forest of quasars, $H(z=2.3)$ was measured at $\sim3.5$ per cent (see \citealt{2012A&A...548A..66P,2013A&A...552A..96B,2013AJ....145...69L,2013JCAP...03..024K,2013JCAP...04..026S}).
Future galaxy surveys plan detections of the BAO at the sub-percent level in the redshift range, $0.7 \lesssim z \lesssim 2$, to bridge the gap between the forementioned results. It constitutes the next step towards the investigation of the accelerated expansion of the Universe.

The measurement of BAO in the power spectrum of galaxies relies on the assumption that `galaxies are tracers of the underlying matter distribution' \citep{1984ApJ...284L...9K}. The validity and the limits of this assumption are not yet completely understood theoretically or observationally. To constrain cosmology with the measurement of the BAO feature in the galaxy clustering, it is necessary to understand the relation between the galaxy distribution and the dark matter field. 
Currently, to bridge the gap between dark matter haloes and galaxies, there are different methods.

The first approach is to use a model that provides how the galaxies are distributed in haloes (halo occupation distribution model, hereafter HOD; see \citet{1999ApJ...520...24D,2000ApJ...543..503M,2000MNRAS.318..203S,2002PhR...372....1C}). The HOD model is fitted on the measure of the galaxy clustering to predict the properties of the haloes that the galaxies inhabit. In particular, HOD predicts the galaxy bias; see \citet{2011ApJ...728..126W,2011ApJ...736...59Z,2012PhRvD..86j3518P} for the HOD applied to galaxies used in BAO surveys.

The alternative method is based on weak lensing (WL). With recent surveys, it has become possible to combine on small areas (a few deg$^2$) different statistical measures of the galaxy distribution: the galaxy clustering, the cosmic shear, and the galaxy-galaxy lensing. Using Cosmic Evolution Survey (COSMOS) data \citep{Capak_2007,2007ApJS..172....1S,Ilbert_2009} \citet{2012ApJ...750...37J} and \citet{2012ApJ...744..159L} led the way in combining these measures to constrain the galaxy--halo relation and the evolution of the galaxy bias. Improved HOD models are being developed to interpret the combination of both measures \citep{2011ApJ...738...45L,2012MNRAS.426..566C,2013MNRAS.430..747M,2013MNRAS.430..725V}.

The application of such analysis on thousands of square degrees by combining WL with galaxy clustering is however currently limited by the depth and the point spread function (PSF) characterization of the photometry \citep{2011arXiv1111.6958H}. 

Another possibility to determine the galaxy bias is the direct comparison with large simulations of dark matter using subhalo abundance matching technique (SHAM; \citealt{2004ApJ...609...35K,2004MNRAS.353..189V,2006ApJ...647..201C,2011ApJ...742...16T}). Applied to the analysis of WiggleZ and SDSS-III/BOSS observations, this technique shows similar results as the HOD (see \citealt{2013MNRAS.tmp.1188N,2012PhRvD..86j3518P,2012MNRAS.423.3018P}).

In this paper, we investigate the clustering amplitude of galaxies selected to be the BAO tracers of future large spectroscopic surveys as SDSS-IV/Extended Baryon Oscillation Spectroscopic Survey (eBOSS), BigBOSS, Dark Energy Spectrometer (DESpec), Prime Focus Spectrograph-Subaru Measurement of Image and Redshifts (PFS-SuMiRe), {\it Euclid} \citep{2011arXiv1110.3193L,bigBOSS_2011,2013MNRAS.428.1498C}. The clustering amplitude being proportional to $\propto (b\sigma_8)^2$, we consider a constant $\sigma_8=0.81$ (value from CMB analysis 7-{\it year Wlikinson Microwave Anisotropy Probe, WMAP7}; \citealt{Komatsu_2011}) in order to derive the value of the bias $b$. For this analysis, we combine an HOD approach with a WL analysis. We obtain rather quantitative rough estimates of the galaxy bias, that should not be taken at face value. 
We make the point for a high clustering amplitude (or a high bias), justifying target selection scheme of future BAO surveys and a more in depth investigation to make forecasts for these surveys accounting for biases at these levels. In fact signal-to-noise ratio (SNR) for BAO studies scales as the density of tracers times the clustering amplitude of these tracers, and hence BAO studies profit from highly clustered tracers.

In section \ref{sec:methods} we examine the galaxy bias and the two different methods used to measure it: the HOD model and the WL aperture statistics method. We also discuss the limits of each measurement. In section \ref{sec:observations} we describe the data on which our measurement is based. In section \ref{sec:Measures} we discuss the bias measurement from clustering and WL for each tracer and the associated errors. Finally, in section \ref{sec:Discussion} we discuss the feasibility of future BAO studies using the described tracers.

Throughout this paper we assume a flat $\Lambda$ cold dark matter ($\Lambda$CDM) cosmology characterized by {\it WMAP}7 parameters $H_0=100 \; {\rm km}\;{\rm s}^{-1}\;{\rm Mpc}^{-1}$, with h=0.71, $\Omega_\mathrm{CDM}=0.226$, $\Omega_\mathrm{b}=0.0455$, $\Omega_\Lambda=0.7285$, $\sigma_8=0.81$ and $n_\mathrm{initial}=0.966$ \citep{Komatsu_2011}. Magnitudes are given in the AB system \citep{1983ApJ...266..713O}. 

%--------------------------------------------------------------------------------------------------------------
%--------------------------------------------------------------------------------------------------------------
%------------------------------			method				  ---------------------------
%--------------------------------------------------------------------------------------------------------------
%--------------------------------------------------------------------------------------------------------------
%\section{Methods to estimate the bias}
\section{Galaxy clustering--weak lensing joint analysis}
\label{sec:methods}
Galaxy clustering and gravitational lensing probe, respectively, the galaxy overdensity field $\delta_\mathrm{g}(\mathbf{r},z)$, which is a discrete random variable function of position $\mathbf{r}$ and redshift $z$, and the matter overdensity field, $\delta_\mathrm{m}(\mathbf{r},z)$, which is a discrete random variable related to the latter by some functional form generically called `bias', eventually allowing for stochasticity \citep{1998ApJ...500L..79T,1999ApJ...520...24D,1999ApJ...518L..69T}.\footnote{For clarity, we drop in the following the space and time dependence from $\delta$s.} As random variables, only the $N$-point correlation functions (or functions thereof) are meaningful and can be measured; by combining galaxy clustering and gravitational lensing one can measure the autocorrelation functions $\langle \delta_\mathrm{g}^2 \rangle$ and $\langle \delta_\mathrm{m}^2 \rangle$ and the cross-correlation $\langle \delta_\mathrm{g} \delta_\mathrm{m}\rangle$ and eventually define the (linear) bias, $b_\mathrm{g}$, and stochasticity, $r$, parameters, after \citet{2002ApJ...577..604H}
\begin{equation}
b^2_g=\frac{\langle \delta_\mathrm{g}^2 \rangle}{\langle \delta_\mathrm{m}^2 \rangle} \; , \;\; r=\frac{\langle \delta_\mathrm{g} \delta_\mathrm{m} \rangle}{\sqrt{\langle \delta_\mathrm{g}^2 \rangle \langle \delta_\mathrm{m}^2 \rangle}}.
\label{bias:eqn:def}
\end{equation}
A positive (negative) value for $r$ indicates correlation (anticorrelation) between the galaxy and the matter fields, while $r=1$ ($r=-1$) indicates total correlation (total anticorrelation).

%\subsection{Halo occupation distribution model}
\subsection{$b_\mathrm{g}$ by clustering, an HOD approach}
To derive the galaxy bias, we need the autocorrelation of matter and of galaxies. The galaxy autocorrelation function $\langle \delta_\mathrm{g}^2 \rangle$ is estimated with \citet{1993ApJ...412...64L} `minimum variance estimator'. Then we use a HOD to interpret the measure of autocorrelation, as it provides an insight into the relation between the galaxies and the haloes they inhabit.

The main ingredients of the model are the following. The halo mass function $n(M,z)$ gives the density of haloes of mass $M$ present at redshift $z$; it is defined as in \citet{1999MNRAS.308..119S}. The large scale halo bias $b_\mathrm{h}(M,z)$ is taken from \citet{2001MNRAS.323....1S} with the parameters and scale dependence described in \citet{2005ApJ...631...41T}. We use the matter autocorrelation from \citet{2003MNRAS.341.1311S}. The parameter of the model is a function that gives the number of galaxies present in a halo of mass $M$, denoted $N(M)$:
\begin{equation}
N(M)=\frac{1}{2}\left[ 1+ \mathrm{erf}\left(\frac{\log M - \log M_\mathrm{min}}{\sigma_{\log M}}\right)\right] \left[ 1+ \left( \frac{M-M_0}{M_1} \right)^\alpha \right],
\label{hod:eqn}
\end{equation}
where $\mathrm{erf}(x)=\frac{2}{\pi} \int_0^x \mathrm{e}^{-t^2} \mathrm{d}t$ is the `error function'.
For a given $N(M)$ the HOD model outputs the corresponding autocorrelation function. A fit on the galaxy autocorrelation function is performed to derive the analytical form of $N(M)$.
During the fit, the cosmological parameters are fixed at their fiducial value ({\it WMAP} 7). The large-scale galaxy bias of a volume-limited sample of galaxies described by $N(M)$ is obtained with 
\begin{equation}
\langle b_\mathrm{g} \rangle = \frac{\int n(M,z) b_\mathrm{h}(M,z) N(M) \mathrm{d}M}{\int n(M,z)N(M)\mathrm{d}M}.
\label{bias:hod:def}
\end{equation}
For more details about the implementation of the HOD we use in this study, see \citet{2002PhR...372....1C,2005ApJ...631...41T,2012A&A...542A...5C}. We compute angular correlation functions with the software \textsc{athena}\footnote{http://www2.iap.fr/users/kilbinge/athena/} and the HOD fits are performed using \textsc{CosmoPMC}\footnote{http://www2.iap.fr/users/kilbinge/CosmoPMC/}; see \citet{2011arXiv1101.0950K,2009A&A...497..677K,2010MNRAS.405.2381K,2012A&A...542A...5C}.

The main limit of the model is the input autocorrelation of matter. This information is taken from \citet{2003MNRAS.341.1311S} and has errors of $\sim8$ per cent at small scales and $<3$ per cent at large scales $k<10h\,\mathrm{Mpc}^{-1}$ in the $\Lambda$CDM paradigm. Also the HOD model is not valid in different cosmological paradigms and can therefore not be used to rule out cosmologies: it can only be used to replace galaxy samples in the galaxy clustering history, which is the aim of this paper.

%\subsection{WL model, aperture statistics}
\subsection{$b_\mathrm{g}$ and $r$ by weak lensing}
Using WL analysis, we derive both $b_\mathrm{g}$ and $r$. For this purpose, we need the three auto and cross-correlations mentioned before. 
The WL measurement is based on the analysis of the distribution of shapes of the galaxies located behind the BAO tracers. Their intrinsic shapes are distorted by the foreground mass around the BAO tracers. The autocorrelation of the ellipticities of the background galaxies contains the information of the matter clustering between the background galaxies and us, $\langle \delta^2_m \rangle$. The cross-correlation of the ellipticities with the position of the foreground galaxies contains the information of the cross-correlation $\langle \delta_\mathrm{g} \delta_\mathrm{m} \rangle$. The angular autocorrelation $\langle \delta^2_g \rangle$ of galaxies is the same as for the HOD.

To be able to compare the auto- and cross-correlation results, the measurements are convolved with Bessel kernels to transform the correlations in aperture correlations. This formalism is described in \citet{1998A&A...333..767S} and applied on RCS data in \citet{2002ApJ...577..604H}, GaBoDS data in \citet{2007A&A...461..861S}, and on COSMOS in \citet{2012ApJ...750...37J}. We use the same routines as in \citet{2012ApJ...750...37J} to measure the aperture autocorrelation of galaxies, denoted $\langle \mathcal{N}_\mathrm{ap}^2(\theta) \rangle$, that of matter, denoted $\langle M_\mathrm{ap}^{2}(\theta)\rangle$, and the cross-correlation, denoted $\langle \mathcal{N}_\mathrm{ap}(\theta) M_\mathrm{ap}(\theta)\rangle$ defined from the power spectrum of the galaxies $P_n$, of the matter $P_\kappa$, and their cross-power spectrum $P_{n\kappa}$.
\begin{equation}
\langle \mathcal{N}_\mathrm{ap}^2(\theta) \rangle = 2 \pi \int_0^\infty \mathrm{d}l\, l P_n(l) \left[ \frac{12J_4(l\theta)}{\pi(l\theta)^2} \right]^2
\end{equation}

\begin{equation}
\langle M_\mathrm{ap}^{2}(\theta)\rangle = 2 \pi \int_0^\infty \mathrm{d}l\, l P_\kappa(l) \left[ \frac{12J_4(l\theta)}{\pi(l\theta)^2} \right]^2
\end{equation}

\begin{equation}
\langle \mathcal{N}_\mathrm{ap}(\theta) M_\mathrm{ap}(\theta)\rangle = 2 \pi \int_0^\infty \mathrm{d}l\, l P_{n\kappa}(l) \left[ \frac{12J_4(l\theta)}{\pi(l\theta)^2} \right]^2
\end{equation}
where $J_4$ is the Bessel function of order 4, see \citet{1998A&A...333..767S,2007A&A...461..861S}.\\
The first output is the galaxy bias in an aperture of an angular radius $\theta$; see equation (\ref{bias:wl:eqn}). The second output is the cross-correlation coefficient $r$, that measures the randomness; see equation (\ref{rrrr:wl:eqn})
 \begin{equation}
b(\theta)=f_1 (\theta,\Omega_\mathrm{m},\Omega_{\Lambda}) \sqrt{\frac{\langle \mathcal{N}_\mathrm{ap}^2(\theta) \rangle }{\langle M_\mathrm{ap}^{2}(\theta)\rangle}},
\label{bias:wl:eqn}
\end{equation}
\begin{equation}
r(\theta)=f_2 (\theta,\Omega_\mathrm{m},\Omega_{\Lambda}) \frac{ \langle \mathcal{N}_\mathrm{ap}(\theta) M_\mathrm{ap}(\theta)\rangle }{\sqrt{ \langle \mathcal{N}_\mathrm{ap}^2(\theta) \rangle \langle M_\mathrm{ap}^{2}(\theta)\rangle }}.
\label{rrrr:wl:eqn}
\end{equation}
The functions $f_{1}$ and $f_{2}$ are normalizations. They correct for the fact that different cosmological volumes are probed by the different statistics. The functions are computed assuming a \citet{2003MNRAS.341.1311S} non-linear power spectrum with unbiased foreground galaxies ($b = r = 1$), that uses the \citet{1998ApJ...496..605E} transfer function, constrained by the {\it WMAP}7 cosmological parameters \citep{Komatsu_2011}.

The WL also uses the \citet{2003MNRAS.341.1311S} non-linear power spectrum and has therefore the same precision limitations as the HOD. Moreover, an additional constraint is that, to be able to extract the WL signal, the sources must be behind the lenses. Because of this point, we are not able to have a significant measure of the matter galaxy cross-correlation by WL for the faintest galaxy samples (see section \ref{sec:observations}).
%----------------------------------------------------------------------------------------------------
%----------------------------------------------------------------------------------------------------
%------------------------------	data							  
%----------------------------------------------------------------------------------------------------
%----------------------------------------------------------------------------------------------------
%\section{Data}
\section{Data}
\label{sec:observations}
In this section, we present the data sets that are used for this analysis. First, we explain the catalogues used to target galaxies tracing BAO, and we detail the selection functions for the BAO tracers. Then we describe the dark matter halo simulation we used. Finally, we describe the catalogues containing the galaxy shape measurements.

%\subsection{Observations}
\subsection{SDSS-Stripe 82 galaxy catalogue}

The SDSS \citep{2000AJ....120.1579Y,2011AJ....142...72E}, delivered under the Data Release 9 (DR9; \citealt{2012ApJS..203...21A}), covers 14 555 deg$^2$ in the five photometric bands \emph{u, g, r, i, z}. It is the largest volume multi-colour extragalactic photometric survey available today. The 3$\sigma$ magnitude depths are: $u=22.0$, $g=22.2$, $r=22.2$, $i=21.3$; see \citet{1996AJ....111.1748F} for the description of the filters, \citet{1998AJ....116.3040G} for the characteristics of the camera, and \citet{Gunn_2006} for the description of the 2.5m telescope located at Apache Point Observatory (New Mexico, USA).

The data set used here for the selection of the different BAO tracers is the SDSS deep Stripe 82 co-add \citep{2011arXiv1111.6619A}. It covers an area of 225 deg$^2$ in one continuous stripe $-1^{\circ}.25<\delta_\mathrm{J2000}<1^{\circ}.25$, $-34^{\circ}<\alpha_\mathrm{J2000}<46^{\circ}$ with five photometric bands \emph{u, g, r, i, z}. We also use the infrared counterpart of this area observed by {\it Wide-field Infrared Survey Explorer (WISE}) at 3.4$\mu$m \citep{2010AJ....140.1868W}. Table \ref{selectionAll:tab} gives an overview of the tracers selected in this analysis, see section \ref{subsec:BAO:trcer:selec}.

% -------------- Table Clustering 2D ==================
\begin{table}
	\caption{Samples of BAO tracers used in this analysis, `b' stands for bright and `f' for faint. $\bar{z}$ is the mean redshift of the sample, $\sigma_z$ is the dispersion of the redshift distribution, $N$ the total amount of tracers in the sample, $N_S$ the density of tracers per deg$^{2}$, `$\bot$ scale' is the transverse scale in $h^{-1}$ Mpc corresponding to 1$^\circ$ at $\bar{z}$.}
	\label{selectionAll:tab}
	\centering
	\begin{tabular}{l c c r r c}
	\hline \hline
	sample & $\bar{z}$ &$\sigma_z$& $N$ & $N_S$ & $\bot$ scale \\
			& &	&	&(deg$^{-2}$) & ($h^{-1}\,$Mpc deg$^{-1}$) \\
	\hline
CMASS & 0.53 & 0.10 & 22k & 98 & 16.0\\

LRG-{\it WISE} b & 0.58 &0.12& 15k & 70 & 17.8 \\
LRG-{\it WISE} f & 0.69 &0.13 & 59k & 275 & 18.5 \\

ELG {\it gri} b & 0.80 & 0.08 & 119k & 530 & 19.2 \\
ELG {\it gri} f & 0.81 &0.15 & 226k & 1007 & 19.3 \\

ELG {\it ugri} b & 0.95 & 0.24 & 156k & 693 & 20.3 \\
ELG {\it ugri} f& 0.94 & 0.22 & 436k & 1938 & 20.2 \\

ELG {\it ugr} b& 1.28 &0.36& 77k & 341  & 21.5\\ 
ELG {\it ugr} f & 1.26 &0.34& 237k & 1052  & 21.5\\
	\hline
	\hline		
	\end{tabular}
\end{table}

\begin{figure}
\begin{center}
\includegraphics[width=88mm]{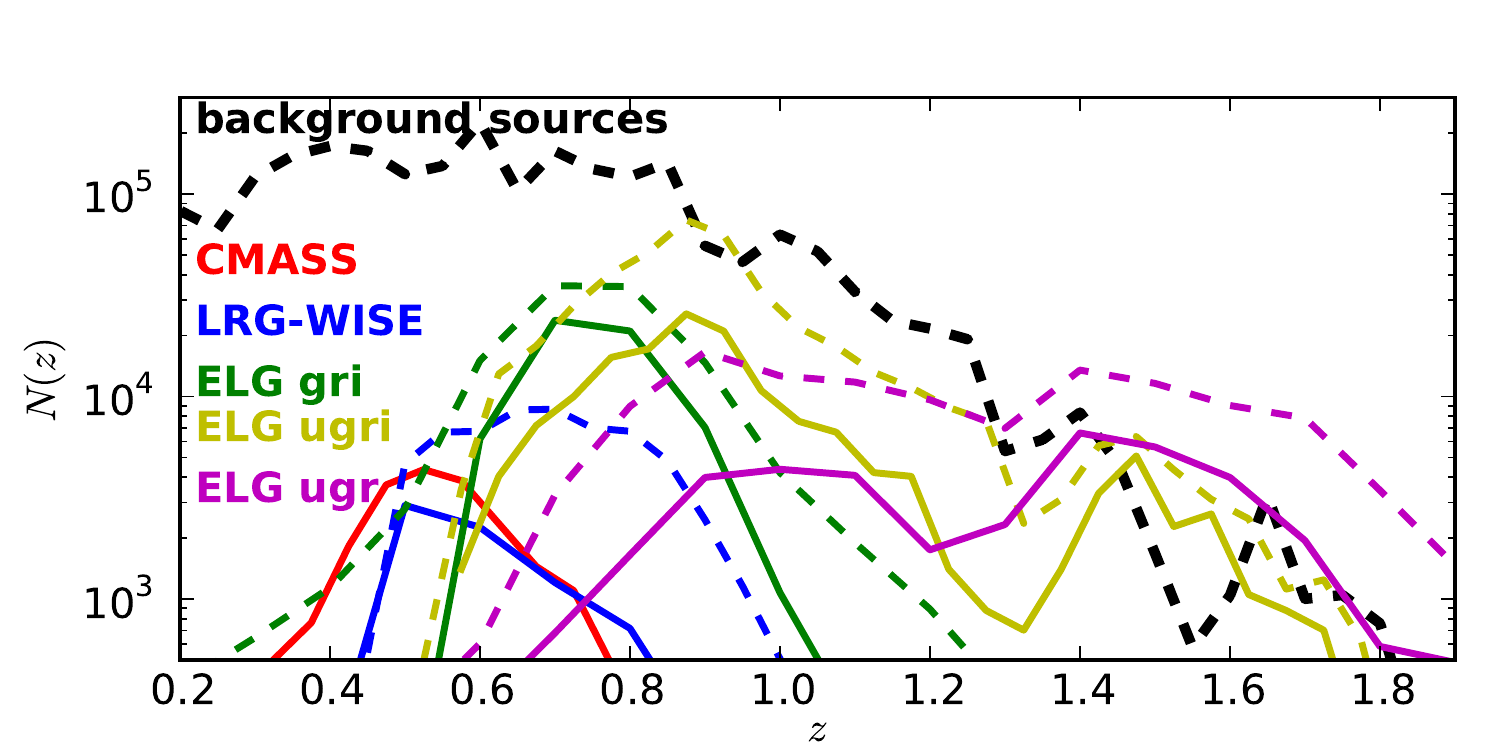}
\caption{Redshift distributions of the samples. For each sample, solid (dashed) lines refer to the bright (faint) sample. 22 000 CMASS (red), 15 000 (59 000) LRG {\it WISE} (blue), 119 000 (226 000) \emph{gri}  ELG (green), 156 000 (436 000) \emph{ugri} (yellow), 77 000 (237 000) \emph{ugr} ELG (magenta). The dashed black line represents the redshift distribution of the background sources used in the WL analysis.}
\label{fig:nz:tracers}
\end{center}
\end{figure}

The photometric redshift distribution of the BAO tracers is obtained using the Canada-France-Hawai Telescope (CFHT)-LS Wide W4 field photometric redshift catalogue T0007. The photometric redshift accuracy is estimated to be $\sigma_z < 0.08 (1+z)$ for $i\leq 24$. The data and cataloguing methods are described in \citet{Ilbert_06} and \citet{Coupon_2009}, and the T0007 release document\footnote{http://www.cfht.hawaii.edu/Science/CFHTLS/}.
The CFHT-LS W4 field overlaps the Stripe 82 on a few square degrees. Both data sets have similar depths in magnitude. From a position match of the targets selected on the stripe 82, we obtain a subcatalogue of the targets with photometric redshifts, from which we infer the redshift distribution of the selections. The redshift distributions are shown in Fig. \ref{fig:nz:tracers}.%Given the errors mentioned above, the redshift bins are set at $dz=0.2$ for the analysis. 

With current data, it is not feasible to obtain a reliable photometric redshift on the complete Stripe 82. Therefore, we cannot construct clean volume-limited samples, as required by the HOD analysis. We can only make magnitude-limited samples. The impact of the galaxies located in the tails of redshift distribution on the angular clustering is discussed in Section \ref{subsec:hodFits}.

% FIGURE ----------  redshift distribution of tracers

%\subsection{BAO Tracers, selections}
\subsection{BAO tracer selections}
\label{subsec:BAO:trcer:selec}

The latest BAO tracers used at $z>0.5$ are SDSS-III Constant Mass galaxies (CMASS) at $z \simeq0.53$ \citep{2013AJ....145...10D} and WiggleZ emission-line galaxies at $z\simeq0.6$ \citep{Drinkwater_2010}. The next stage BAO experiments plan to use luminous red galaxies (LRG) and emission line galaxies (ELG) at $z>0.7$ \citep{bigBOSS_2011}. The target selection of ELG tracers is extensively discussed in \citet{2013MNRAS.428.1498C}.

\begin{figure*}
\begin{center}
\includegraphics[width=180mm]{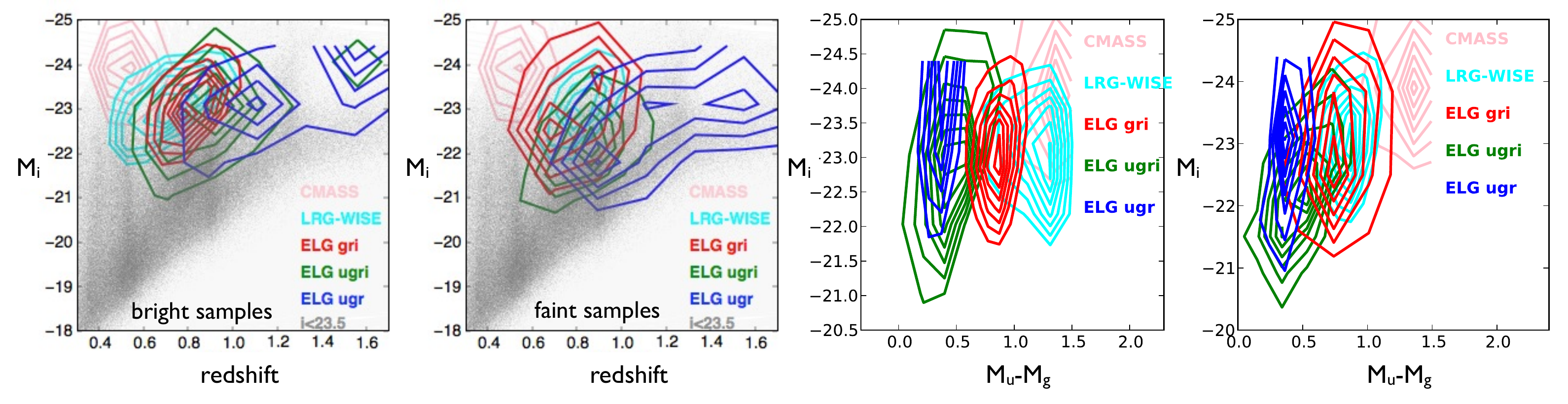}
\caption{Absolute magnitude $M_i$ versus redshift and rest-frame $M_u-M_g$. All galaxies with $i<23.5$ (grey), CMASS (pink), LRG {\it WISE} (light blue), \emph{gri} ELG (red), \emph{ugri} ELG (green), \emph{ugr} ELG (blue). 
First panel shows the bright selections are quite luminous with $M_i\sim-23$.
The second panel shows the faint selections are less luminous than the bright selections with $M_i\sim-22$. 
The third panel shows the bright selections span the complete rest-frame colour range from \emph{ugr} ELG (blue) to the CMASS (pink). 
The fourth panel shows the faint selections. Their rest frame colours are similar to that of the bright selections, except for the LRG-{\it WISE} faint galaxies that are bluer than the bright sample. 
The \emph{ugri} ELG overlap with the \emph{gri} and \emph{ugr} ELG selections, which is expected as the selection encompasses galaxies from both other selections.
This figure was made with the same CFHT W4 matched sample as for the redshift distributions.}
\label{fig:Miz:tracers}
\end{center}
\end{figure*}

The selection of the tracers used in this analysis is only based on photometric criteria, depth and colour. We do not use the photometric redshifts for the selection as their quality varies strongly across the stripe. The selection criteria target the most luminous galaxies in incremental redshift intervals to $z=2$; 
see Fig. \ref{fig:Miz:tracers} which compares all samples in absolute magnitude $M_i$ versus redshift or rest-frame $M_u-M_g$. 
The cleanest way to select BAO tracers would be to select by stellar mass, but with current data this is not possible.

% FIGURE ----------  MI vs. redshift

%\subsection{CMASS at $z\simeq0.53$}
\subsubsection{CMASS at $z\simeq0.5$}
The CMASS tracer is the SDSS-III/BOSS experiment BAO tracer \citep{2013AJ....145...10D}. The current BAO detection using the Data Release 9 \citep{2012ApJS..203...21A} (a third of the observation plan) with the CMASS tracers has a $6.7\sigma$ significance \citep{2012MNRAS.427.3435A}. Its selection is described in detail in \citet{2013AJ....145...10D}. The parent catalogue of CMASS selection on Stripe 82 contains 22 034 tracers, i.e. $\sim98$ deg$^{-2}$. We use the complete CMASS selection, not only the galaxies confirmed by spectroscopy, in order to avoid fibre collision issues. The redshift distribution is shown by the red solid line in Fig \ref{fig:nz:tracers}. The mean redshift is 0.53 with a dispersion of 0.1. We use snapshot number 62 of the multidark simulation. The density of the tracer at $\bar{z}$ is $2.64\times 10^{-4} h^3$Mpc$^{-3}$, which corresponds to a circular velocity cut of $V_\mathrm{cir}=384.1$ km s$^{-1}$. This value of $V_{cir}$ is consistent with the study done by \citet{2013MNRAS.tmp.1188N}. The halo mass distribution peaks at $\log_{10}{M_{200}/M\odot}=13.17_{-0.19}^{+0.34}$. The minimum satellite fraction expected is 5 per cent.

%\subsubsection{LRG WISE-selected at $z\simeq0.67$}
\subsubsection{LRG-{\it WISE} at $z\simeq0.65$}
This selection is the extension in redshift of the red CMASS selection. We name this sample the `luminous red galaxies selected with {\it WISE}' (LRG-{\it WISE}) (see \citealt{bigBOSS_2011}, Prakash et al., in preparation). On Stripe 82, it selects 15 643 galaxies, i.e. a density of 70 deg$^{-2}$ at a mean redshift of 0.58 for the bright sample with $i<20$ and $\sim59000$ galaxies at a mean redshift of 0.69 for the faint sample with $i<21$, i.e. a density of 262 deg$^{-2}$. The selection is as follows:
\begin{itemize}
\item we use dereddened \emph{r, i, z} and 3.4$\mu$m ({\it WISE} 1) magnitudes;
\item SDSS $z$-band flag must be true, and {\it WISE} channel 1 flag must be 0;
\item objects with extreme astrometric uncertainties were rejected;
\item bright selection: ($i < 20 \; \mathrm{or} \; z < 20$) $and$ $r-i >1$ $and$ $r-W1 > 2(r-i) -0.5$ $and$ $r>16$ $and$ $r<22.5$;
\item faint selection: ($i < 21 \; \mathrm{or} \; z < 21$) $and$ $r-i >1$ $and$ $r-W1 > 2(r-i) -0.5$ $and$ $r>16$ $and$ $r<22.5$.
\end{itemize}
The redshift distribution of the bright (faint) samples are shown with a solid (dashed) blue line in Fig. \ref{fig:nz:tracers}. Their densities are $1.57$ and $5.06\times 10^{-4} h^3$Mpc$^{-3}$, respectively.  We use the multidark snapshot 58 for the bright and the faint samples SHAM. The samples correspond to cuts at $V_\mathrm{cir}=431.5$ and $323.6$ km s$^{-1}$, respectively. The halo mass distributions peak at $\log_{10}{M_{200}/M\odot}=13.28_{-0.17}^{+0.31}$ and $12.92_{-0.2}^{+0.36}$, respectively. They constitute an upper limit for the halo mass these tracer inhabit. The minimum satellite fractions expected are 4 and 7 per cent, respectively.

%\subsubsection{ELG \emph{gri} at $z\simeq0.78$}
\subsubsection{ELG \emph{gri} at $z\simeq0.8$}
The \emph{gri} ELG selection \citep{2013MNRAS.428.1498C} is an intermediate selection between the reddest galaxies selected in SDSS-III and the bluest ones selected by the $ugr$ and $ugri$ schemes (see below), identifies emission-line galaxies with emission lines a small Balmer break. On Stripe 82, it selects $\approx119\,000$ and $226\,000$ galaxies for the bright and faint samples, which correspond to a projected density of about 530 and 1000 deg$^{-2}$ at mean redshift $z \simeq 0.8$ and 0.81, respectively. The bright sample is limited by $19<i<21.5$, and the faint sample by $19<i<22$. The densities are respectively 8.16 and 13.37 $\times 10^{-4} \,h^3$Mpc$^{-3}$. We use multidark snapshot number 56 to perform the SHAM. The samples match with a circular velocity cut of 292.2 and 250.7 km s$^{-1}$. The halo mass distributions peak at $\log_{10}{M_{200}/M\odot}=12.77_{-0.21}^{+0.37}$ and $12.34_{-0.25}^{+0.44}$, respectively. They constitute an upper limit for the halo mass these tracer inhabit. The minimum satellite fractions expected are 8 and 10 per cent, respectively.

%\subsection{ELG \emph{ugr} at $z\simeq1.23$}
\subsubsection{ELG \emph{ugr} at $z\simeq1.2$}
The \emph{ugr} ELG selection \citep{2013MNRAS.428.1498C} selects the blue tail of the colour distribution aiming for galaxies with strong emission lines at higher redshift ($z\simeq1.2$). The \emph{ugr} redshift distribution is flatter and extends to higher redshifts. The bright sample is limited by $20<g<22.5$, and the faint sample by $20<g<23$. For the bright and faint sample it selects, $77\,000$ and $237\,000$ galaxies, corresponding to a density of 341 and 1052 deg$^{-2}$, respectively. Their mean redshift are at 1.28 and 1.26. We use multidark snapshot number 48. The bright sample density, $1.45\times 10^{-4} h^3$Mpc$^{-3}$, corresponds to a cut at 434.5 km s$^{-1}$, which gives an upper limit of $\log_{10}{M_{200}/M\odot }\lesssim 13.21_{-0.16}^{+0.29}$. The faint sample density, $5.76\times 10^{-4} h^3$Mpc$^{-3}$, corresponds to a cut at 315.0 km s$^{-1}$, which gives an upper limit of $\log_{10}{M_{200}/M\odot}\lesssim 12.81_{-0.19}^{+0.34}$. The minimum satellite fractions expected are 3 and 7 per cent, respectively.

%\subsection{ELG \emph{ugri} at $z\simeq0.96$}
\subsubsection{ELG \emph{ugri} at $z\simeq0.95$: building SDSS-IV/eBOSS}
The \emph{ugri} ELG selection is designed to be the BAO tracer sample for SDSS-IV/eBOSS project (eBOSS collaboration, in preparation). It is a mix of the two previous \emph{gri} and \emph{ugr} ELG selections. The selection criteria are
\begin{itemize}
\item $20<g<22.5$ for the bright; $20<g<23$ for the faint
\item $-0.5<u-r<0.7(g-i)+0.1$ and $g-i>-0.5$
\end{itemize}
The mean redshifts are 0.95 and 0.94. The bright sample contains $156\,000$ galaxies and the faint sample $436\,000$ (i.e. densities of 693 and 1938 deg$^{-2}$). We use multidark snapshot number 52 for the SHAM. The densities are respectively 5.44 and 16.45 $\times 10^{-4} h^3$Mpc$^{-3}$, corresponding to mass upper limits of $\log_{10}{M_{200}/M\odot}\lesssim 12.84_{-0.19}^{+0.35}$ and $\lesssim 12.48_{-0.23}^{+0.4}$. The minimum satellite fractions expected are 7 and 11 per cent, respectively.

%\subsection{Multidark simulation, halo abundance matching}
\subsection{Multidark-Stripe 82 halo catalogue}

We use the Multidark dark matter simulation to characterize qualitatively our samples to the first order using SHAM. To perform the match, we use snapshots of the Multidark simulation corresponding to the mean redshift of our tracers. Every snapshot occupies a volume of 1 $h^{-3}$Gpc$^3$ filled with haloes obtained with bound density maximum algorithm with a cut-off at 200 times the critical density ($\log_{10}{M_\mathrm{200}/M\odot}$). These data sets are taken from the data base Multidark\footnote{http://www.multidark.org/} \citep{2011arXiv1109.0003R}. 

In each snapshot, we select the dark matter halo population that corresponds to the number density of galaxies with $\bar{z}-\sigma_z<z<\bar{z}+\sigma_z$ selected on the Stripe 82. This links each galaxy sample to a halo sample. The density, velocity cuts and characteristic masses and satellite fractions are given for each sample in Table \ref{tab:description:ham}. 

The SHAM brings two components to this analysis. 
Under the hypothesis that the galaxy sample is complete in mass. The abundance matched subhalo sample has a mass distribution characterized by the three quartiles of $\log_{10}{M_\mathrm{200}/M\odot}$ and a satellite fraction $f_\mathrm{sat}$.
If the galaxy sample is not complete in mass, the mass distribution of the subhaloes the galaxies live in will be shifted towards lower masses and the satellite fraction will increase.
As we cannot demonstrate that the BAO tracers selected are complete in mass and that they are tracing the most massive structure at their redshift, we consider the subhalo mass scales obtained with the SHAM procedure are an upper boundary, and that the satellite fraction is a lower limit to the true values.
Moreover, it provides the mass distribution of the subhaloes. 
Given that we measure clustering over $\sim200$ deg$^2$ and WL on $\sim150$ deg$^2$ and not the full sky, 
we check the mean halo mass deduced from the maximum likelihood HOD model is between the first and third quartile of the sub-halo mass distribution given by SHAM.

From the qualitative numbers obtained by the SHAM, the LRG-{\it WISE} faint sample appears to be located in massive subhaloes of $\sim10^{13}M\odot$ with a low satellite fraction, similarly to the CMASS tracers. Concerning the ELG, the satellite fraction are higher and the subhalo masses are lower: $\lesssim10^{12.8}M\odot$ for the bright samples and $\lesssim10^{12.6}M\odot$ for the faint samples.

% -------------- Table Clustering 2D ==================
\begin{table}
	\caption{Abundance matching of the samples of BAO tracers used in this analysis, `b' stands for bright and `f' for faint. The density of tracers is in units of $10^{-4}\,h^{3}$Mpc$^{-3}$. $V_\mathrm{cir}$ or $V_\mathrm{max}$ is the velocity cut applied to the multidark simulation snapshot to match the density $d$. The obtained sample is characterized by the three quartiles of its distributions given in the columns $\log_{10}{M_\mathrm{200}/M\odot}$. $f_\mathrm{sat}$ is the fraction of satellites in the halo catalogue.}
	\label{tab:description:ham}
	\centering
	\setlength{\extrarowheight}{3pt}
	\begin{tabular}{l r c c c c r}
	\hline \hline
	Sample & Density& \multicolumn{1}{c}{$V_\mathrm{cir}$} & \multicolumn{1}{c}{$\log_{10}{M_\mathrm{200}/M\odot}$}&$f_\mathrm{sat}$\\ %&mass4
			&$d$&(km s$^{-1})$&$M_{d}$&(per cent)\\
	\hline
CMASS   & 2.64 & 384& $13.17_{-0.19}^{+0.34}$&5\\

LRG-{\it WISE} b  &  1.57   & 431 &  $13.28_{-0.17}^{+0.31}$&4\\ 
LRG-{\it WISE} f   &  5.06  	&324 &  $12.92_{-0.2}^{+0.36}$ &7\\ 

ELG {\it gri} b  &  8.16 	&292 &  $12.77_{-0.21}^{+0.37}$ &8\\
ELG {\it gri} f  &  13.37  	&251 &  $12.59_{-0.23}^{+0.41}$ &10\\

ELG {\it ugri} b  & 5.44 & 317 &  $12.84_{-0.19}^{+0.35}$ &7 \\
ELG {\it ugri} f &  16.45  & 238 &  $12.48_{-0.23}^{+0.4}$ &11 \\
ELG {\it ugr} b & 1.45& 435 &  $13.21_{-0.16}^{+0.29}$ &3 \\
ELG {\it ugr} f & 5.76	&315 &  $12.81_{-0.19}^{+0.34}$ &7\\
	\hline
	\hline		
	\end{tabular}
\end{table}

%\subsection{CFHT-Stripe 82 weak-lensing catalog}
\subsection{CFHT-Stripe 82 weak-lensing catalogue}

The existing comprehensive data sets available on the SDSS Stripe-82 area were complemented with high-quality $i$'-band observations at CFHT. The main purpose of this one-band Megaprime@CFHT (see \citealt{2003SPIE.4841...72B}) follow-up survey is to enrich already existing multi-colour data with deep optical observations suitable for WL studies. The CFHT Stripe-82 program (CFHT/Stripe 82) is a large collaborative effort between the Canadian/French and Brazilian communities\footnote{The survey CFHT identification numbers are 10BB09, 10BC22, and 10BF23}. The observations encompass 169 MegaPrime new pointings resulting in an effective survey area of about 150 deg$^2$. This area comprises 169 -- 4(due to fringing) = 165 pointings CFHT/Stripe 82 and eight CFHT-LS Wide pointings, i.e. a total of 173 pointings. All the data were obtained under excellent seeing conditions, and the final co-added images show an image quality between 0.4 and 0.8 arcsec with a median of 0.59 arcsec. Each pointing was obtained in four dithered observations with an exposure time of 410s, each resulting in a 5-$\sigma$ limiting magnitude in a 2 arcsec diameter aperture of about 24.

The data processing, which starts with the Elixir preprocessed images obtained from CADC and ends with co-added images and all necessary quantities to perform weak gravitational lensing studies, follows closely the descriptions given in \citet{2012arXiv1210.8156E}.

We exploit the opportunity that the CFHTLS includes the {\it Hubble Space Telescope HST} COSMOS survey field, and verify the calibration of our shear measurement pipeline for ground-based data against an independent analysis of much higher resolution space-based data. We stack the subsets of the CFHTLS DEEP2 imaging to the same depth as the CFHT Stripe 82 survey and analyse it using KSB method \citep{1995ApJ...449..460K}. 
The KSB method is used to measure the shapes of all the selected galaxies. Our implementation of KSB is based on the KSBf90\footnote{\tt http://www.roe.ac.uk/$\sim$heymans/KSBf90/Home.html} pipeline \citep{2006MNRAS.371..750H}. This has been generically tested on simulated images containing a known shear signal as part of the Shear Testing Programme (STEP; \citet{2006MNRAS.368.1323H,2007MNRAS.376...13M}) and the Gravitational lensing Accuracy Testing challenge (GREAT08; \citet{2010MNRAS.405.2044B}; GREAT10: \citet{2012MNRAS.423.3163K}). The systematic errors of this pipeline are well controlled in all cases. This pipeline has also been used on the shear measurement of CFHTLS-Wide W1 fields \citep{2012ApJ...748...56S}. Finally, we match galaxies in the CFHT/Stripe 82 catalogue to those of the HST COSMOS catalogue \citep{2010ApJ...709...97L} to calibrate the measurements. 

Here is the description of the calibration tests we made on the lensing catalogue. The multiplicative calibration component is $(m_1,m_2)=(0.13,0.14)$ and the additive component is $(c_1,c_2)=(0.0003,0.0015)$. These coefficients are obtained from simulations and connect through the linear relation between the observed galaxy ellipticities and the true ones 
\begin{equation}
e^\mathrm{obs} =(1+m)[\gamma+e^\mathrm{true}] +c
\end{equation}
They mainly depend on the algorithm that measures the shapes. 
If the PSF anisotropy is small, the shear $\gamma$ can be recovered to first-order from the observed ellipticity $e^{\rm obs}$ of each galaxies via
\begin{equation} \label{eqn:weight}
\gamma=P_{\gamma}^{-1}\left(e^{\rm obs}-\frac{P^{\rm sm}}{P^{\rm sm*}}e^{*}\right),
\end{equation}
where asterisks indicate quantities that should be measured from the PSF model interpolated to the position of the galaxy, $P^{\rm sm}$ is the smear polarisability, and $P_{\gamma}$ is the correction to the shear polarisability that includes the smearing with the isotropic component of the PSF. The ellipticities are constructed from a combination of an object weighted quadrupole moments, and the other quantities involve higher order shape moments. All definitions are taken from \citet{1997ApJ...475...20L} with the assumption $\kappa \sim 0$, i.e. the reduced shear reads $g=\frac{\gamma}{1-\kappa}\sim \gamma$.

In order to check for significant additive systematics we performed four tests.
\begin{enumerate}
%ITEM
\item We test the average shear $\langle \gamma \rangle$ across all 173 pointings of the CFHT/Stripe 82 field. Fig. \ref{fig:average} demonstrates that the average shear is almost consistent with zero as expected, for galaxies of all sizes, magnitudes and SNR. As in \citet{2012MNRAS.427..146H}, $\langle g_2\rangle$ is clearly dependent on the SNR (red points). We quantify the additive component of the calibration correction with a simple relation $\langle g_2 \rangle=a+b\,SNR$. We find a best fit reduced $\chi^2= 0.54$ with $a=-0.002, b=0.000118$. The corrected average shear is almost consistent with zero for galaxies of all sizes, magnitudes and SNR. Furthermore, the average shear of each pointing has also been tested, see Fig. \ref{fig:g1}. We find that the mean shear $\langle g_1 \rangle$ of pointing S82p24p is around $-0.0126 \pm 0.0017$, which is far from zero compared to other observation pointings. Systematic error residual signal may exist in this pointing. We remove the related galaxies in the final catalogue.

\begin{figure}
\includegraphics[width=80mm]{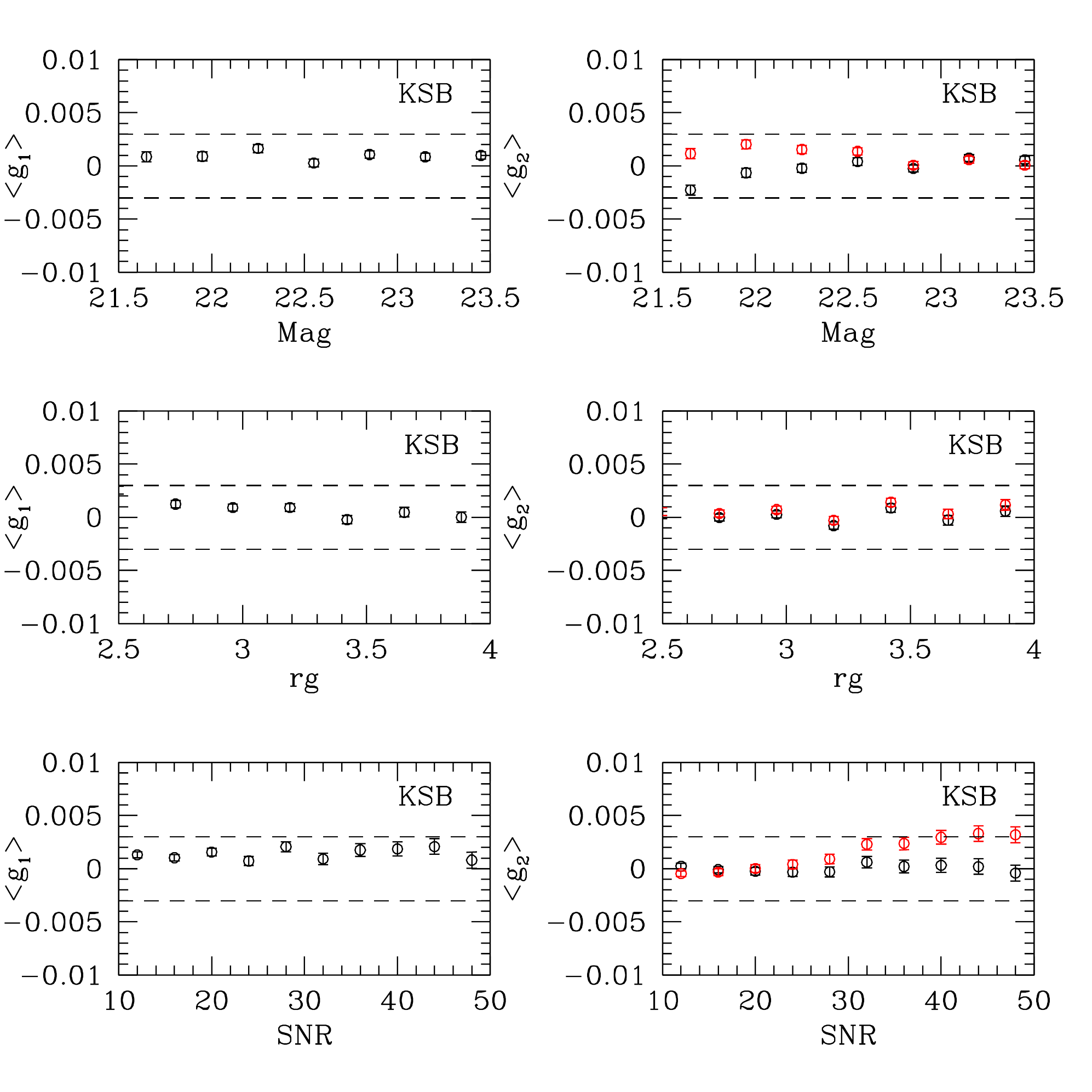}
\caption{Average shear measurements $\langle g_1\rangle$ (left) and $\langle g_2\rangle$ (right) of KSB from galaxies in $i'$-band observations of the entire CFHT/Stripe 82 field. In the absence of additive systematics, these should be consistent with zero. In practice, they always remain within the dashed lines that indicate an order of magnitude lower than the $1$--$10$ per cent shear signal around clusters. Top, middle and bottom panels show trends as a function of galaxy magnitude, size and SNR. The black points denote the average shear from all the galaxies in the two shear catalogues. The red points denote $\langle g_2\rangle$ before the correction described in (i).}
\label{fig:average}
\end{figure}

\begin{figure}
\begin{center}
\includegraphics[width=80mm]{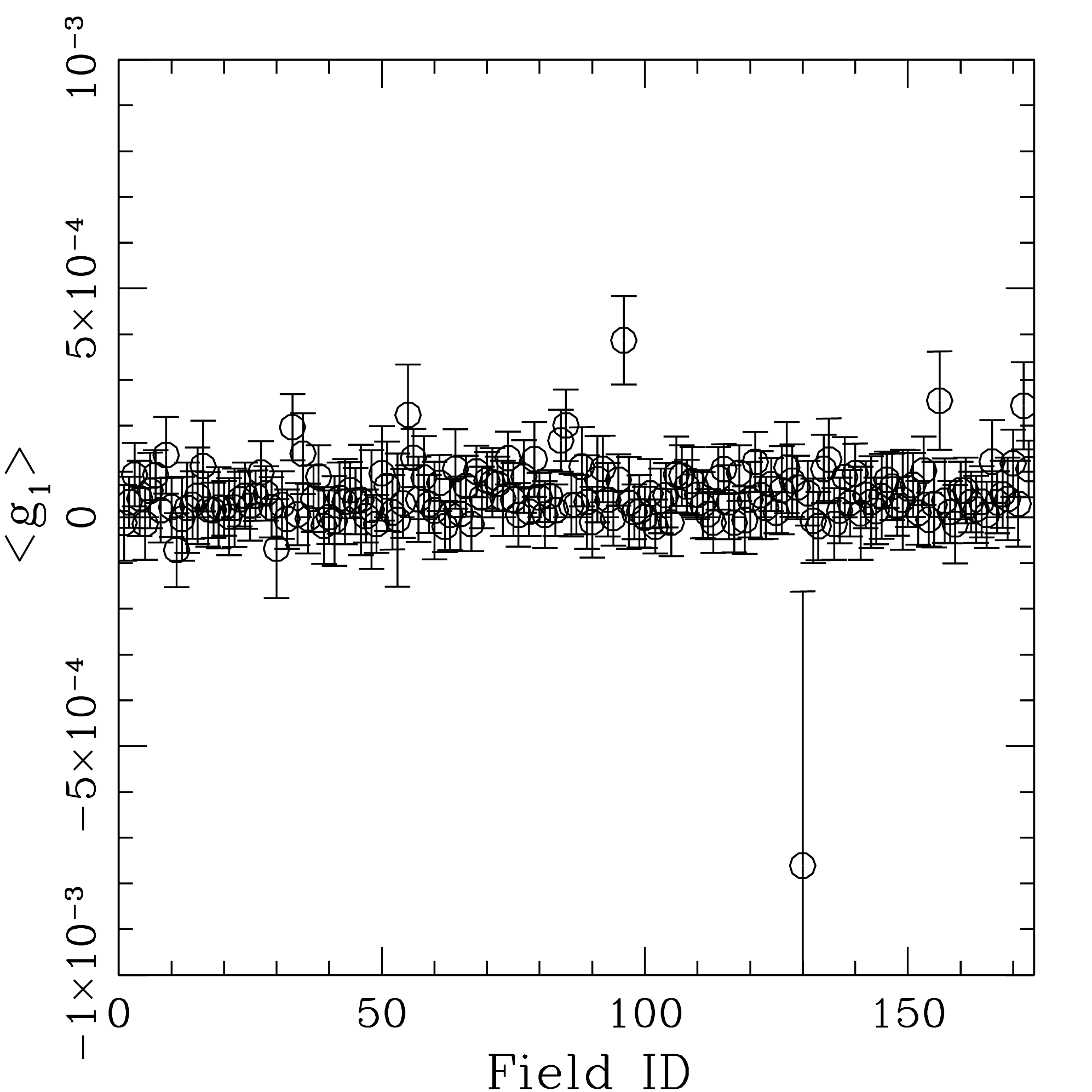}
\caption{Average shear $\langle g_1 \rangle$ of $173$ pointings, in which S82p24p field is biased from zero.}
\label{fig:g1}
\end{center}
\end{figure}
%ITEM
\item Star--galaxy cross-correlation (residual systematics due to
imperfect PSF).  For the total 173 pointings, we find our PSF correction is well within requirements  for our analysis. Fig. \ref{fig:sg} shows the correlation $\xi_\mathrm{sys}$ between the corrected shapes of galaxies and the uncorrected shapes of the stars. We normalize the star-galaxy ellipticity correlation by the uncorrected star-galaxy ellipticity correlation to assess its impact on shear measurements \citep{2003MNRAS.344..673B}; see Fig. \ref{fig:delta_xi}. 
Though, we find $\xi_\mathrm{sys}$ is obviously higher than $\xi_+$ in one pointing S82p35p at scale $\theta>3 \rm arcmin$, which reveals the presence of a systematic error residual signal in this pointing. We remove the related galaxies in the final catalogue (only the field with $\delta_{\xi}<0$ is excluded).
\begin{figure}
\begin{center}
\includegraphics[width=70mm]{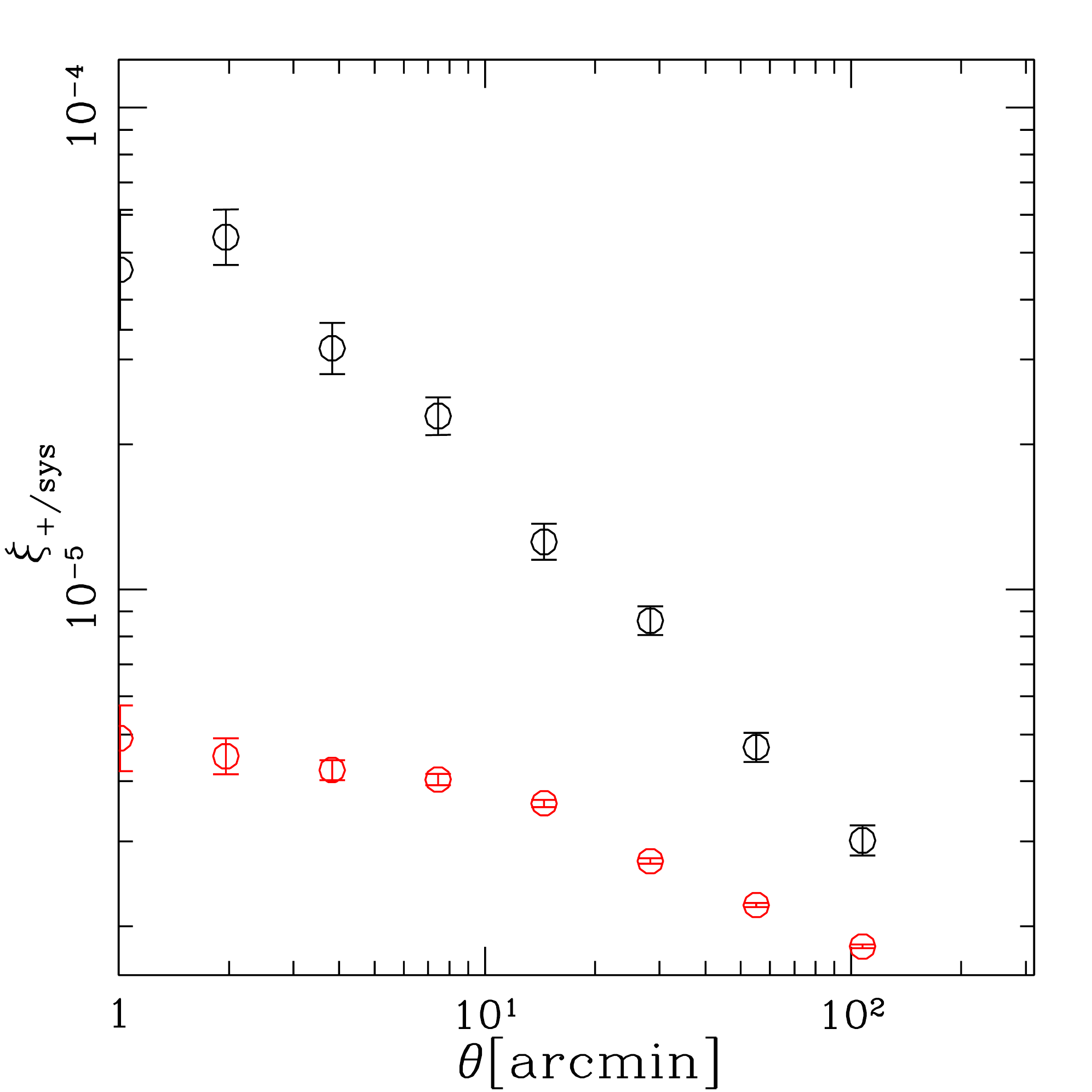}
\caption{The cross-correlation between shear measurements and stellar ellipticities as a function of the separation between galaxies and stars, averaged throughout the CFHTLS Stripe 82 field. The black and red points are shear correlation function and star-galaxy cross-correlation, respectively. If all residual influence of the observational PSF has been successfully removed from the galaxy shape measurements, the red points should be consistent with zero.\label{fig:sg}}
\end{center}
\end{figure}

\begin{figure}
\begin{center}
\includegraphics[width=80mm]{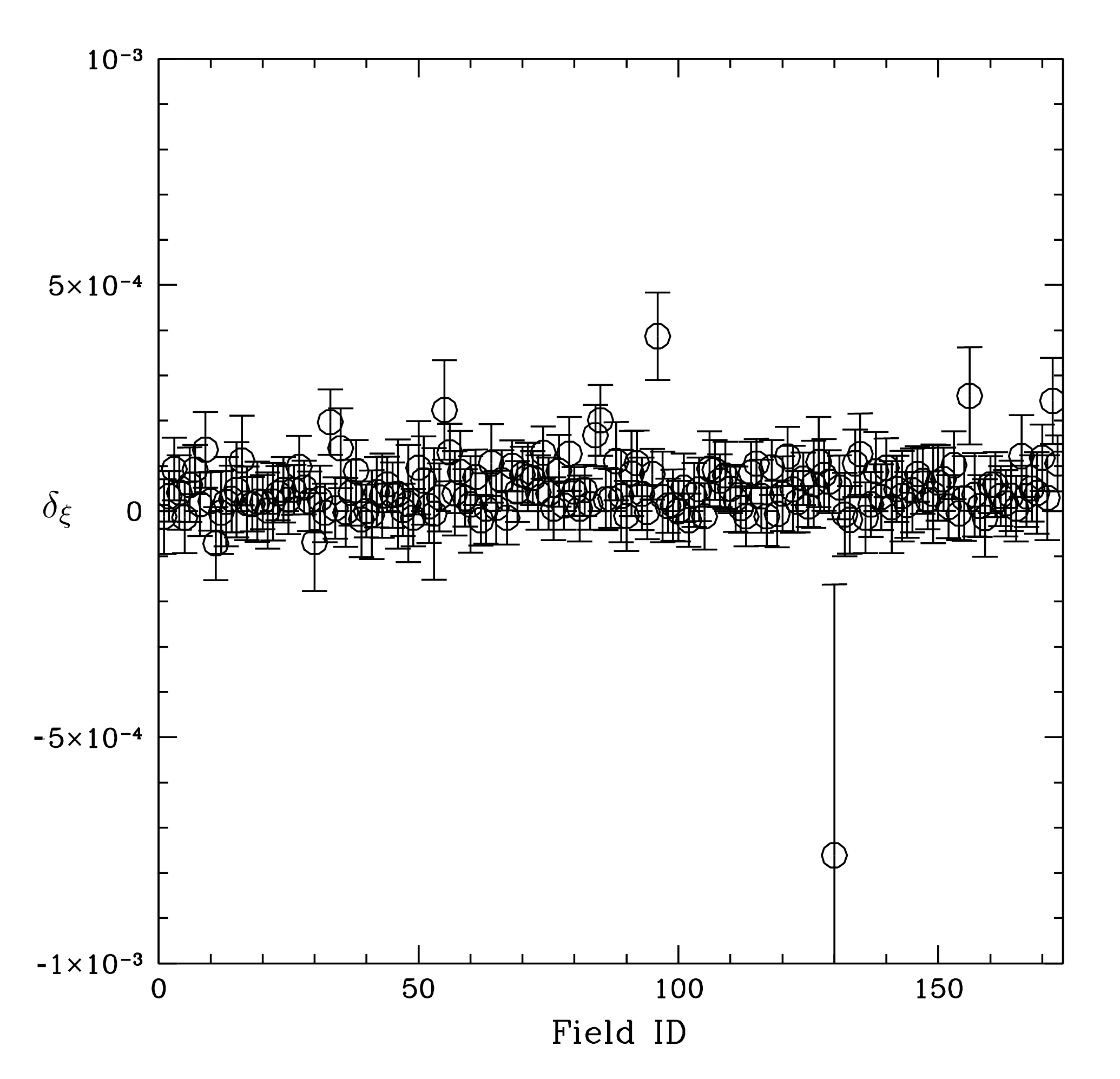}
\caption{Difference $\delta_{\xi}$ between shear autocorrelation $\xi_+$ and star--galaxy cross-correlation $\xi_{\rm sys}$ at scale $\theta=3 \rm arcmin$. The difference delta of the S82p35p pointing is higher than others. 
\label{fig:delta_xi}}
\end{center}
\end{figure}

%ITEM
\item Two-point shear correlation functions. We calculate the two-point shear correlation functions of the shear catalogue and compare them to an analytical prediction. The comparison between the analytical prediction of halo model and the measured $\xi_{+}$ is good; see Fig. \ref{fig:2pt}. Note the cross-correlation $\xi_{\rm tx}$ between the tangential and rotated shear should be zero. The theory appears to be systematically higher than the observation. This is due to the photometric redshift sample incompleteness.
\begin{figure}
\begin{center}
\includegraphics[width=80mm]{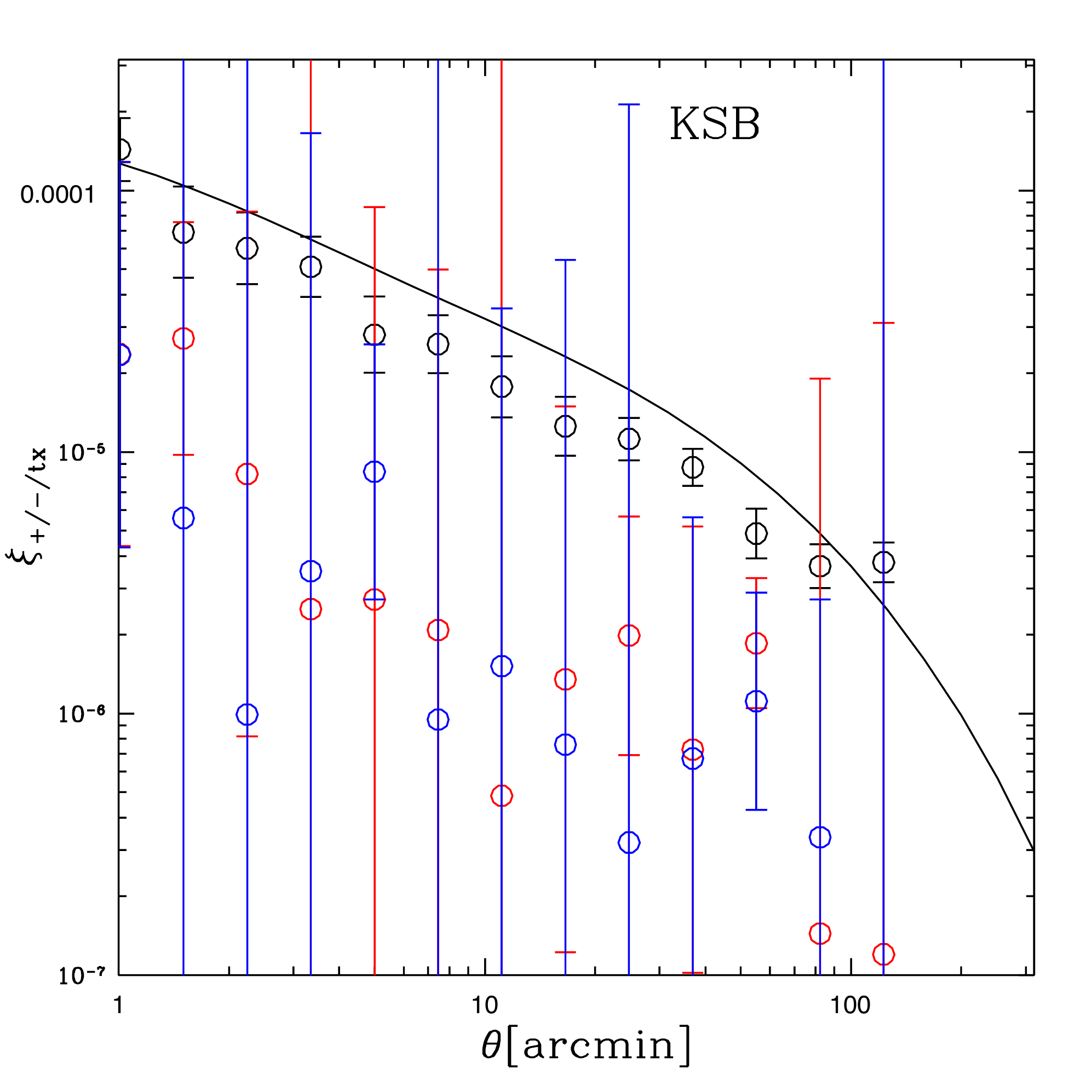}
\caption{The two-point shear correlation functions of the shear catalogue:
$\xi_+$ (black points), $\xi_-$ (red points), and $\xi_{\rm tx}$ (blue points). Solid lines show the analytical predictions of $\xi_+$. Note the cross-correlation $\xi_{\rm tx}$ between the tangential and rotated shear should be zero. The solid line is the theoretical prediction for a {\it WMAP}7 cosmology.
\label{fig:2pt}}
\end{center}
\end{figure}

\item $E,B$-mode signals. We test for the (nonphysical) $B$-mode signal. The $B$-mode signal corresponds to the imaginary component of $P_{\kappa}(\ell)$. We rotate all galaxy shears through $45^\circ$ and remeasure the $E$-mode signal. 
We measure that the $B$-mode signals are around zero except for small scales; see Fig. \ref{fig:map2}.%Pure gravitational fields produce zero $B$-mode for isolated clusters and only tiny $B$-modes through coupling between multiple systems along adjacent lines of sight. 
\begin{figure}
\begin{center}
\includegraphics[width=70mm]{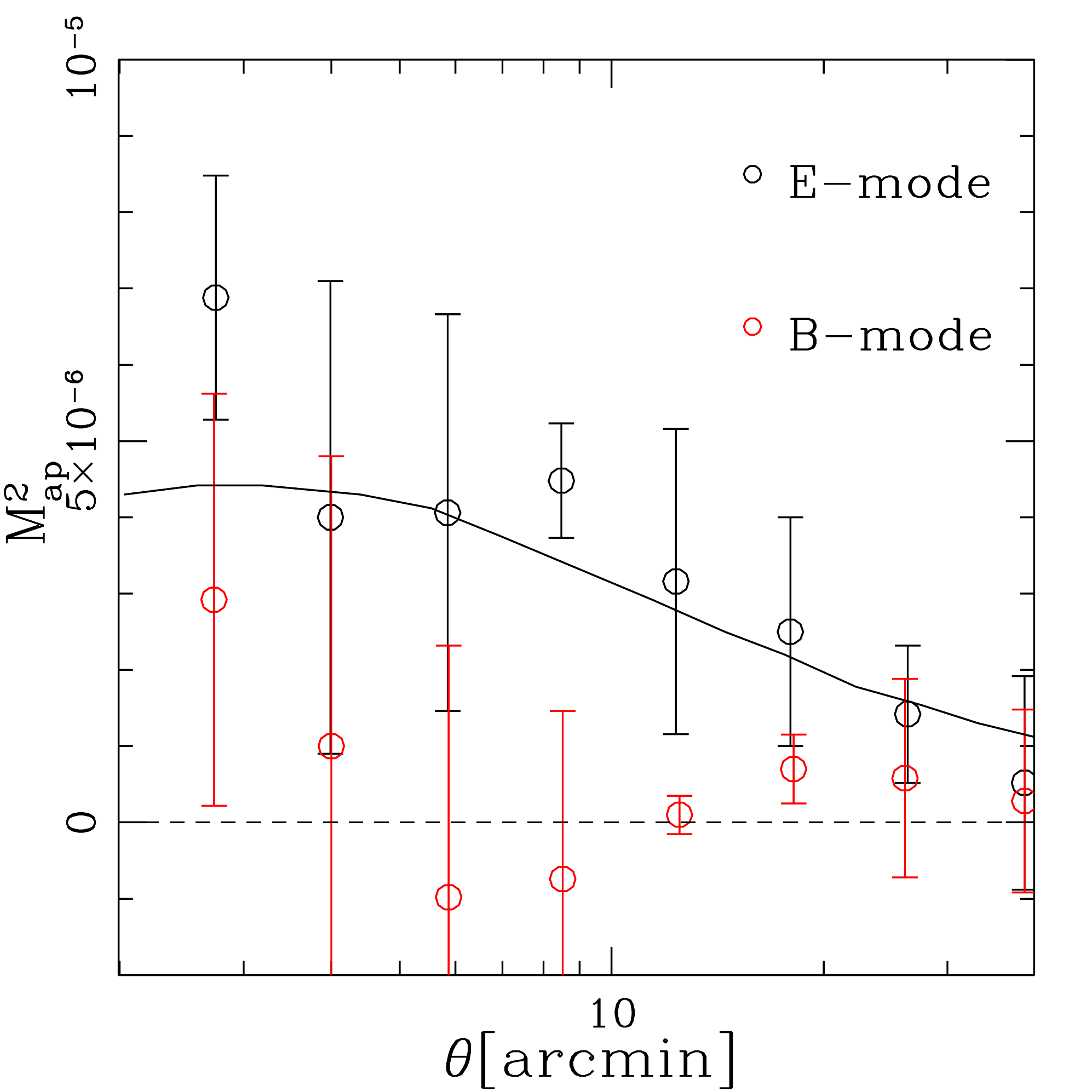}
\caption{The aperture mass from the entire CFHT/Stripe 82 field. Red points show the B-mode, black points show the E-mode. The error bars of the E-mode include statistical noise added in quadrature to the non-Gaussian cosmic variance. Only statistical uncertainty contributes to the error budget for the B-mode. The errorbars are $1\sigma$ confidence level. Within $2\sigma$ errors the B-modes are consistent with zero. The solid line is the theoretical prediction for a {\it WMAP} 7-cosmology.\label{fig:map2}}
\end{center}
\end{figure}
\end{enumerate}
For the lensing catalogue, we only select galaxies fainter than the foreground samples by 1 mag in $i$ band, so that we have no overlap between the two samples. The complete redshift distribution of the sources used for the WL analysis is displayed in Fig. \ref{fig:nz:tracers} with a thick dashed black line labelled `background sources'.

%--------------------------------------------------------------------------------------------------------------
%--------------------------------------------------------------------------------------------------------------
%------------------------------			measures				  ---------------------------
%--------------------------------------------------------------------------------------------------------------
%--------------------------------------------------------------------------------------------------------------
%\section{Measures}
\section{Measurements}
\label{sec:Measures}
In this section, we present the measurements made through clustering and WL analysis of the data. The correlation function measurements are displayed in Figs \ref{clustering:all}-\ref{clustering:all4}.
Figs \ref{bias_all:fig} and \ref{bias_all2:fig} show the galaxy bias measurement. Table \ref{hod:All:tab} summarizes the details of the measurements.

\subsection{Angular clustering measure and HOD fits}
\label{subsec:hodFits}
\begin{figure*}
\begin{center}
\includegraphics[width=40mm]{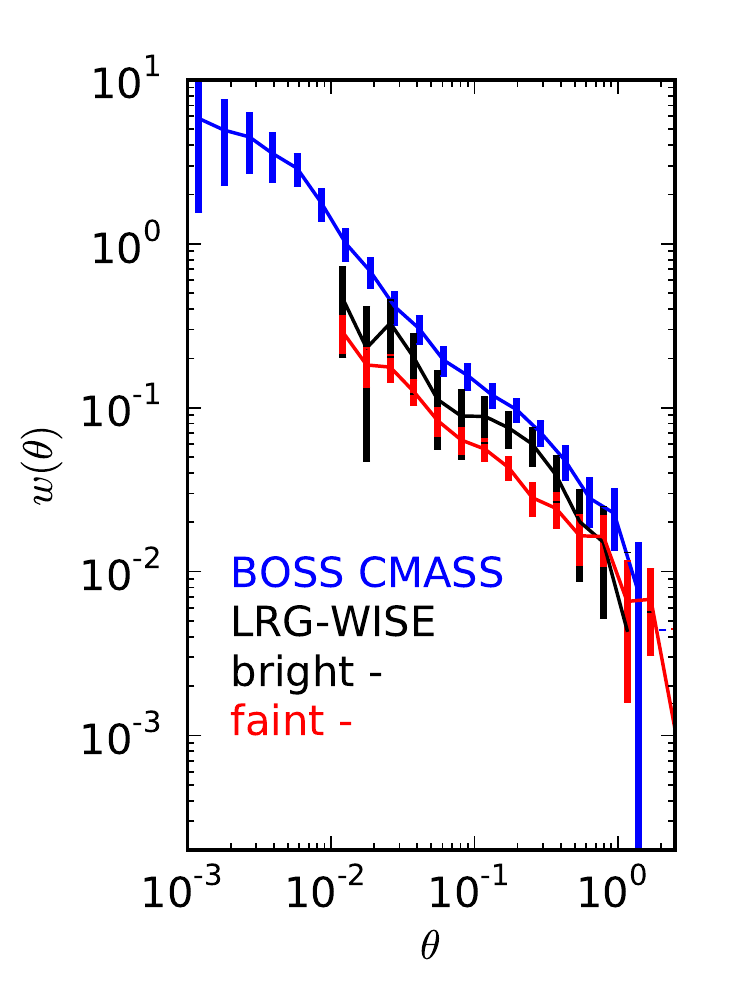}
\includegraphics[width=40mm]{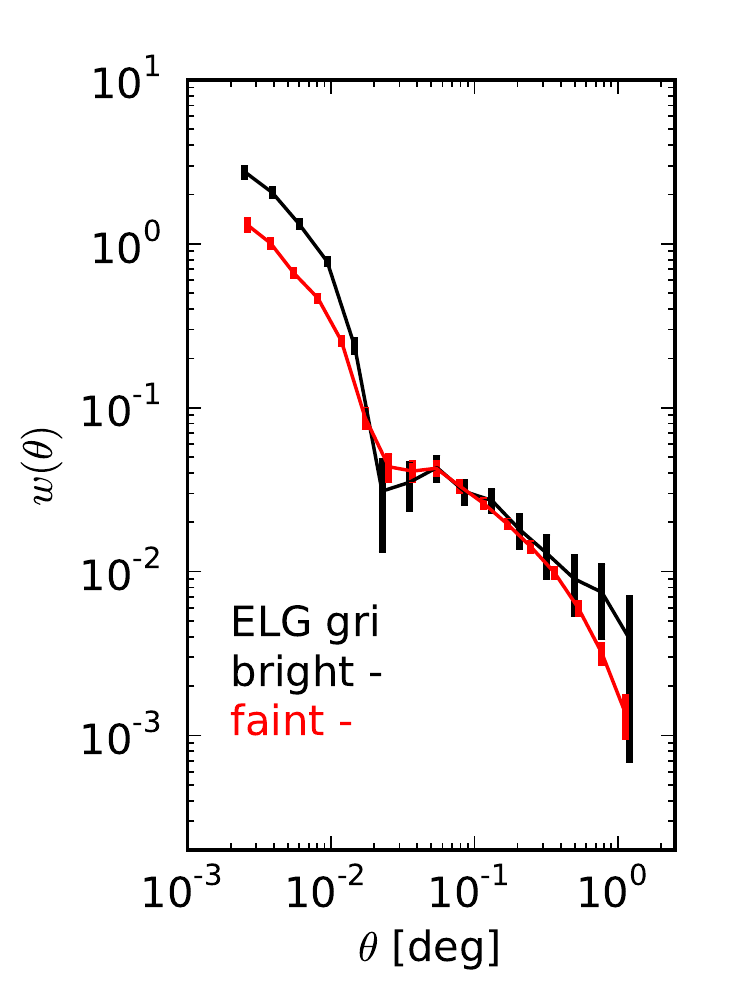}
\includegraphics[width=40mm]{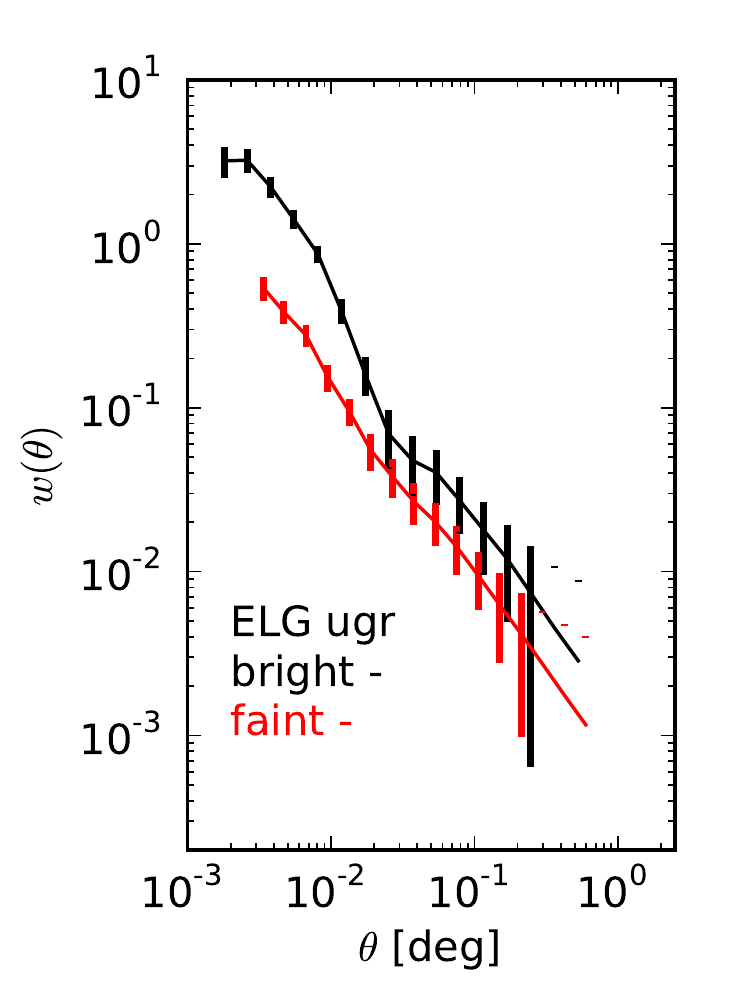}
\includegraphics[width=40mm]{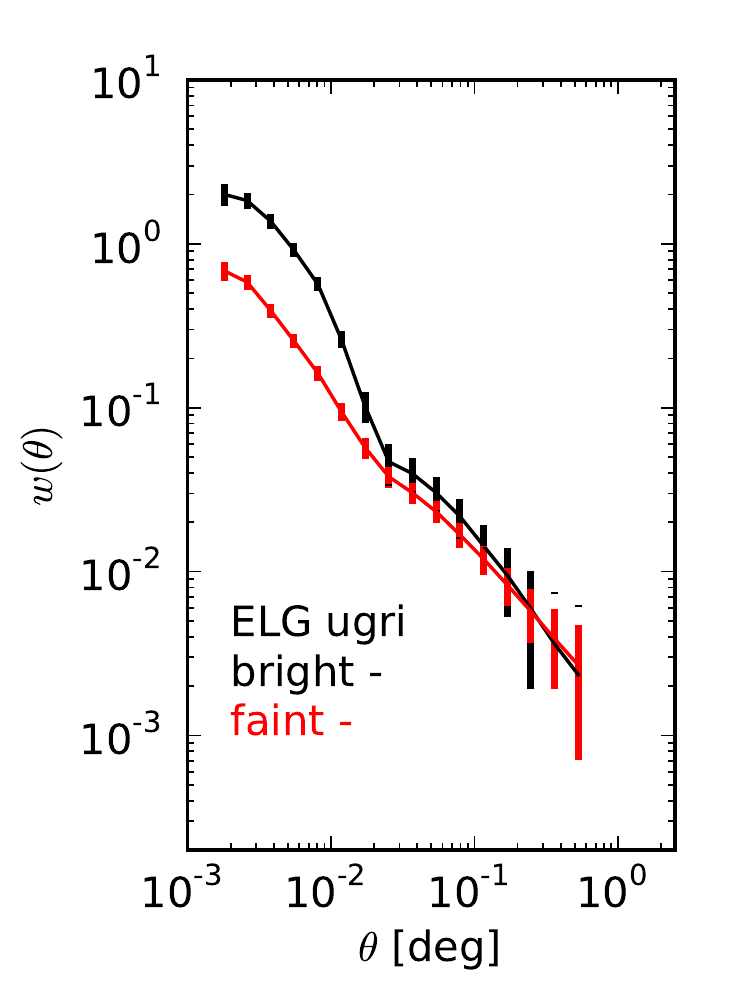}
\caption{Angular clustering of the selected tracers. The wide angular span of the clustering allows the  HOD analysis, except for the LRG-{\it WISE}, that are limited to $\theta>0.01^\circ$.}
\label{clustering:all}
\end{center}
\end{figure*}

We measure the angular clustering, denoted $w(\theta)$, for angles in the range 0.001$^\circ$ to 1.5$^\circ$. In this manner we avoid the signal due to blended sources. In the case of LRG-{\it WISE}, the clustering analysis starts at 0.01$^\circ$. We use between 15 and 20 angular bins regularly log-spaced. The errors and the covariance matrix are issued from 112 bootstrap realizations of the angular clustering measure. Fig. \ref{clustering:all} shows a comparison of the angular clustering of all previously defined samples. 

The most precise estimations of the angular correlation functions are between 0.01$^\circ$ and 0.1$^\circ$, due to a balance between the density of BAO tracers and the geometry of the survey. For the sparse tracers (the bright selections at low redshift), below 0.01$^\circ$, there are only a few pairs of tracers. For the densest tracers, the correlation function is precise down to 0.001$^\circ$. For all tracers above 0.1$^\circ$, given the geometry of the survey (a long stripe), the correlation function measurements are covariant. We correct from the integral constraint as in \citet{2012psa..book.....W}.

The main trends in the angular clustering measurements follow expectations. The lower redshift samples are more clustered than higher redshift samples. The bright samples are more clustered than the faint samples in particular at small scales.

To interpret the angular clustering, we use the HOD parametrization given in (\ref{hod:eqn}). From $N(M)$ we deduce the angle averaged 3D correlation function, that we project using the Limber equation \citep{1954ApJ...119..655L,2007A&A...473..711S} to obtain the angular autocorrelation function $w_\mathrm{HOD}$. The $\chi^2$ accounting for the accuracy of the fit is defined by equation (\ref{hod:chi2:def}):
\begin{equation}
\chi^2=(w_\mathrm{observed}-w_\mathrm{HOD})^\mathrm{T}  C^{-1}  (w_\mathrm{observed}-w_\mathrm{HOD}) + \frac{(N^\mathrm{gal}_\mathrm{mes}-N^\mathrm{gal}_\mathrm{mes})^2}{\sigma^2_\mathrm{Ngal}}.
\label{hod:chi2:def}
\end{equation}

The main source of error is the uncertainty on the number of galaxies representing the redshift distribution of each population. This uncertainty lies between 5 and 20 per cent for the bright tracers (CMASS and ELG bright). For the faint tracers it is between 30 and 40 per cent (error denoted $\sigma_\mathrm{Ngal}$). 
The second source of errors is the statistical errors on the measurement of $w(\theta)$, which is accounted for by the covariance matrix $C$.
Another source of errors is the cosmic variance, but it is small compared to other uncertainty sources. The fractional error on the measured density due to cosmic variance is $\sigma_\mathrm{cv} < 0.1$; see \citet{2002ApJ...564..567N}. We account for these when computing the error on the HOD derived parameters; see Table \ref{hod:All:tab}.

We perform additional tests to search for systematic errors. We test if the angular clustering measured is representative of the redshift distribution. We compute the angular clustering on only the tracers that have a reliable photoZ. We find a discrepancy on the angular clustering smaller by a factor of three than the statistical error on the angular clustering measurement. We therefore assume the angular clustering measurement is representative of the redshift distribution.

The validity of the use of an HOD model can be discussed. $N(M)$ is valid to predict the clustering when applied to a volume-limited sample. Current data does not allow us to construct a clean volume-limited sample, as we do not have good photometric redshifts for all the tracers. We cannot cut a clean sample with redshift $z\in\bar{z}\pm\sigma_z$. 
We quantify the impact of the wide redshift distribution on the angular clustering on large scale, where it matters for the determination of the galaxy bias, using the Multidark simulation. We compare the angular clustering of haloes SHAM selected with a top-hat redshift distribution (1 if $z\in\bar{z}\pm\sigma_z$, else 0) and the observed redshift distribution. The large-scale angular clustering $\theta>0.08^\circ$ is unchanged. The small-scale angular clustering does change. Therefore we assume the large-scale angular clustering measured on the full sample is representative of the sample located in $\bar{z}\pm\sigma_z$, and that the bias, within the error bars will be correct. 
The values of the galaxy bias derived by the HOD method are given in Table \ref{hod:All:tab} and plotted in Figs \ref{bias_all:fig} and \ref{bias_all2:fig}.
This shows the major flaw of this HOD parametrization that cannot account for a galaxy population with a wide redshift distribution. A hint to try and model this would be to introduce a redshift dependence for the parameters describing $N(M)$. We further discuss this issue in section \ref{sec:Discussion}.

\subsection{Weak lensing measurements}
\begin{figure*}
\begin{center}
\includegraphics[width=50mm]{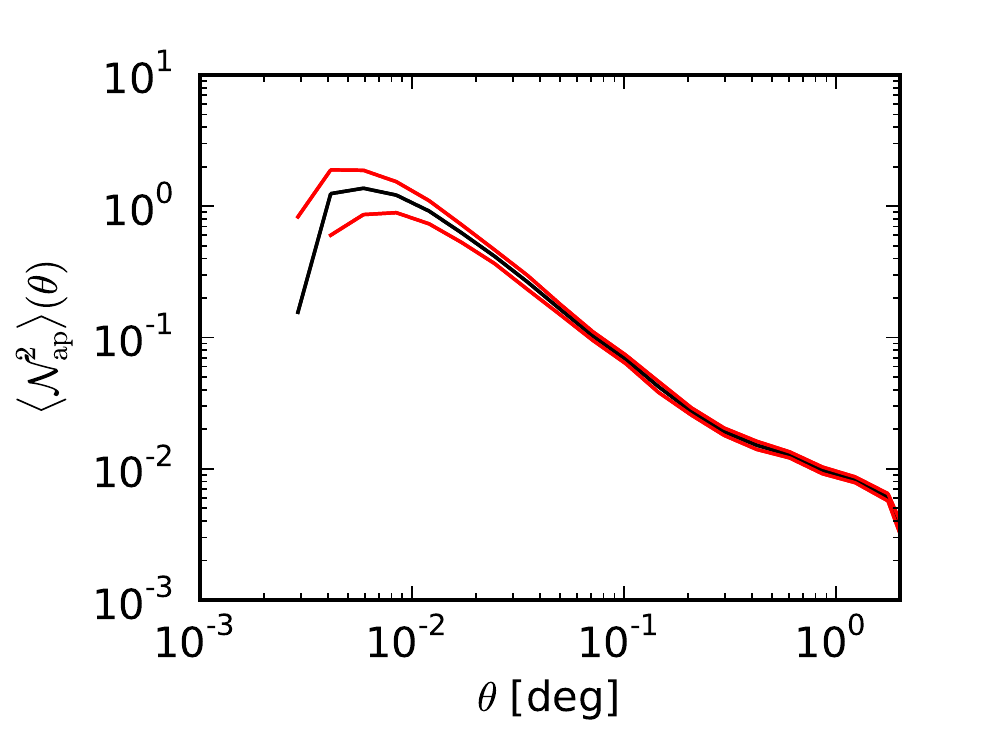}
\includegraphics[width=50mm]{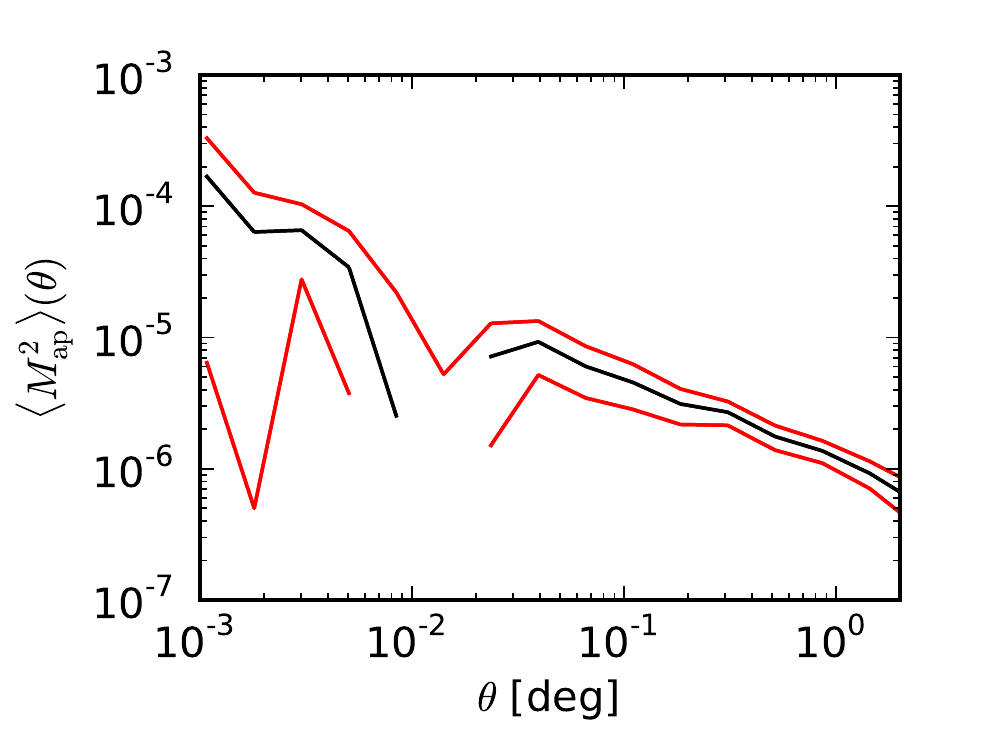}
\includegraphics[width=50mm]{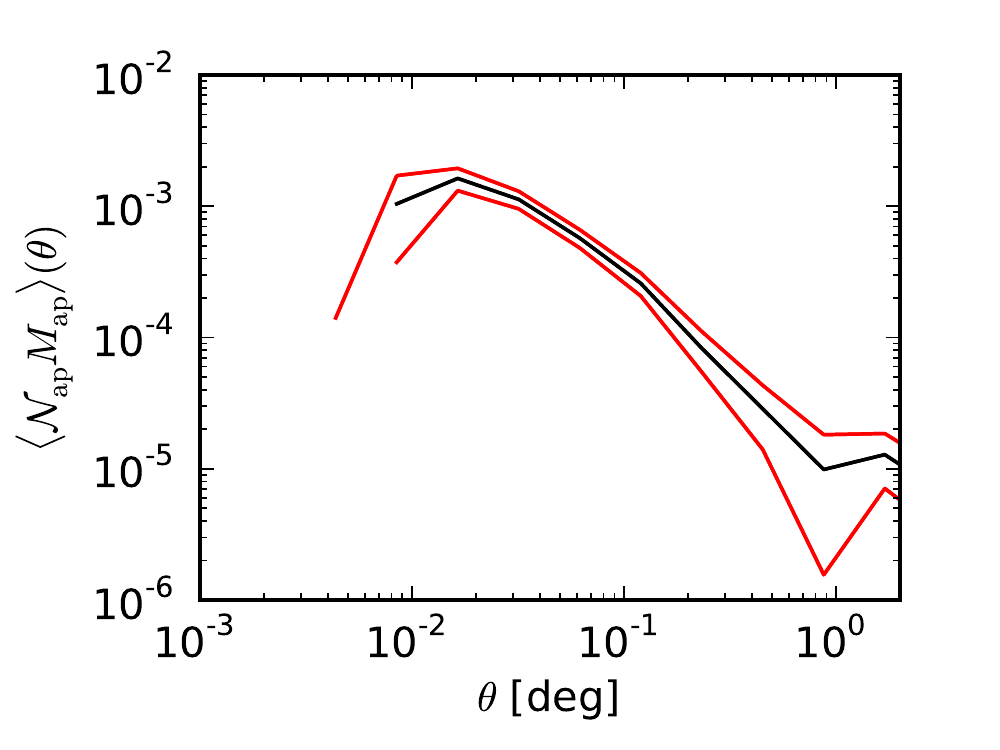}
\caption{WL measurements for the BOSS-CMASS sample. From the left to the right the galaxy autocorrelation $\mathcal{N}^2(\theta)$, the matter autocorrelation $M_\mathrm{ap}^{2}(\theta)$, and the galaxy--matter cross-correlation $\mathcal{N}(\theta)M_\mathrm{ap}(\theta)$. The black line represents the mean value and the red lines the $1\sigma$ error contours. Measurements of $\mathcal{N}^2(\theta)$ and $M_\mathrm{ap}^{2}(\theta)$ are clean between $0.05^\circ$ and $1^\circ$. $\mathcal{N}(\theta)M_\mathrm{ap}(\theta)$ is clean between $0.02^\circ$ and $0.2^\circ$. We thus can deduce $b_\mathrm{g}$ and $r$.}
\label{clustering:all1}
\end{center}
\end{figure*}

\begin{figure*}
\begin{center}
\includegraphics[width=50mm]{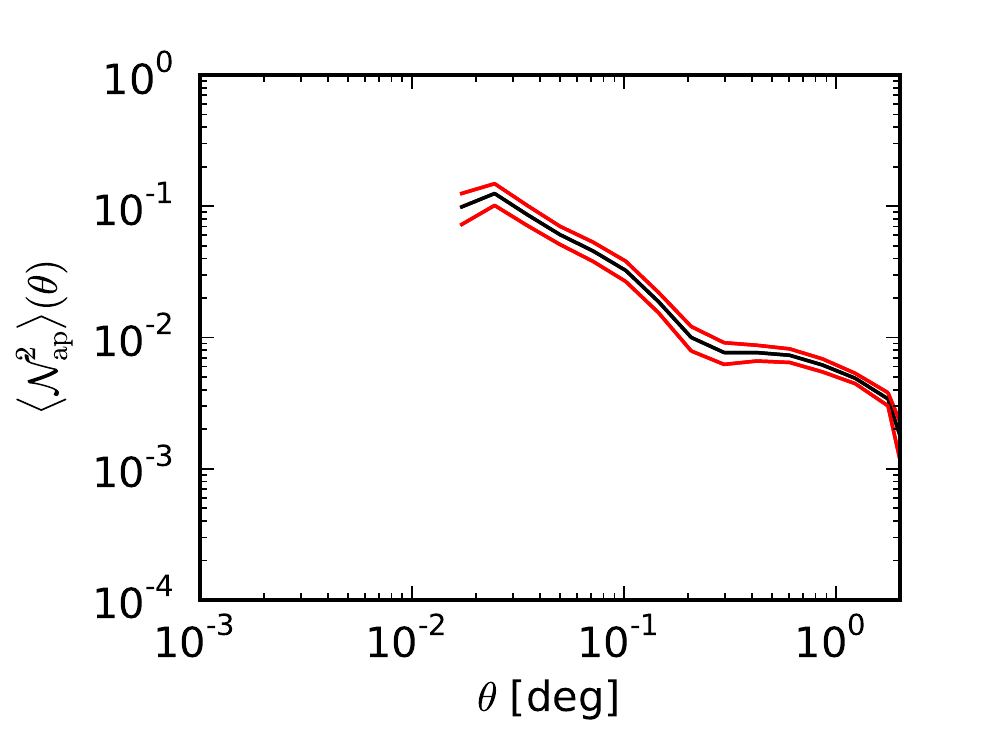}
\includegraphics[width=50mm]{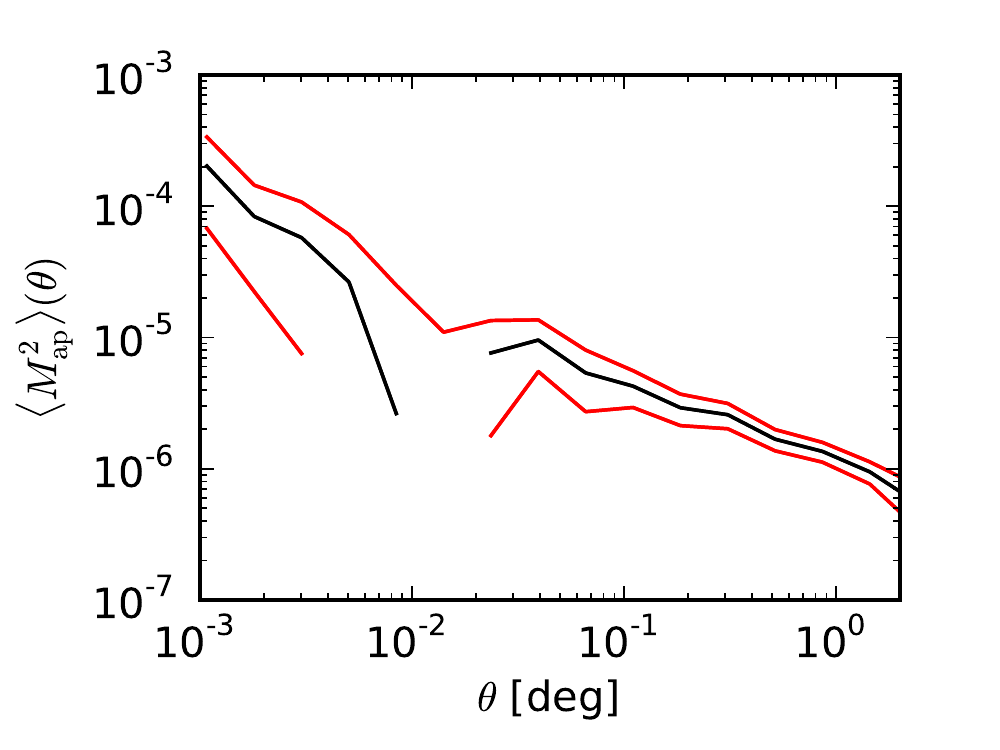}
\includegraphics[width=50mm]{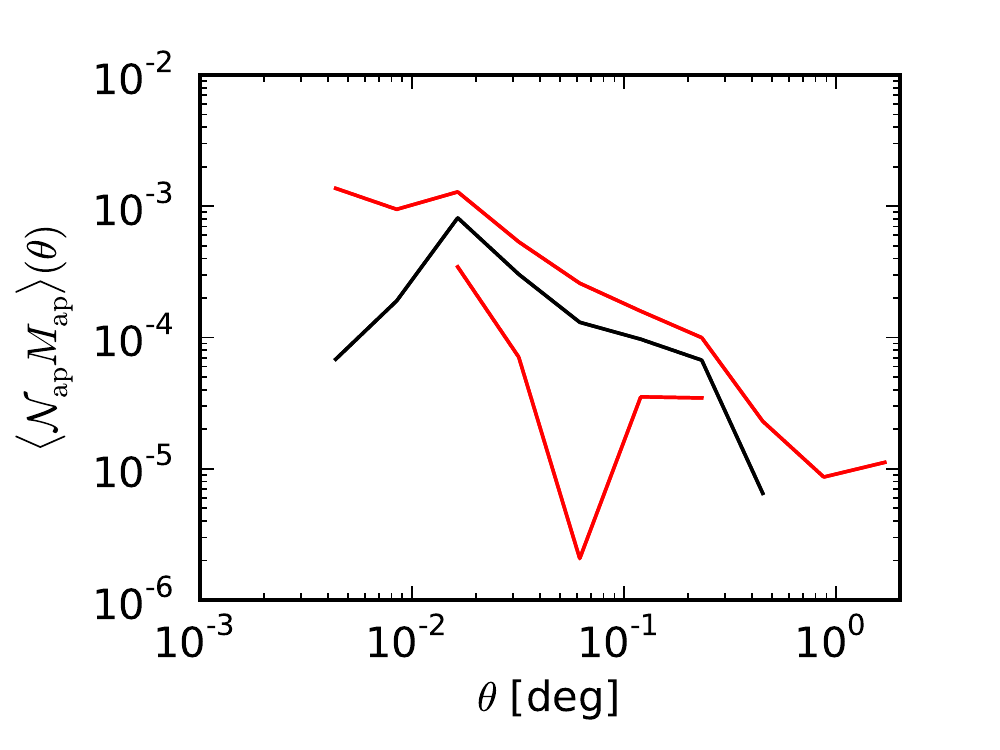}\\
\includegraphics[width=50mm]{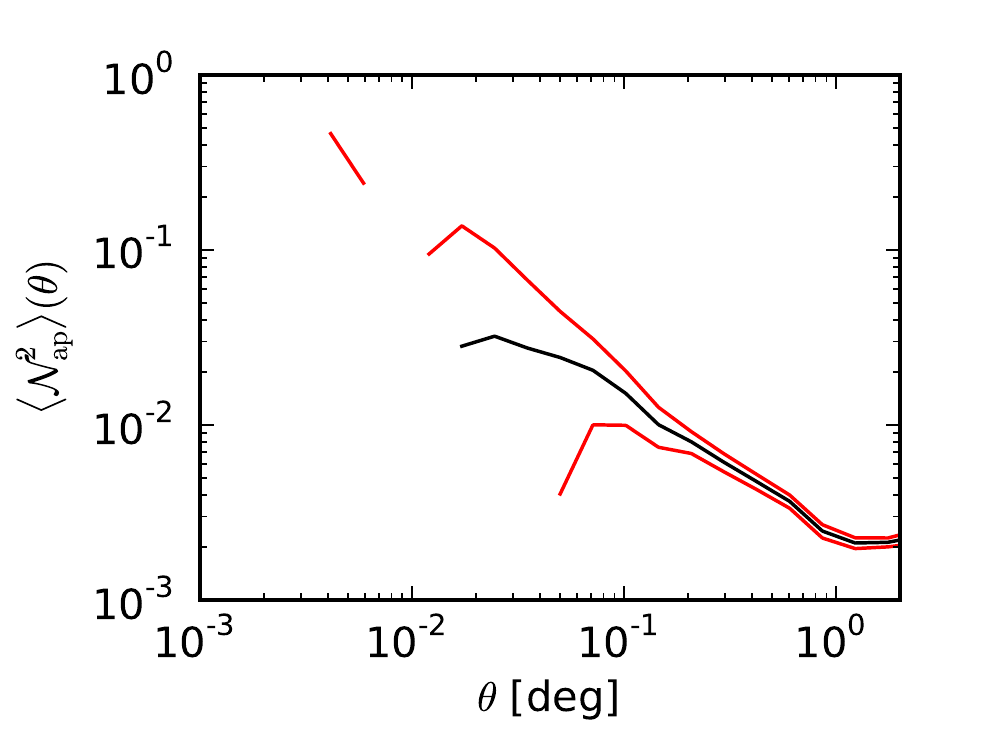}
\includegraphics[width=50mm]{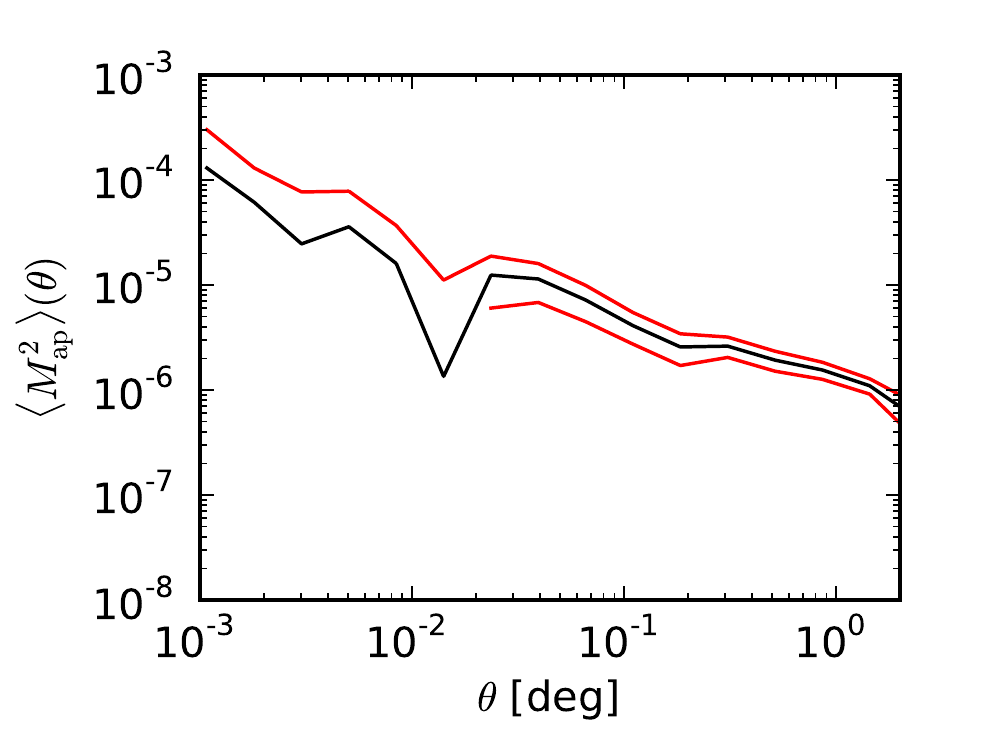}
\includegraphics[width=50mm]{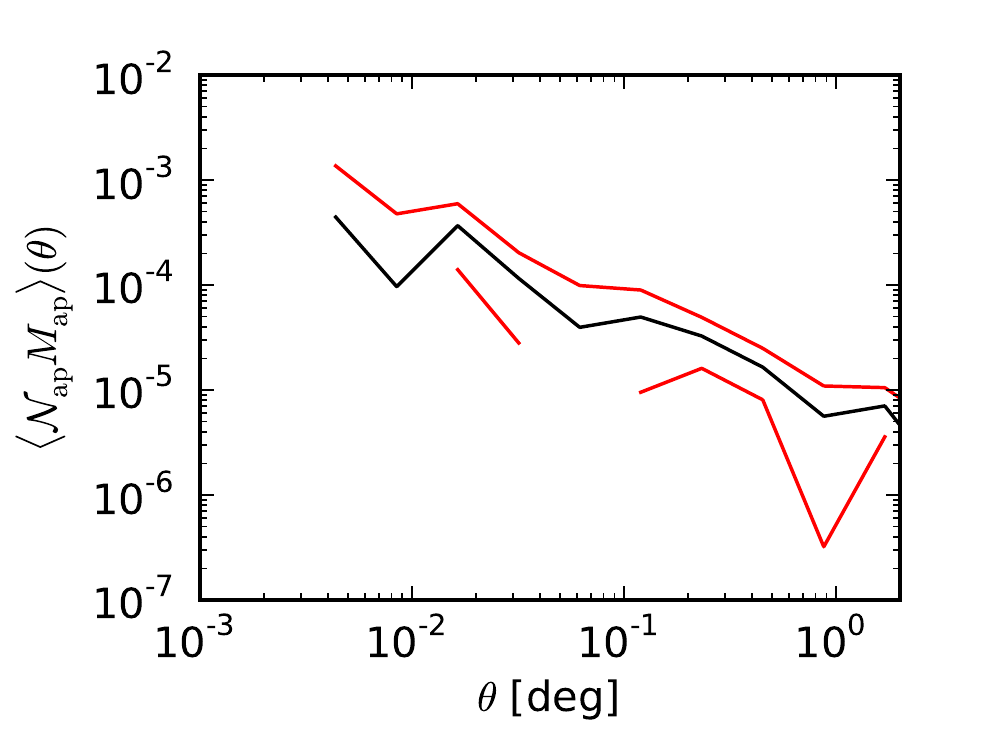}
\caption{WL measurements for the LRG-{\it WISE} bright (top row) faint (bottom row) sample. From the left to the right the galaxy autocorrelation $\mathcal{N}^2(\theta)$, the matter autocorrelation $M_\mathrm{ap}^{2}(\theta)$, and the galaxy--matter cross-correlation $\mathcal{N}(\theta)M_\mathrm{ap}(\theta)$. The black line represents the mean value and the red lines the $1\sigma$ error contours. Measurements of $M_\mathrm{ap}^{2}(\theta)$ and $\mathcal{N}^2(\theta)$ are clean for $\theta>0.06^\circ$ for both samples. $\mathcal{N}(\theta)M_\mathrm{ap}(\theta)$ is not.}
\label{clustering:all12}
\end{center}
\end{figure*}

\begin{figure*}
\begin{center}
\includegraphics[width=50mm]{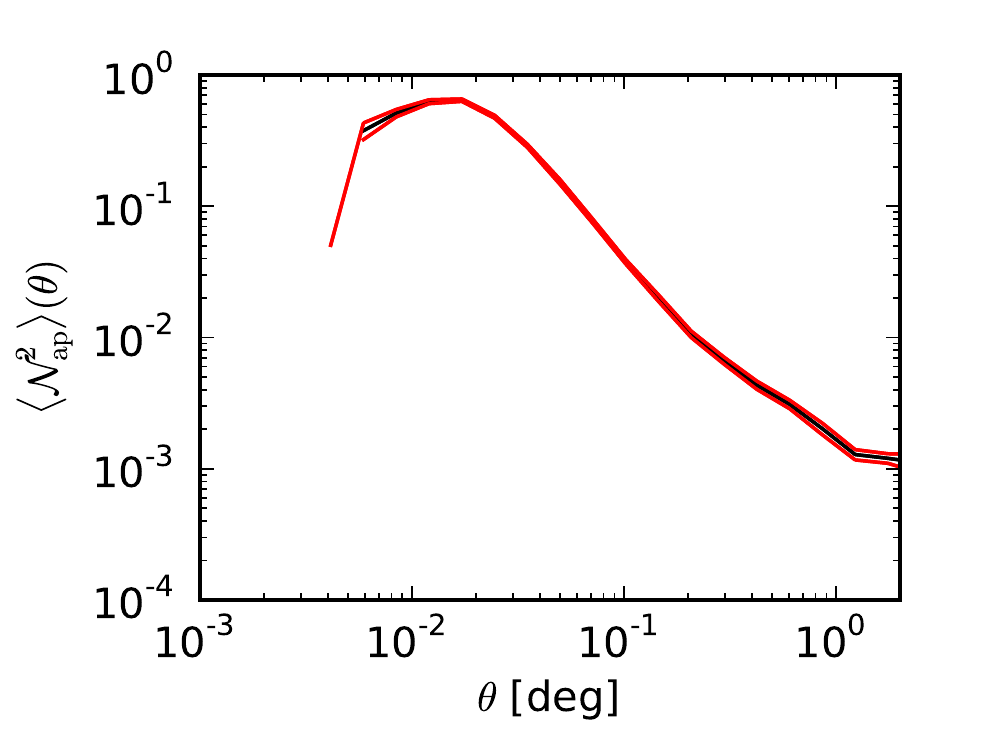}
\includegraphics[width=50mm]{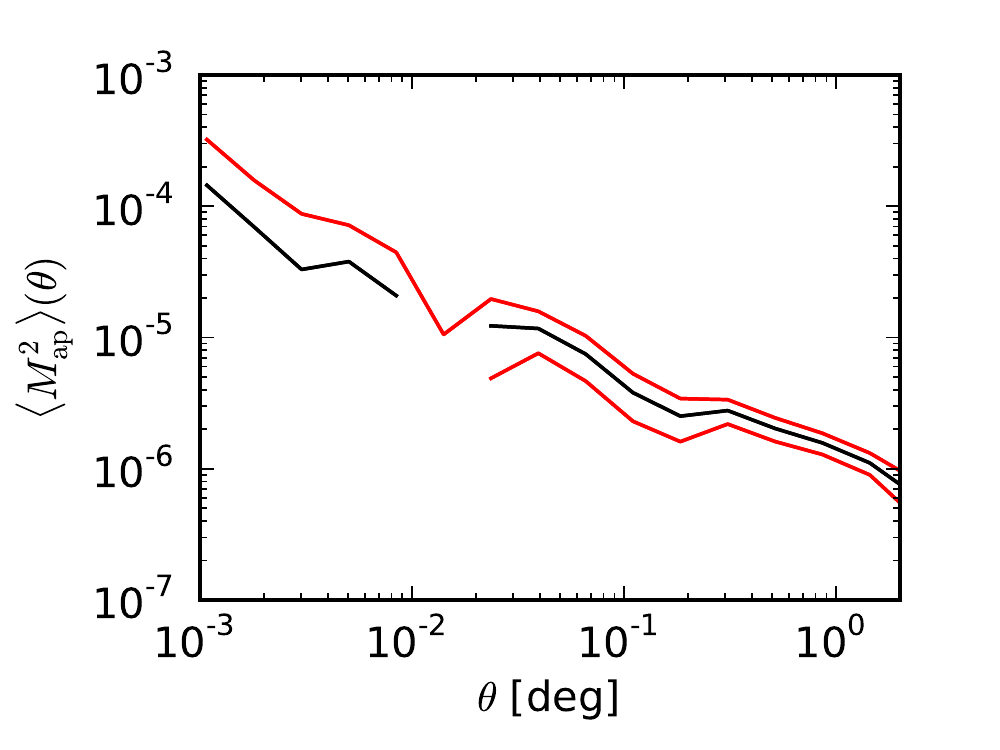}
\includegraphics[width=50mm]{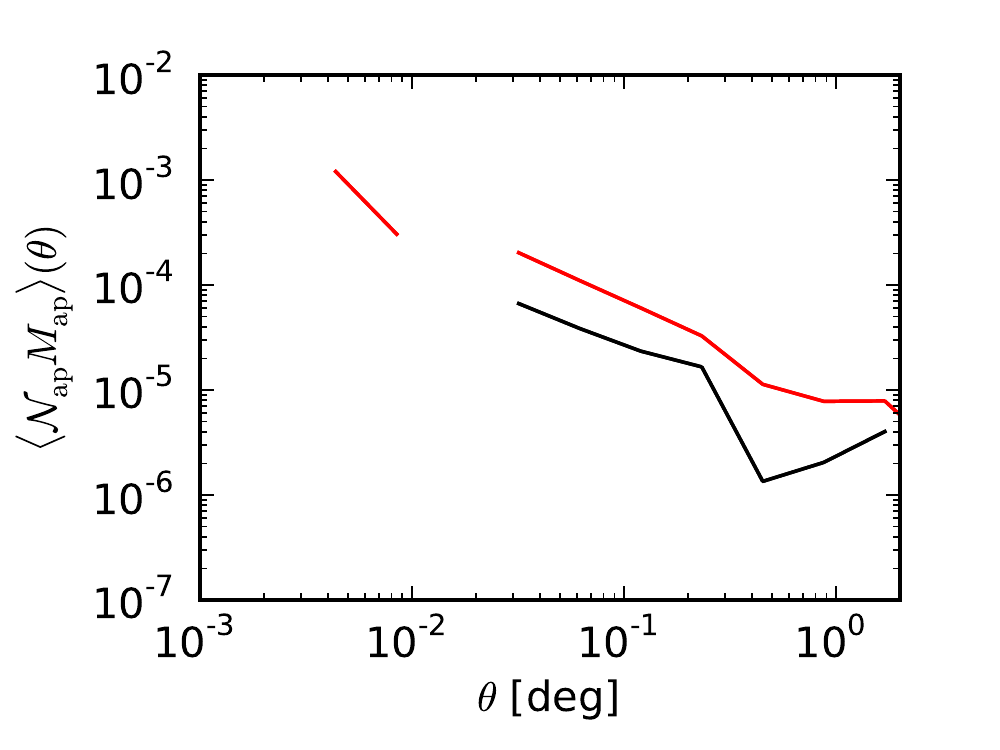}\\
\includegraphics[width=50mm]{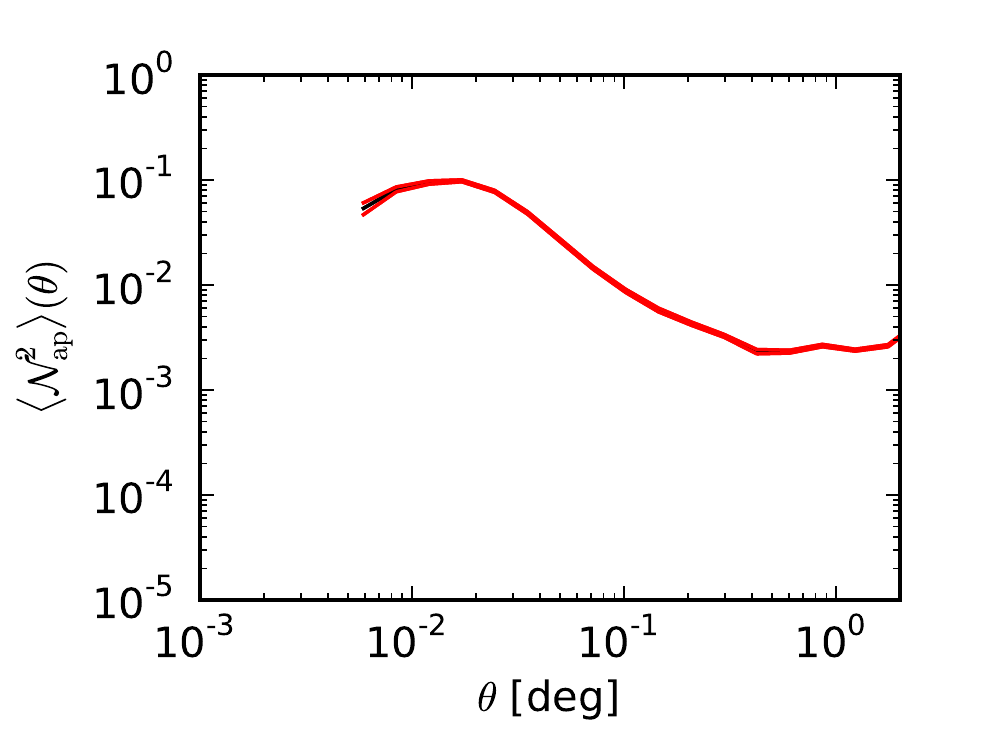}
\includegraphics[width=50mm]{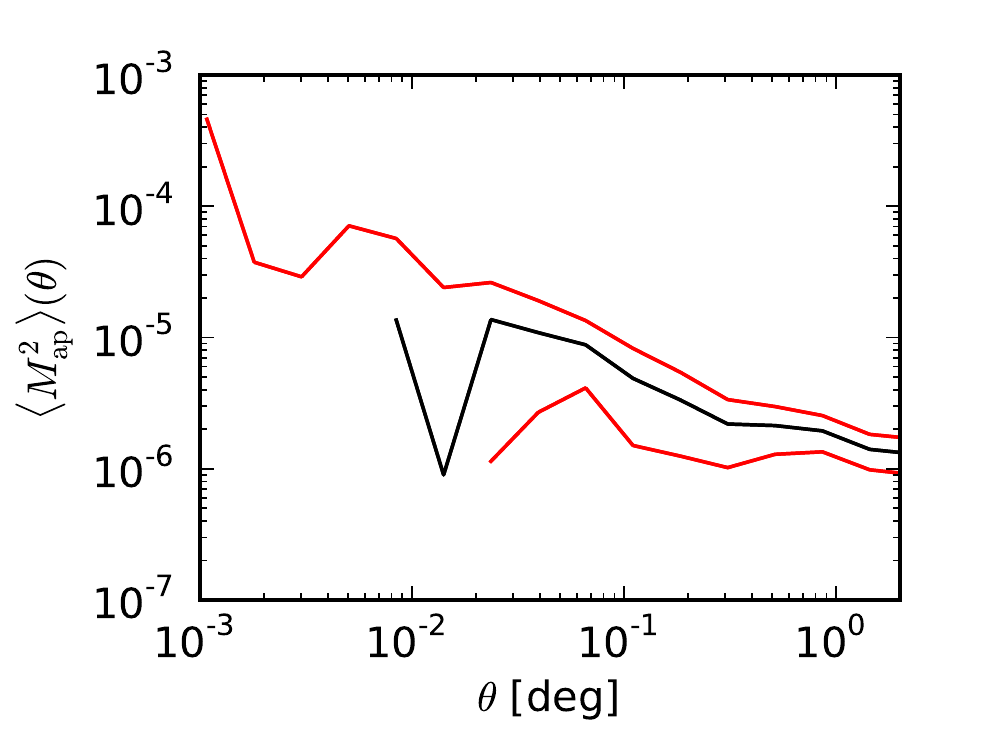}
\includegraphics[width=50mm]{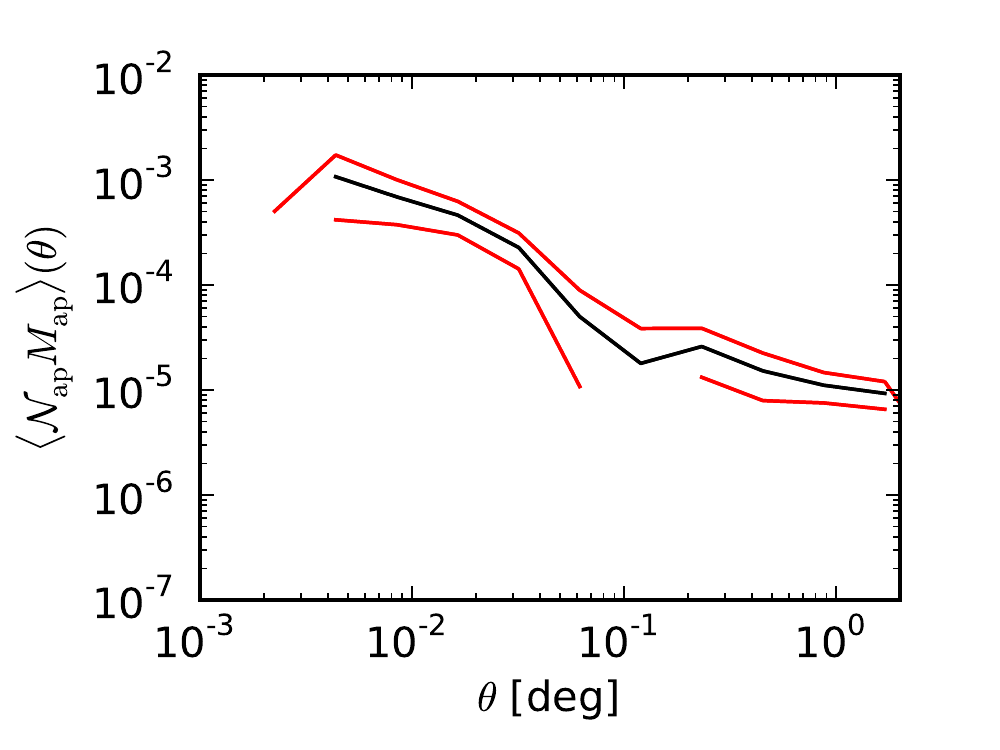}
\caption{WL measurements for the \emph{gri} ELG bright (top row) faint (bottom row) sample. From the left to the right the galaxy autocorrelation $\mathcal{N}^2(\theta)$, the matter autocorrelation $M_\mathrm{ap}^{2}(\theta)$, and the galaxy--matter cross-correlation $\mathcal{N}(\theta)M_\mathrm{ap}(\theta)$. The black line represents the mean value and the red lines the $1\sigma$ error contours. Measurements are clean between $0.1^\circ$ and $1^\circ$ for the $M_\mathrm{ap}^{2}(\theta)$ and $\mathcal{N}^2(\theta)$ of the bright sample and $\mathcal{N}^2(\theta)$ of the faint sample. The other lensing measurements are not robust.}
\label{clustering:all2}
\end{center}
\end{figure*}

\begin{figure*}
\begin{center}
\includegraphics[width=50mm]{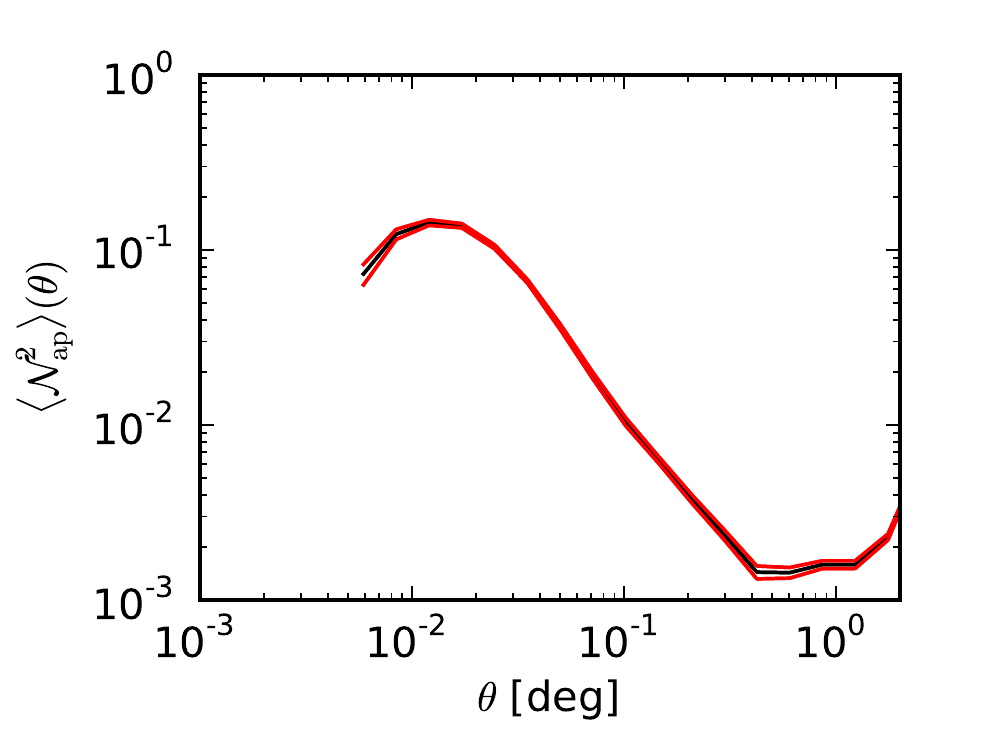}
\includegraphics[width=50mm]{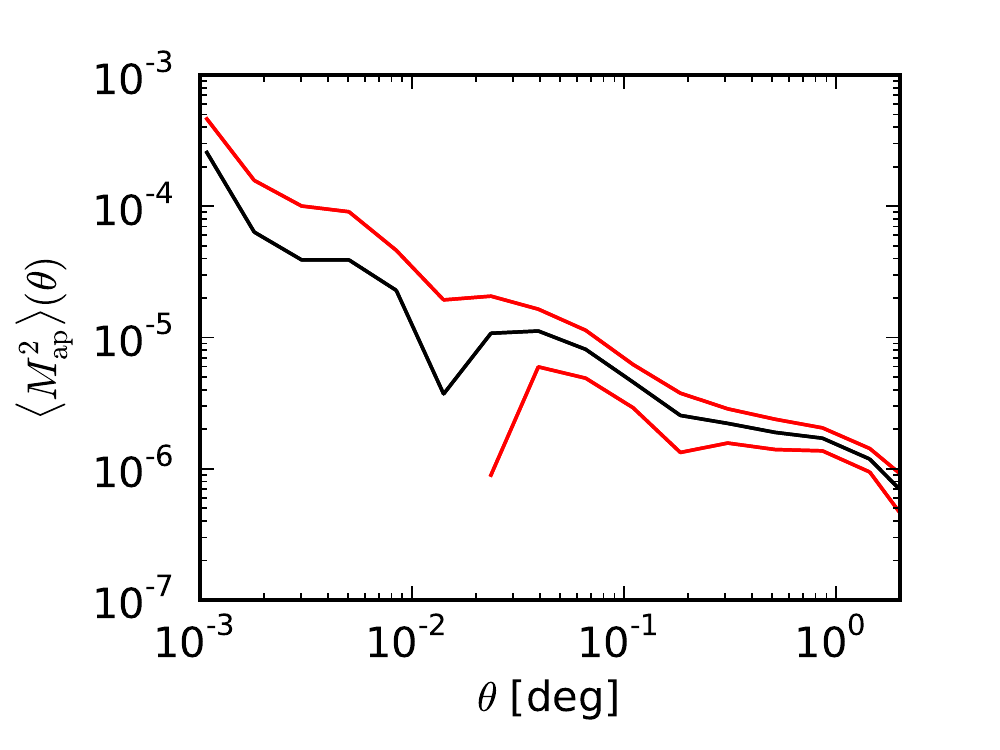}
\includegraphics[width=50mm]{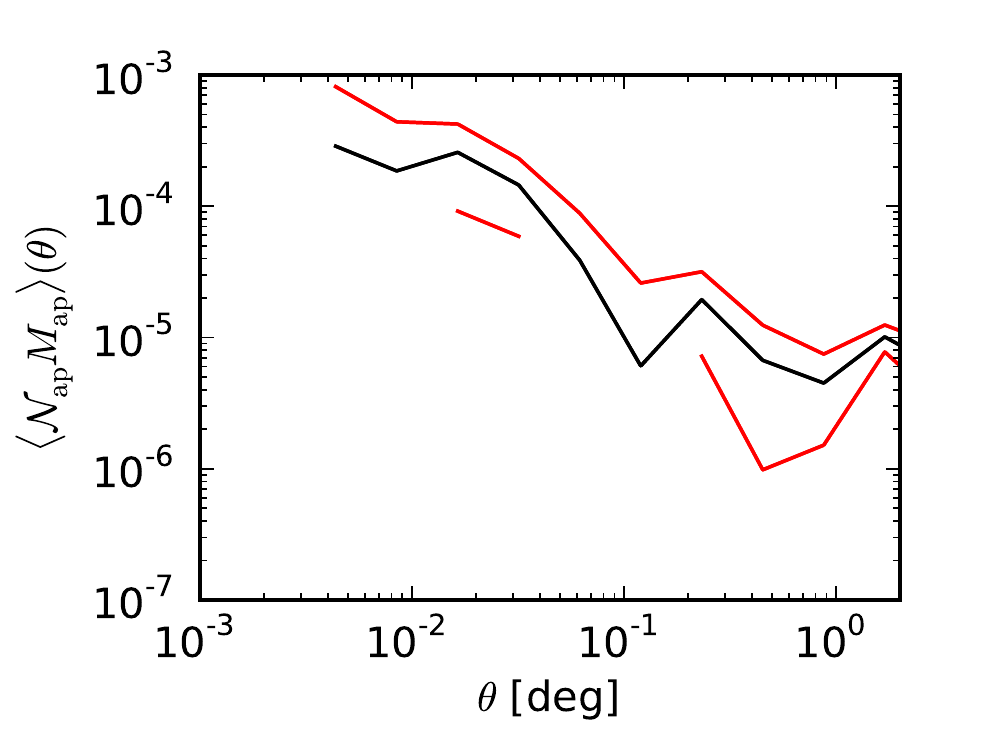}\\
\includegraphics[width=50mm]{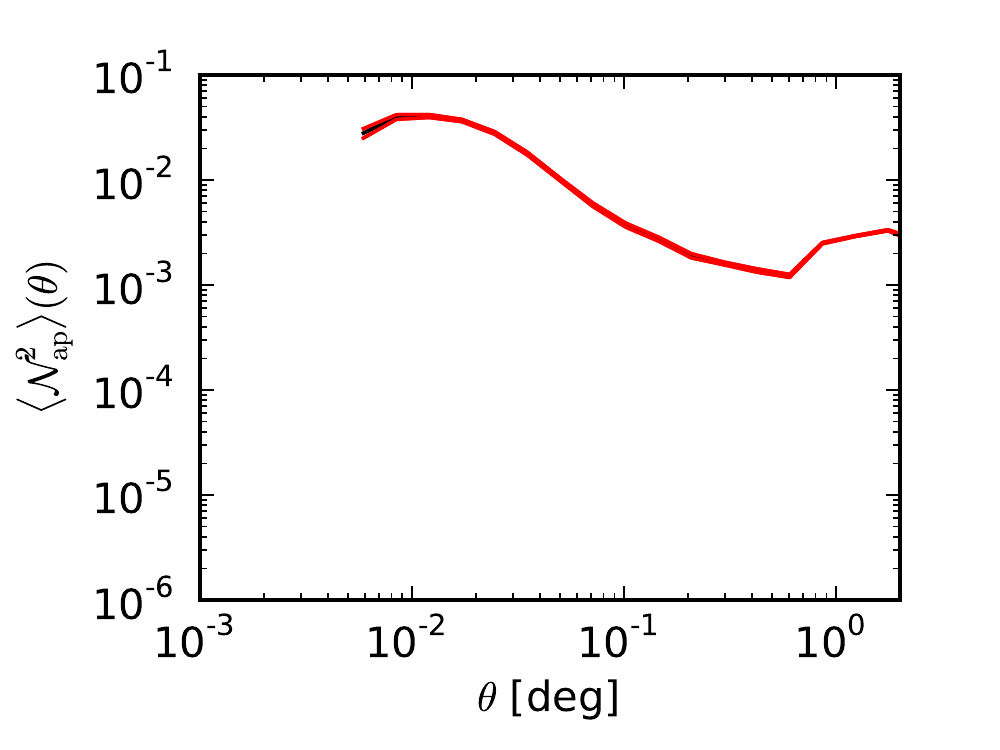}
\includegraphics[width=50mm]{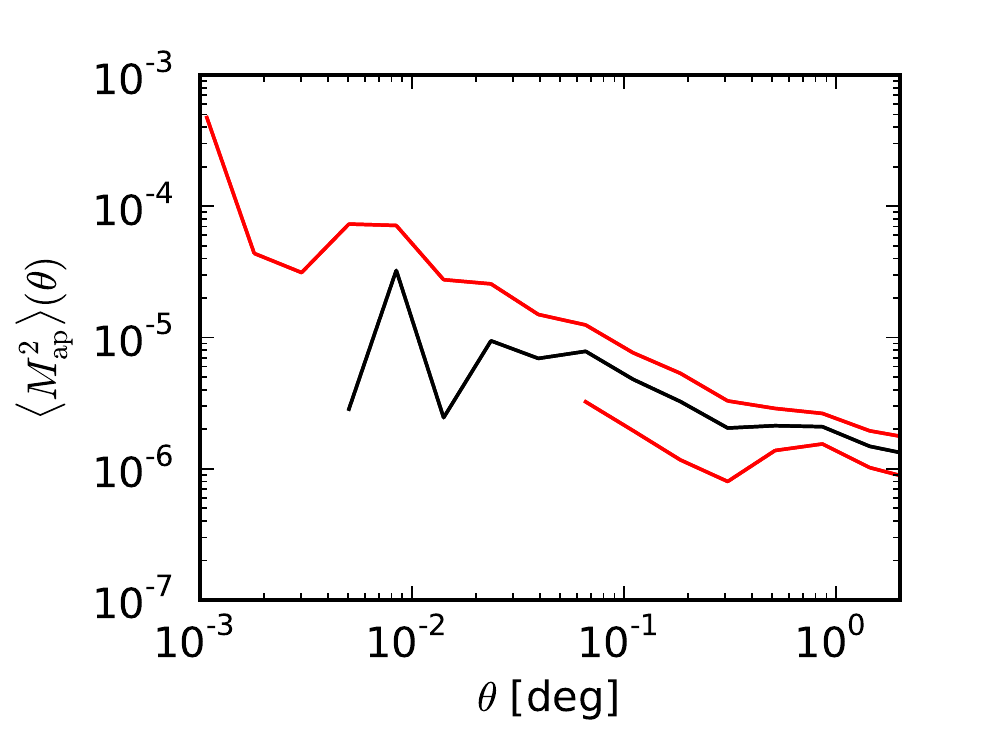}
\includegraphics[width=50mm]{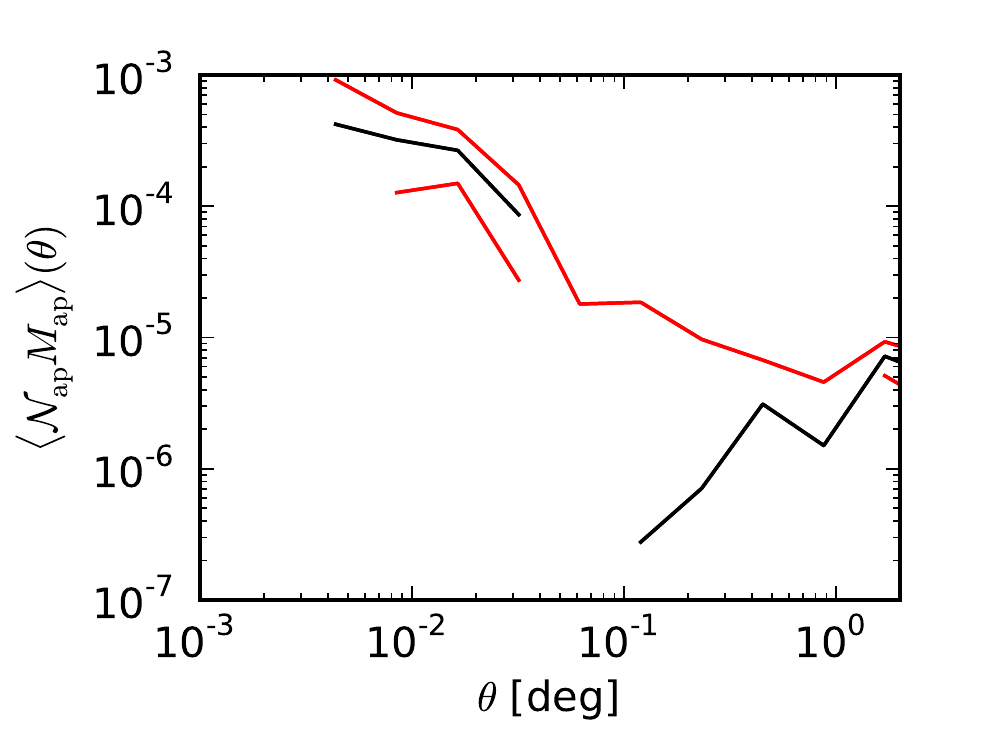}
\caption{WL measurements for the \emph{ugri} ELG bright (top row) faint (bottom row) sample. From the left to the right the galaxy autocorrelation $\mathcal{N}^2(\theta)$, the matter autocorrelation $M_\mathrm{ap}^{2}(\theta)$, and the galaxy--matter cross-correlation $\mathcal{N}(\theta)M_\mathrm{ap}(\theta)$. The black line represents the mean value and the red lines the $1\sigma$ error contours. Measurements are clean between $0.05^\circ$ and $1^\circ$ for the $M_\mathrm{ap}^{2}(\theta)$ and $\mathcal{N}^2(\theta)$ of the bright sample and $\mathcal{N}^2(\theta)$ of the faint sample. The other lensing measurements are not robust.}
\label{clustering:all4}
\end{center}
\end{figure*}

We measure $\mathcal{N}(\theta)M_\mathrm{ap}(\theta)$, $\mathcal{N}^2(\theta)$, $M_\mathrm{ap}^{2}(\theta)$ with the same routines as in \citet{2012ApJ...750...37J}. The SNR obtained on the bias is $\sim3$ for tracers with $z<1$ and of $\sim1.5$ for $z>1$. Because of this low SNR measurement, we exclude the \emph{ugr} ELG sample at $z\simeq1.2$ from the WL analysis. The measurements are shown in Figs \ref{clustering:all1} (CMASS), \ref{clustering:all12} (LRG-{\it WISE}), \ref{clustering:all2} (\emph{gri} ELG) and \ref{clustering:all4} (\emph{ugri} ELG). We derive the galaxy bias for all these samples; see Table \ref{hod:All:tab}. The cross-correlation $\mathcal{N}(\theta)M_\mathrm{ap}(\theta)$ is measured with precision only for the lower redshift CMASS sample. Therefore $r$ is computed only for this sample.

The main source of error arises from the error on the shape measurement; see the data description section. The uncertainty on the redshift distribution can also potentially introduce some systematic errors. 
All statistical errors are propagated with bootstrap as described in \citet{2012ApJ...750...37J}. For each foreground and background galaxy catalogue, 100 samples are randomly drawn and processed to compute the aperture auto- and cross-correlation functions as well as the bias and correlation coefficient. The background source catalogues are cut at 1 mag fainter than the foreground. Systematic error on $M_\mathrm{ap}^{2}(\theta)$ on large scale due to the lack of information at small scale is below $1$ per cent \citep{2006A&A...457...15K}. 

%\subsection{cross-correlation coefficient and galaxy bias}
\subsection{Cross-correlation coefficient and galaxy bias}

Measurements of the galaxy clustering as a function of the luminosity were performed in the local universe by \citep{2008ApJ...672..153C}, \citep{2009MNRAS.392.1080S}, \citep{2011ApJ...736...59Z} and \citep{2013ApJ...767..122G}. They showed the clustering amplitude is greater for more luminous samples. Thus we expect to measure high values of the galaxy bias for all samples. There is no straightforward existing relation that can predict the expected galaxy bias. 

\subsubsection{Cross-correlation coefficient}
We measure the cross-correlation coefficient only for the CMASS sample, $r=1.16\pm0.35$. This measurement is a first, and implies the CMASS galaxies are fully correlated with the matter field. This strongly supports current BAO analysis made on this sample \citep{2012MNRAS.427.3435A}.

With current data, it is not possible to constrain the cross-correlation coefficient of the other galaxy samples. 

\subsubsection{Galaxy bias}

\begin{figure}
\begin{center}
\includegraphics[width=80mm]{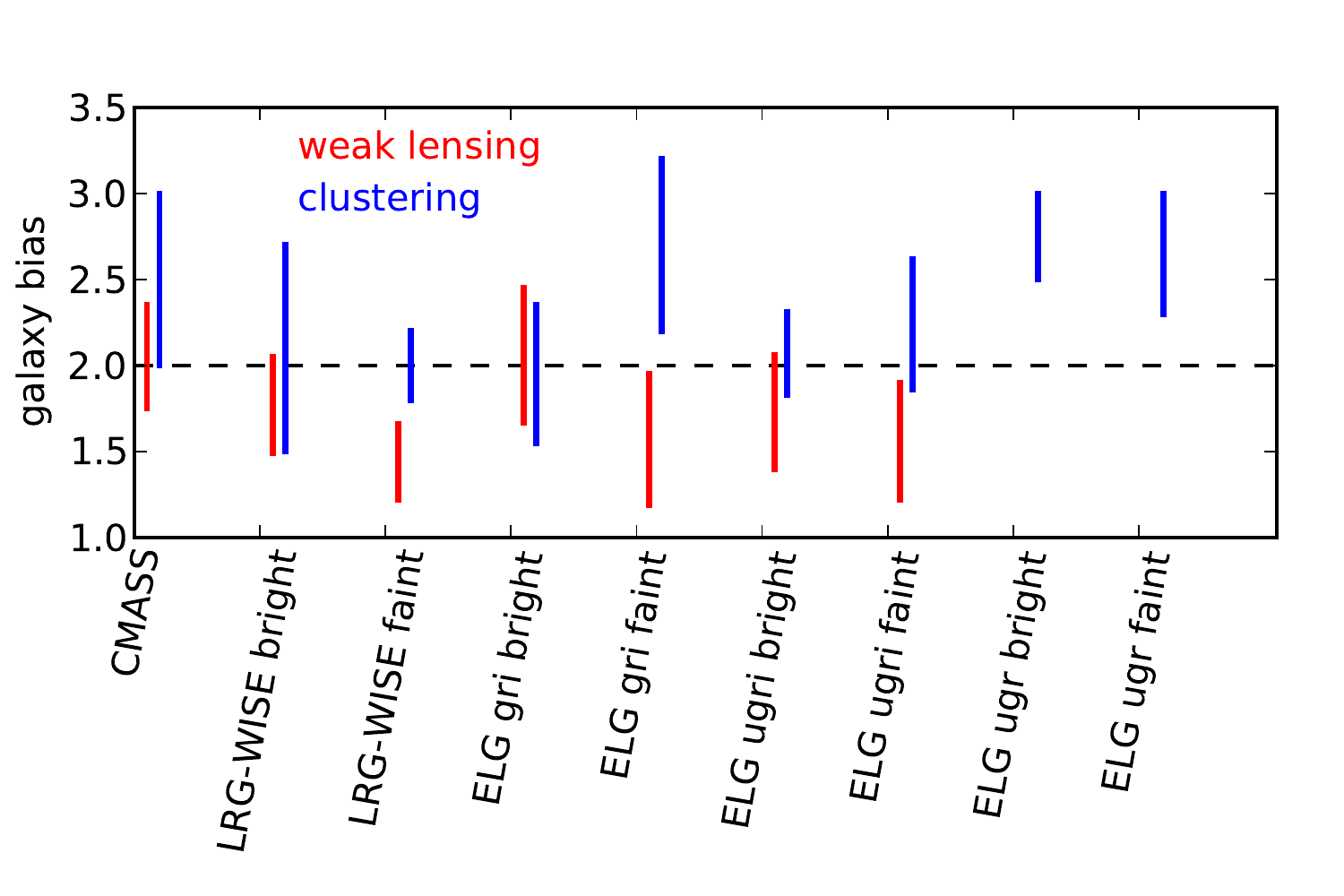}
\caption{Large-scale galaxy bias of the different BAO tracers ordered by mean redshift. The bright samples are more biased than the faint samples. The HOD bias is not in agreement for the faint samples and tends to be greater than the WL estimation.}
\label{bias_all:fig}
\end{center}
\end{figure}

\begin{figure*}
\begin{center}
\includegraphics[width=180mm]{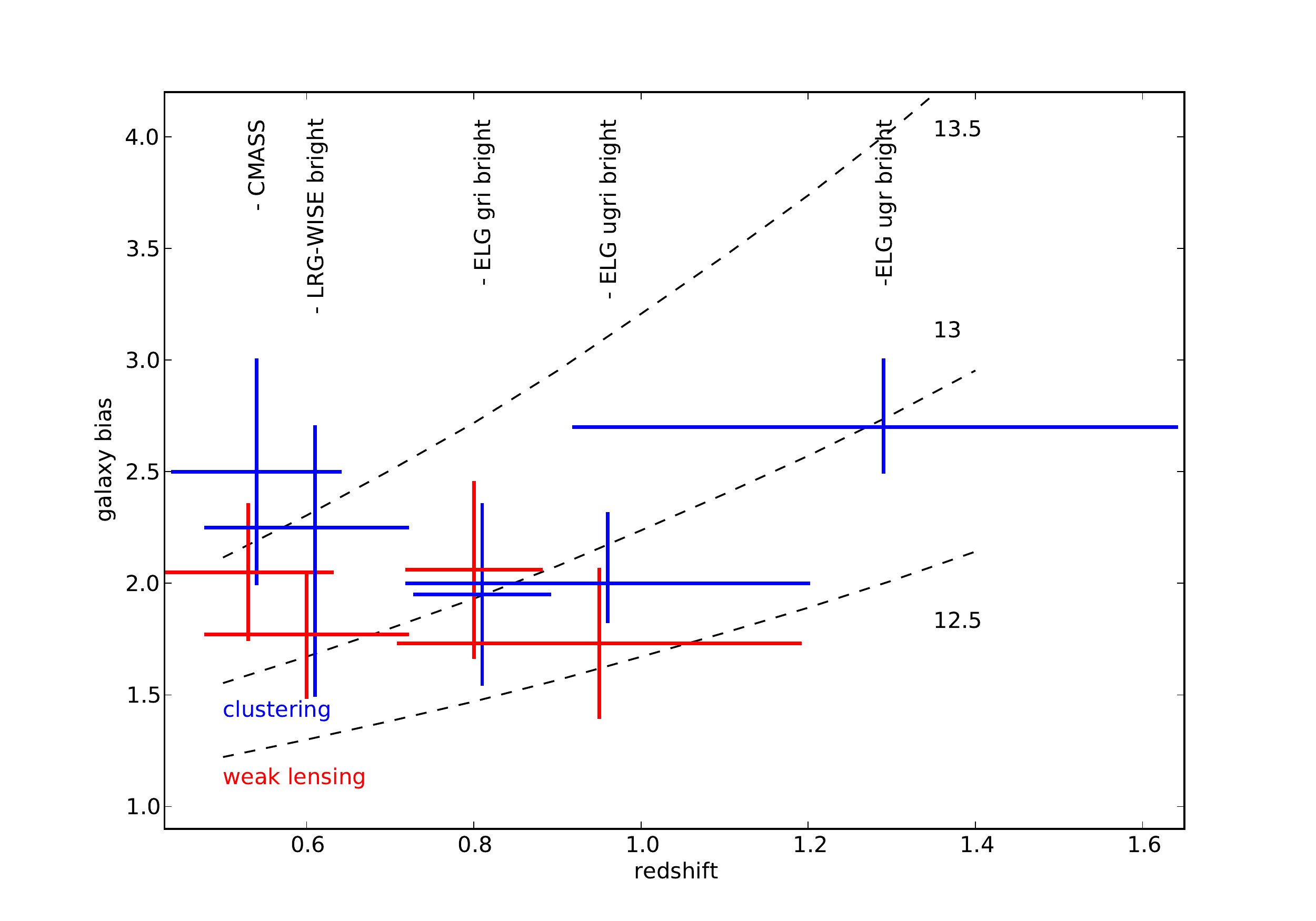}
\caption{Large-scale galaxy bias of the different BAO tracers vs. redshift. The interquartile are shown as error bars (25 per cent - 75 per cent intervals). The dashed black lines correspond to the bias of haloes of mass $\log_{10}{M_\mathrm{200}/M\odot}$ of 12.5, 13 and 13.5 from bottom to the top, after \citet{2005ApJ...631...41T}. The figure shows the ELGs inhabit haloes of mass between $10^{12.5}$ and $10^{13}M\odot$. The comparison with the SHAM halo mass values (Table \ref{tab:description:ham}) shows these selections are tracing the most massive haloes at their mean redshift.}
\label{bias_all2:fig}
\end{center}
\end{figure*}

For the CMASS, LRG-{\it WISE} bright, \emph{gri} ELG bright, and \emph{ugri} ELG bright samples, the measurements of the two methods are in agreement; see Fig. \ref{bias_all:fig}. 
We find the bias is above $1.5$ for the all ELG selections, which is consistent with the measurements made by \citet{2012A&A...542A...5C} and \citet{2013ApJ...767...89M} on their brightest samples. We consider both measurements give a trend about the size of dark matter haloes these galaxies inhabit; see Fig. \ref{bias_all2:fig}. Comparing the values of the bias with the curves of the bias for a constant halo mass after \citet{2005ApJ...631...41T} and with the mass distributions obtained via SHAM shows all these samples are tracing the most massive haloes at their respective redshifts.
From Fig. \ref{bias_all:fig} we note that the bias from the HOD seems systematically higher than that of WL. 
The bias of the HOD corresponds to the bias of a sample that is complete in mass (represented by $N(M)$). Though the assumption of completeness in mass does not really hold for clue colour-selected samples.
The bias output by the WL does not need this assumption, and is thus lower than the value obtained via the HOD and probably closer to the reality for the bright samples where SNR is sufficient.
We think this is the primary reason to this systematic discrepancy. With current data, it is not possible to estimate precisely stellar masses and understand this discrepancy in greater details.

For the LRG-{\it WISE} faint, \emph{gri} ELG faint and \emph{ugri} ELG faint samples, there is a systematic disagreement between the methods: the clustering analysis outputs a higher bias than the WL analysis; see Fig. \ref{bias_all:fig}.
Thus for the LRG-{\it WISE} faint, \emph{gri} ELG faint and \emph{ugri} ELG faint samples, we find the combination of current surveys "CFHT/Stripe 82 + CFHT-LS + SDSS-S82 deep co-add" is too shallow to derive a reliable value of the bias. We need a fair spectroscopic sample of such tracers to improve the photometric redshift distribution accuracy, and a deeper survey for the shape of background galaxies.

For the \emph{ugr} ELG bright (and faint), with current data, the WL analysis is not feasible. The galaxy bias derived using only the clustering information is although of prime interest. It is consistent with the fact that the \emph{ugr} ELG bright sample occupies the most massive haloes at $z\sim1.2$.

%\subsubsection{Satellite fraction, halo mass and HOD}
\subsubsection{Satellite fraction, halo mass and HOD}
The shape of the angular clustering for the ELG samples suggests a high satellite fraction. Indeed there is a strong bump for $\theta<0.02^\circ$. The HOD models fail to reproduce trustworthy satellite fractions for the ELG samples; see Table \ref{hod:All:tab}. For the CMASS sample, that is approximately complete in mass and a volume-limited sample, HOD and SHAM methods are in agreement and obtain 5 per cent of satellites. For the $gri$ ELG sample, which is the reddest of the ELG samples, the HOD and SHAM satellite fractions are in agreement, but remain below 10 per cent. Given the observed small-scale bump, we expected satellite fraction of the order of 15-25 per cent. For the $ugri$ and the $ugr$ ELG samples, the satellite fractions do not really make sense. They are not consistent between the bright and the faint sample, which is puzzling. It shows the HOD models have a high degeneracy when they fit the small angular scales of a sample with a wide redshift distribution.
A similar effect is seen for the mean halo mass predicted by the HOD; it is always 0.5 dex higher than the expectations set by the SHAM. The clustering curves of the brightest galaxies in each redshift bin therefore constitute a new challenge to halo occupation models.
 
\begin{landscape}
% -------------- Table Clustering 2D ==================
\begin{table}
	\label{hod:All:tab}
	\caption{Results from the clustering and the WL analysis. The number of degrees of freedom is $\sim18$ for the HOD fits. The values of the $\chi^2$ show current HOD models fail in explaining angular clustering of the ELGs. The column $\leftrightarrow  M_\mathrm{d}$ states the relation $M_\mathrm{halo}-M_\mathrm{d}$. $b_\mathrm{clustering}$ is the galaxy bias obtained with the HOD, $\langle b_\mathrm{WL} \rangle$ with the WL, and $\langle r_\mathrm{WL} \rangle$ is cross-correlation coefficient. The column satellite fraction compares the results from the SHAM and the HOD.}
	\setlength{\extrarowheight}{3pt}
	\begin{tabular}{l c c c c c c c c c c c c c c c c c c c c c}
	\hline \hline
	Sample &  \multicolumn{5}{c}{Maximum likelihood HOD parameter} & $\chi^2/$dof & \multicolumn{2}{c}{Halo mass}&\multicolumn{2}{c}{Satellite fraction}&$b_\mathrm{clustering}$ & $\langle b_\mathrm{WL} \rangle$&$\langle r_\mathrm{WL} \rangle$  \\
	 & $log(\frac{M_{min}}{M\odot})$ & $log(\frac{M_1}{M\odot})$ & $log(\frac{M_0}{M\odot})$ & $\sigma_{logM}$ & $\alpha$&&$\langle M_{halo} \rangle$&$\leftrightarrow M_\mathrm{d}$ &HOD& SHAM& \\
	\hline

CMASS & 13.433 &14.576 &10.73 &0.242 &1.790 &0.3012&13.779&$>$& 5 & 5 &$2.5\pm0.5$&$2.05 \pm 0.3$&$1.16\pm0.35$\\

LRG-{\it WISE} bright & -&-&-&-&-&-&-&-&-&4&$2.25^{+0.5}_{-0.7}$&$1.77 \pm 0.28$&-\\
LRG-{\it WISE} faint & -&-&-&-&-&-&-&-&-&7&$2.0\pm0.2$&$1.44 \pm 0.22$&-\\ 

ELG {\it gri} bright &13.51 & 14.51& 11.07& 0.89& 0.81 &16.05&13.219&$=$& 6 & 8 &1.95$\pm0.4$&$2.06 \pm 0.39$&-\\ 

ELG {\it gri} faint & 13.256& 14.931 &11.994& 0.268& 0.691 & 3.379 &13.558&$>$ & 9 & 10 &  $2.5^{+0.7}_{-0.3}$ &$1.57 \pm 0.38$&-\\

ELG {\it ugri} bright  & 12.41 &12.043 &12.422 &0.58 & 0.64 &2.88&13.157&$=$& 52 & 7&$2.0^{+0.31}_{-0.17}$&$1.73 \pm 0.33$ &-\\
ELG {\it ugri} faint& 13.049& 14.443 &13.089 &0.69& 0.54&6.95&13.084&$>$& 5 &11 & $2.1^{+0.5}_{-0.2}$&$1.56 \pm 0.34$&-\\

ELG {\it ugr} bright&13.179& 13.409 &13.074 &0.63&0.62&1.54&13.241&$<$& 16&3 &$2.7^{+0.3}_{-0.2}$ &-&-\\ %$1.88 \pm 0.36$
ELG {\it ugr} faint & 13.103& 14.386 &12.928 &0.592 &0.779&6.53&13.126&$=$&4 & 7&$2.5^{+0.5}_{-0.2}$&-&-\\ %$2.01 \pm 0.42$
	\hline

	\end{tabular}
\end{table}
\end{landscape}

%--------------------------------------------------------------------------------------------------------------
%--------------------------------------------------------------------------------------------------------------
%------------------------------			DISCUSSION       -----------------------------------
%--------------------------------------------------------------------------------------------------------------
%--------------------------------------------------------------------------------------------------------------
%\section{Discussion}

\section{Discussion}
\label{sec:Discussion}
The measurements presented in this paper are meaningful only for the  CMASS, LRG-{\it WISE} bright, \emph{gri} ELG bright, \emph{ugri}  ELG bright, and \emph{ugr} ELG bright (clustering only) samples. The following discussion is therefore restricted to these tracers unless otherwise stated (we set aside the ELG faint samples).

%\subsection{Are these selections suitable for future BAO studies ?}
\subsection{Are these selections suitable for future BAO studies ?}
The SNR in BAO goes as the number density ($\bar{n}$) times the power spectrum, which amplitude scales as the square of the galaxy bias $b^2$ for a constant $\sigma_8$. Thus $\mathrm{SNR_{BAO}}\propto \bar{n}P \propto \bar{n} b^2$. Thus the higher the bias (or the amplitude, it is equivalent), the smaller is the number density required to a reach a given SNR, and thus the faster the survey. For the same SNR, the density required using tracers with $b=1.5$ (respectively $b=2$) is 2.25 (respectively 4) times smaller than using tracers with a bias $b=1$.
A BAO survey using these tracers has thus a real advantage compared to selecting galaxies with lower absolute luminosity that have a lower clustering amplitude. 
These numbers constitute a zeroth order value of the galaxy bias that can be used in simulation to predict the accuracy of the detection and anticipate the requirements on geometry and density for surveys as SDSS-IV/eBOSS, DESspec, BigBOSS, PFS-SuMiRe, and Euclid. A check could be performed with n-body simulations by trying to reproduce the angular clustering measurements from this paper. 

The primary role of the target selection, to select highly-biased galaxies, is here fulfilled.
There is, however, a trade-off. The galaxy bias is the source of the dominating systematic on the BAO measurement; see \citet{2011ApJ...734...94M} and references therein. With current reconstruction algorithms, and a knowledge of the bias better than 10 per cent, simulations show it is possible to reduce the systematic errors down to 0.15 per cent. This study enables to start calibrating simulations on observed data to understand how to cope with this systematic effect.

\subsection{Sensitivity to input cosmological parameters}
In this study, we use {\it WMAP}7 fiducial values for cosmological parameters, yet the latest parameters obtained by {\it Planck} differ slightly, see \citet{2013arXiv1303.5076P}. The mean change of the angular correlation function of matter on linear scales (between $0.05^\circ$ and $0.5^\circ$) at $z\sim0.8$ with the redshift distribution of the bright $gri$ ELG is of $\sim7$ per cent. The {\it WMAP} parameters predict a matter clustering stronger by $\sim7$ per cent than that of {\it Planck}. Therefore with {\it Planck} parameters, the galaxy bias would increase by $\sim3.5$ per cent compared to the values we measure here.

\subsection{Complementarity of the lower redshift selections}
The redshift distribution of the CMASS, LRG-{\it WISE} and \emph{gri} ELG are overlapping in $0.6<z<0.7$; see Fig. \ref{fig:nz:tracers}. 
The match between the catalogues shows that 20 per cent of the tracers are common between the LRG-{\it WISE} bright sample and the CMASS. There is no common tracers between the $gri$ ELG and the LRG-{\it WISE} bright or the CMASS.
These three samples are therefore complementary to map the large scale structures in this redshift zone, and particularly to perform cross-correlation analysis.

\subsection{ELG $ugri$ selection, an intermediate selection}
The \emph{ugri} ELG colour selection is a hybrid between the \emph{ugr} and the \emph{gri} selections. The clustering analysis shows the $w(\theta)$ of the \emph{ugri} is located between that of the \emph{ugr} and the $gri$ $w(\theta)$. It is thus consistent.

\subsection{Hints on luminosity, stellar mass, halo mass, HOD modelling and $N$-body simulations}
\label{trends:zz}
\citet{2011ApJ...728...46W} showed that stellar mass and halo mass are correlated for $1<z<2$, and \citet{2006MNRAS.368...21L} reported that on large scales the galaxy clustering amplitude dependence on stellar mass is positively correlated to the luminosity. Our selections target the most luminous galaxies (in their redshift range). We therefore expect to track high stellar masses and high halo masses. The combination of the clustering analysis from this work and of the estimates of stellar masses from \citet{2013MNRAS.428.1498C} demonstrate that these galaxies trace haloes with large masses ($M_{halo} \sim 10^{12.5} - 10^{13}M\odot$), i.e. we comply with the general picture.

This analysis is complementary to similar analyses based on faint and deep `pencil beam' surveys, that investigate the clustering properties of average luminosity galaxies, see \citet{2012ApJ...744..159L} and \citet{2013ApJ...767...89M}. In fact our target selections aim for galaxies that are not in large numbers in such surveys. Given the trends in these papers, our results are in good agreement with what galaxy bias is expected for the brightest galaxies at redshift $z=1$. 

These clustering measurements are the first of its kind and are of prime interest to improve our knowledge of the halo occupation distribution of such galaxy populations. In this work, we assumed the angular clustering measured was representative of the clustering of a volume-limited sample at the mean redshift of the sample in order to derive its large scale bias. This assumption is not simple to test with current data, in particular with current photometric redshifts. Though this might be one of the reasons why the HOD model fails to describe in details the observed angular clustering at small scales, in particular the satellite fractions. In fact the properties of the selected galaxies evolve with redshift i.e. the lower redshift ones are less luminous than the higher redshift ones. The HOD model used here does not have parameters that evolve with redshift, it is a fixed halo population convolved with the redshift distribution. These considerations are oriented towards the construction of a mock catalogue.

\section{Conclusion}

We measured, for the first time, the angular clustering and the cosmic shear of the future samples selected for BAO studies. We have validated the use of this method on the CMASS sample as our measurement agrees with previous results. Moreover, we show that the CMASS sample is well correlated to the dark matter field. 

We demonstrate that the bright ELG samples are highly biased and therefore suited to detect BAO. The galaxies to be targeted for the near future SDSS-IV/eBOSS BAO experiment will provide a high SNR in the galaxy power spectrum. 

This work provides a strong basis to develop simulations and mock catalogues of these galaxy populations in order to understand the details of the relation ELG - dark matter haloes.

When pushing the method further in depth (ELG faint samples), we are limited by the knowledge of the redshift distribution of the tracers and by the depth of galaxy shape catalogues for WL. Both methods disagree when the redshift distribution is poorly known, i.e., with an uncertainty of about 40 per cent. Within the next few years, deeper observations and improved photometric redshifts should provide the necessary data to describe with precision the relation between galaxies and matter probed by the faint target selection schemes.

\section*{Acknowledgements}
Funding for SDSS-III has been provided by the Alfred P. Sloan Foundation, the Participating Institutions, the National Science Foundation, and the U.S. Department of Energy Office of Science. The SDSS-III web site is http://www.sdss3.org/.

SDSS-III is managed by the Astrophysical Research Consortium for the Participating Institutions of the SDSS-III Collaboration including the University of Arizona, the Brazilian Participation Group, Brookhaven National Laboratory, University of Cambridge, Carnegie Mellon University, University of Florida, the French Participation Group, the German Participation Group, Harvard University, the Instituto de Astrofisica de Canarias, the Michigan State/Notre Dame/JINA Participation Group, Johns Hopkins University, Lawrence Berkeley National Laboratory, Max Planck Institute for Astrophysics, Max Planck Institute for Extraterrestrial Physics, New Mexico State University, New York University, Ohio State University, Pennsylvania State University, University of Portsmouth, Princeton University, the Spanish Participation Group, University of Tokyo, University of Utah, Vanderbilt University, University of Virginia, University of Washington, and Yale University. 

The BOSS French Participation Group is supported by Agence Nationale de la Recherche under grant ANR-08-BLAN-0222.

Based on observations obtained with MegaPrime/MegaCam, a joint project of CFHT and CEA/DAPNIA, at the Canada-France-Hawaii Telescope (CFHT) which is operated by the National Research Council (NRC) of Canada, the Institut National des Science de l'Univers of the Centre National de la Recherche Scientifique (CNRS) of France, and the University of Hawaii. This work is based in part on data products produced at TERAPIX and the Canadian Astronomy Data Centre as part of the Canada-France-Hawaii Telescope Legacy Survey, a collaborative project of NRC and CNRS.

This publication makes use of data products from the Wide-field Infrared Survey Explorer, which is a joint project of the University of California, Los Angeles, and the Jet Propulsion Laboratory/California Institute of Technology, funded by the National Aeronautics and Space Administration.

The MultiDark Database used in this paper and the web application providing online access to it were constructed as part of the activities of the German Astrophysical Virtual Observatory as result of a collaboration between the Leibniz-Institute for Astrophysics Potsdam (AIP) and the Spanish MultiDark Consolider Project CSD2009-00064. The Bolshoi and MultiDark simulations were run on the NASA's Pleiades supercomputer at the NASA Ames Research Center.

We also thank the Laborat\'orio Interinstitucional de e-Astronomia (LIneA) operated jointly by the Centro Brasileiro de Pesquisas F\'isicas (CBPF), the Laborat\'orio Nacional de Computa\c c\~ao Cient\'ifica (LNCC), and the Observat\'orio Nacional (ON) and funded by the Ministry of Science, Technology and Inovation (MCTI) of Brazil. 
\bibliographystyle{mn2e}
\bibliography{biblio.bib}

\begin{thebibliography}{94}
\expandafter\ifx\csname natexlab\endcsname\relax\def\natexlab#1{#1}\fi

\bibitem[{{Ahn} {et~al}\mbox{.}(2012){Ahn}, {Alexandroff}, {Allende Prieto},
  {Anderson}, {Anderton}, {Andrews}, {Aubourg}, {Bailey}, {Balbinot}, {Barnes},
  \& et~al.}]{2012ApJS..203...21A}
{Ahn} C.~P. {et~al.}, 2012, \apjs, 203, 21

\bibitem[{{Anderson} {et~al}\mbox{.}(2012){Anderson}, {Aubourg}, {Bailey},
  {Bizyaev}, {Blanton}, {Bolton}, {Brinkmann}, {Brownstein}, {Burden},
  {Cuesta}, {da Costa}, {Dawson}, {de Putter}, {Eisenstein}, {Gunn}, {Guo},
  {Hamilton}, {Harding}, {Ho}, {Honscheid}, {Kazin}, {Kirkby}, {Kneib},
  {Labatie}, {Loomis}, {Lupton}, {Malanushenko}, {Malanushenko}, {Mandelbaum},
  {Manera}, {Maraston}, {McBride}, {Mehta}, {Mena}, {Montesano}, {Muna},
  {Nichol}, {Nuza}, {Olmstead}, {Oravetz}, {Padmanabhan},
  {Palanque-Delabrouille}, {Pan}, {Parejko}, {P{\^a}ris}, {Percival},
  {Petitjean}, {Prada}, {Reid}, {Roe}, {Ross}, {Ross}, {Samushia},
  {S{\'a}nchez}, {Schlegel}, {Schneider}, {Sc{\'o}ccola}, {Seo}, {Sheldon},
  {Simmons}, {Skibba}, {Strauss}, {Swanson}, {Thomas}, {Tinker}, {Tojeiro},
  {Maga{\~n}a}, {Verde}, {Wagner}, {Wake}, {Weaver}, {Weinberg}, {White}, {Xu},
  {Y{\`e}che}, {Zehavi}, \& {Zhao}}]{2012MNRAS.427.3435A}
{Anderson} L. {et~al.}, 2012, \mnras, 427, 3435

\bibitem[{{Annis} {et~al}\mbox{.}(2011){Annis}, {Soares-Santos}, {Strauss},
  {Becker}, {Dodelson}, {Fan}, {Gunn}, {Hao}, {Ivezic}, {Jester}, {Jiang},
  {Johnston}, {Kubo}, {Lampeitl}, {Lin}, {Lupton}, {Miknaitis}, {Seo}, {Simet},
  \& {Yanny}}]{2011arXiv1111.6619A}
{Annis} J. {et~al.}, 2011, ArXiv e-prints

\bibitem[{{Bacon} {et~al}\mbox{.}(2003){Bacon}, {Massey}, {Refregier}, \&
  {Ellis}}]{2003MNRAS.344..673B}
{Bacon} D.~J., {Massey} R.~J., {Refregier} A.~R., {Ellis} R.~S., 2003, \mnras,
  344, 673

\bibitem[{{Beutler} {et~al}\mbox{.}(2011){Beutler}, {Blake}, {Colless},
  {Jones}, {Staveley-Smith}, {Campbell}, {Parker}, {Saunders}, \&
  {Watson}}]{2011MNRAS.416.3017B}
{Beutler} F. {et~al.}, 2011, \mnras, 416, 3017

\bibitem[{{Blake} {et~al}\mbox{.}(2011){Blake}, {Davis}, {Poole}, {Parkinson},
  {Brough}, {Colless}, {Contreras}, {Couch}, {Croom}, {Drinkwater}, {Forster},
  {Gilbank}, {Gladders}, {Glazebrook}, {Jelliffe}, {Jurek}, {Li}, {Madore},
  {Martin}, {Pimbblet}, {Pracy}, {Sharp}, {Wisnioski}, {Woods}, {Wyder}, \&
  {Yee}}]{Blake_2011B}
{Blake} C. {et~al.}, 2011, \mnras, 415, 2892

\bibitem[{{Boulade} {et~al}\mbox{.}(2003){Boulade}, {Charlot}, {Abbon}, {Aune},
  {Borgeaud}, {Carton}, {Carty}, {Da Costa}, {Deschamps}, {Desforge},
  {Eppell{\'e}}, {Gallais}, {Gosset}, {Granelli}, {Gros}, {de Kat}, {Loiseau},
  {Ritou}, {Rouss{\'e}}, {Starzynski}, {Vignal}, \&
  {Vigroux}}]{2003SPIE.4841...72B}
{Boulade} O. {et~al.}, 2003, in Society of Photo-Optical Instrumentation
  Engineers (SPIE) Conference Series, Vol. 4841, Society of Photo-Optical
  Instrumentation Engineers (SPIE) Conference Series, {Iye} M., {Moorwood}
  A.~F.~M., eds., pp. 72--81

\bibitem[{{Bridle} {et~al}\mbox{.}(2010){Bridle}, {Balan}, {Bethge}, {Gentile},
  {Harmeling}, {Heymans}, {Hirsch}, {Hosseini}, {Jarvis}, {Kirk}, {Kitching},
  {Kuijken}, {Lewis}, {Paulin-Henriksson}, {Sch{\"o}lkopf}, {Velander},
  {Voigt}, {Witherick}, {Amara}, {Bernstein}, {Courbin}, {Gill}, {Heavens},
  {Mandelbaum}, {Massey}, {Moghaddam}, {Rassat}, {R{\'e}fr{\'e}gier}, {Rhodes},
  {Schrabback}, {Shawe-Taylor}, {Shmakova}, {van Waerbeke}, \&
  {Wittman}}]{2010MNRAS.405.2044B}
{Bridle} S. {et~al.}, 2010, \mnras, 405, 2044

\bibitem[{{Busca} {et~al}\mbox{.}(2013){Busca}, {Delubac}, {Rich}, {Bailey},
  {Font-Ribera}, {Kirkby}, {Le Goff}, {Pieri}, {Slosar}, {Aubourg}, {Bautista},
  {Bizyaev}, {Blomqvist}, {Bolton}, {Bovy}, {Brewington}, {Borde}, {Brinkmann},
  {Carithers}, {Croft}, {Dawson}, {Ebelke}, {Eisenstein}, {Hamilton}, {Ho},
  {Hogg}, {Honscheid}, {Lee}, {Lundgren}, {Malanushenko}, {Malanushenko},
  {Margala}, {Maraston}, {Mehta}, {Miralda-Escud{\'e}}, {Myers}, {Nichol},
  {Noterdaeme}, {Olmstead}, {Oravetz}, {Palanque-Delabrouille}, {Pan},
  {P{\^a}ris}, {Percival}, {Petitjean}, {Roe}, {Rollinde}, {Ross}, {Rossi},
  {Schlegel}, {Schneider}, {Shelden}, {Sheldon}, {Simmons}, {Snedden},
  {Tinker}, {Viel}, {Weaver}, {Weinberg}, {White}, {Y{\`e}che}, \&
  {York}}]{2013A&A...552A..96B}
{Busca} N.~G. {et~al.}, 2013, \aap, 552, A96

\bibitem[{{Cacciato} {et~al}\mbox{.}(2012){Cacciato}, {Lahav}, {van den Bosch},
  {Hoekstra}, \& {Dekel}}]{2012MNRAS.426..566C}
{Cacciato} M., {Lahav} O., {van den Bosch} F.~C., {Hoekstra} H., {Dekel} A.,
  2012, \mnras, 426, 566

\bibitem[{{Capak} {et~al}\mbox{.}(2007){Capak}, {Aussel}, {Ajiki}, {McCracken},
  {Mobasher}, {Scoville}, {Shopbell}, {Taniguchi}, {Thompson}, {Tribiano},
  {Sasaki}, {Blain}, {Brusa}, {Carilli}, {Comastri}, {Carollo}, {Cassata},
  {Colbert}, {Ellis}, {Elvis}, {Giavalisco}, {Green}, {Guzzo}, {Hasinger},
  {Ilbert}, {Impey}, {Jahnke}, {Kartaltepe}, {Kneib}, {Koda}, {Koekemoer},
  {Komiyama}, {Leauthaud}, {Le Fevre}, {Lilly}, {Liu}, {Massey}, {Miyazaki},
  {Murayama}, {Nagao}, {Peacock}, {Pickles}, {Porciani}, {Renzini}, {Rhodes},
  {Rich}, {Salvato}, {Sanders}, {Scarlata}, {Schiminovich}, {Schinnerer},
  {Scodeggio}, {Sheth}, {Shioya}, {Tasca}, {Taylor}, {Yan}, \&
  {Zamorani}}]{Capak_2007}
{Capak} P. {et~al.}, 2007, \apjs, 172, 99

\bibitem[{{Coil} {et~al}\mbox{.}(2008){Coil}, {Newman}, {Croton}, {Cooper},
  {Davis}, {Faber}, {Gerke}, {Koo}, {Padmanabhan}, {Wechsler}, \&
  {Weiner}}]{2008ApJ...672..153C}
{Coil} A.~L. {et~al.}, 2008, \apj, 672, 153

\bibitem[{{Comparat} {et~al}\mbox{.}(2013){Comparat}, {Kneib}, {Escoffier},
  {Zoubian}, {Ealet}, {Lamareille}, {Mostek}, {Steele}, {Aubourg}, {Bailey},
  {Bolton}, {Brownstein}, {Dawson}, {Ge}, {Ilbert}, {Leauthaud}, {Maraston},
  {Percival}, {Ross}, {Schimd}, {Schlegel}, {Schneider}, {Thomas}, {Tinker}, \&
  {Weaver}}]{2013MNRAS.428.1498C}
{Comparat} J. {et~al.}, 2013, \mnras, 428, 1498

\bibitem[{{Conroy} {et~al}\mbox{.}(2006){Conroy}, {Wechsler}, \&
  {Kravtsov}}]{2006ApJ...647..201C}
{Conroy} C., {Wechsler} R.~H., {Kravtsov} A.~V., 2006, \apj, 647, 201

\bibitem[{{Cooray} \& {Sheth}(2002)}]{2002PhR...372....1C}
{Cooray} A., {Sheth} R., 2002, \physrep, 372, 1

\bibitem[{{Coupon} {et~al}\mbox{.}(2009){Coupon}, {Ilbert}, {Kilbinger},
  {McCracken}, {Mellier}, {Arnouts}, {Bertin}, {Hudelot}, {Schultheis}, {Le
  F{\`e}vre}, {Le Brun}, {Guzzo}, {Bardelli}, {Zucca}, {Bolzonella}, {Garilli},
  {Zamorani}, {Zanichelli}, {Tresse}, \& {Aussel}}]{Coupon_2009}
{Coupon} J. {et~al.}, 2009, \aap, 500, 981

\bibitem[{{Coupon} {et~al}\mbox{.}(2012){Coupon}, {Kilbinger}, {McCracken},
  {Ilbert}, {Arnouts}, {Mellier}, {Abbas}, {de la Torre}, {Goranova},
  {Hudelot}, {Kneib}, \& {Le F{\`e}vre}}]{2012A&A...542A...5C}
{Coupon} J. {et~al.}, 2012, \aap, 542, A5

\bibitem[{{Dawson} {et~al}\mbox{.}(2013){Dawson}, {Schlegel}, {Ahn},
  {Anderson}, {Aubourg}, {Bailey}, {Barkhouser}, {Bautista}, {Beifiori},
  {Berlind}, {Bhardwaj}, {Bizyaev}, {Blake}, {Blanton}, {Blomqvist}, {Bolton},
  {Borde}, {Bovy}, {Brandt}, {Brewington}, {Brinkmann}, {Brown}, {Brownstein},
  {Bundy}, {Busca}, {Carithers}, {Carnero}, {Carr}, {Chen}, {Comparat},
  {Connolly}, {Cope}, {Croft}, {Cuesta}, {da Costa}, {Davenport}, {Delubac},
  {de Putter}, {Dhital}, {Ealet}, {Ebelke}, {Eisenstein}, {Escoffier}, {Fan},
  {Filiz Ak}, {Finley}, {Font-Ribera}, {G{\'e}nova-Santos}, {Gunn}, {Guo},
  {Haggard}, {Hall}, {Hamilton}, {Harris}, {Harris}, {Ho}, {Hogg}, {Holder},
  {Honscheid}, {Huehnerhoff}, {Jordan}, {Jordan}, {Kauffmann}, {Kazin},
  {Kirkby}, {Klaene}, {Kneib}, {Le Goff}, {Lee}, {Long}, {Loomis}, {Lundgren},
  {Lupton}, {Maia}, {Makler}, {Malanushenko}, {Malanushenko}, {Mandelbaum},
  {Manera}, {Maraston}, {Margala}, {Masters}, {McBride}, {McDonald}, {McGreer},
  {McMahon}, {Mena}, {Miralda-Escud{\'e}}, {Montero-Dorta}, {Montesano},
  {Muna}, {Myers}, {Naugle}, {Nichol}, {Noterdaeme}, {Nuza}, {Olmstead},
  {Oravetz}, {Oravetz}, {Owen}, {Padmanabhan}, {Palanque-Delabrouille}, {Pan},
  {Parejko}, {P{\^a}ris}, {Percival}, {P{\'e}rez-Fournon},
  {P{\'e}rez-R{\`a}fols}, {Petitjean}, {Pfaffenberger}, {Pforr}, {Pieri},
  {Prada}, {Price-Whelan}, {Raddick}, {Rebolo}, {Rich}, {Richards}, {Rockosi},
  {Roe}, {Ross}, {Ross}, {Rossi}, {Rubi{\~n}o-Martin}, {Samushia},
  {S{\'a}nchez}, {Sayres}, {Schmidt}, {Schneider}, {Sc{\'o}ccola}, {Seo},
  {Shelden}, {Sheldon}, {Shen}, {Shu}, {Slosar}, {Smee}, {Snedden}, {Stauffer},
  {Steele}, {Strauss}, {Streblyanska}, {Suzuki}, {Swanson}, {Tal}, {Tanaka},
  {Thomas}, {Tinker}, {Tojeiro}, {Tremonti}, {Vargas Maga{\~n}a}, {Verde},
  {Viel}, {Wake}, {Watson}, {Weaver}, {Weinberg}, {Weiner}, {West}, {White},
  {Wood-Vasey}, {Yeche}, {Zehavi}, {Zhao}, \& {Zheng}}]{2013AJ....145...10D}
{Dawson} K.~S. {et~al.}, 2013, \aj, 145, 10

\bibitem[{{Dekel} \& {Lahav}(1999)}]{1999ApJ...520...24D}
{Dekel} A., {Lahav} O., 1999, \apj, 520, 24

\bibitem[{{Drinkwater} {et~al}\mbox{.}(2010){Drinkwater}, {Jurek}, {Blake},
  {Woods}, {Pimbblet}, {Glazebrook}, {Sharp}, {Pracy}, {Brough}, {Colless},
  {Couch}, {Croom}, {Davis}, {Forbes}, {Forster}, {Gilbank}, {Gladders},
  {Jelliffe}, {Jones}, {Li}, {Madore}, {Martin}, {Poole}, {Small}, {Wisnioski},
  {Wyder}, \& {Yee}}]{Drinkwater_2010}
{Drinkwater} M.~J. {et~al.}, 2010, \mnras, 401, 1429

\bibitem[{{Eisenstein} \& {Hu}(1998)}]{1998ApJ...496..605E}
{Eisenstein} D.~J., {Hu} W., 1998, \apj, 496, 605

\bibitem[{{Eisenstein} {et~al}\mbox{.}(2011){Eisenstein}, {Weinberg}, {Agol},
  {Aihara}, {Allende Prieto}, {Anderson}, {Arns}, {Aubourg}, {Bailey},
  {Balbinot}, \& et~al.}]{2011AJ....142...72E}
{Eisenstein} D.~J. {et~al.}, 2011, \aj, 142, 72

\bibitem[{{Erben} {et~al}\mbox{.}(2012){Erben}, {Hildebrandt}, {Miller}, {van
  Waerbeke}, {Heymans}, {Hoekstra}, {Kitching}, {Mellier}, {Benjamin}, {Blake},
  {Bonnett}, {Cordes}, {Coupon}, {Fu}, {Gavazzi}, {Gillis}, {Grocutt}, {Gwyn},
  {Holhjem}, {Hudson}, {Kilbinger}, {Kuijken}, {Milkeraitis}, {Rowe},
  {Schrabback}, {Semboloni}, {Simon}, {Smit}, {Toader}, {Vafaei}, {van Uitert},
  \& {Velander}}]{2012arXiv1210.8156E}
{Erben} T. {et~al.}, 2012, ArXiv e-prints

\bibitem[{Frieman {et~al}\mbox{.}(2008)Frieman, Turner, \&
  Huterer}]{doi:10.1146/annurev.astro.46.060407.145243}
Frieman J.~A., Turner M.~S., Huterer D., 2008, Annual Review of Astronomy and
  Astrophysics, 46, 385

\bibitem[{{Fukugita} {et~al}\mbox{.}(1996){Fukugita}, {Ichikawa}, {Gunn},
  {Doi}, {Shimasaku}, \& {Schneider}}]{1996AJ....111.1748F}
{Fukugita} M., {Ichikawa} T., {Gunn} J.~E., {Doi} M., {Shimasaku} K.,
  {Schneider} D.~P., 1996, \aj, 111, 1748

\bibitem[{{Gunn} {et~al}\mbox{.}(1998){Gunn}, {Carr}, {Rockosi}, {Sekiguchi},
  {Berry}, {Elms}, {de Haas}, {Ivezi{\'c}}, {Knapp}, {Lupton}, {Pauls},
  {Simcoe}, {Hirsch}, {Sanford}, {Wang}, {York}, {Harris}, {Annis}, {Bartozek},
  {Boroski}, {Bakken}, {Haldeman}, {Kent}, {Holm}, {Holmgren}, {Petravick},
  {Prosapio}, {Rechenmacher}, {Doi}, {Fukugita}, {Shimasaku}, {Okada}, {Hull},
  {Siegmund}, {Mannery}, {Blouke}, {Heidtman}, {Schneider}, {Lucinio}, \&
  {Brinkman}}]{1998AJ....116.3040G}
{Gunn} J.~E. {et~al.}, 1998, \aj, 116, 3040

\bibitem[{{Gunn} {et~al}\mbox{.}(2006){Gunn}, {Siegmund}, {Mannery}, {Owen},
  {Hull}, {Leger}, {Carey}, {Knapp}, {York}, {Boroski}, {Kent}, {Lupton},
  {Rockosi}, {Evans}, {Waddell}, {Anderson}, {Annis}, {Barentine}, {Bartoszek},
  {Bastian}, {Bracker}, {Brewington}, {Briegel}, {Brinkmann}, {Brown}, {Carr},
  {Czarapata}, {Drennan}, {Dombeck}, {Federwitz}, {Gillespie}, {Gonzales},
  {Hansen}, {Harvanek}, {Hayes}, {Jordan}, {Kinney}, {Klaene}, {Kleinman},
  {Kron}, {Kresinski}, {Lee}, {Limmongkol}, {Lindenmeyer}, {Long}, {Loomis},
  {McGehee}, {Mantsch}, {Neilsen}, {Neswold}, {Newman}, {Nitta}, {Peoples},
  {Pier}, {Prieto}, {Prosapio}, {Rivetta}, {Schneider}, {Snedden}, \&
  {Wang}}]{Gunn_2006}
{Gunn} J.~E. {et~al.}, 2006, \aj, 131, 2332

\bibitem[{{Guo} {et~al}\mbox{.}(2013){Guo}, {Zehavi}, {Zheng}, {Weinberg},
  {Berlind}, {Blanton}, {Chen}, {Eisenstein}, {Ho}, {Kazin}, {Manera},
  {Maraston}, {McBride}, {Nuza}, {Padmanabhan}, {Parejko}, {Percival}, {Ross},
  {Ross}, {Samushia}, {S{\'a}nchez}, {Schlegel}, {Schneider}, {Skibba},
  {Swanson}, {Tinker}, {Tojeiro}, {Wake}, {White}, {Bahcall}, {Bizyaev},
  {Brewington}, {Bundy}, {da Costa}, {Ebelke}, {Malanushenko}, {Malanushenko},
  {Oravetz}, {Rossi}, {Simmons}, {Snedden}, {Streblyanska}, \&
  {Thomas}}]{2013ApJ...767..122G}
{Guo} H. {et~al.}, 2013, \apj, 767, 122

\bibitem[{{Heymans} {et~al}\mbox{.}(2006{\natexlab{a}}){Heymans}, {Van
  Waerbeke}, {Bacon}, {Berge}, {Bernstein}, {Bertin}, {Bridle}, {Brown},
  {Clowe}, {Dahle}, {Erben}, {Gray}, {Hetterscheidt}, {Hoekstra}, {Hudelot},
  {Jarvis}, {Kuijken}, {Margoniner}, {Massey}, {Mellier}, {Nakajima},
  {Refregier}, {Rhodes}, {Schrabback}, \& {Wittman}}]{2006MNRAS.368.1323H}
{Heymans} C. {et~al.}, 2006{\natexlab{a}}, \mnras, 368, 1323

\bibitem[{{Heymans} {et~al}\mbox{.}(2012){Heymans}, {Van Waerbeke}, {Miller},
  {Erben}, {Hildebrandt}, {Hoekstra}, {Kitching}, {Mellier}, {Simon},
  {Bonnett}, {Coupon}, {Fu}, {Harnois D{\'e}raps}, {Hudson}, {Kilbinger},
  {Kuijken}, {Rowe}, {Schrabback}, {Semboloni}, {van Uitert}, {Vafaei}, \&
  {Velander}}]{2012MNRAS.427..146H}
{Heymans} C. {et~al.}, 2012, \mnras, 427, 146

\bibitem[{{Heymans} {et~al}\mbox{.}(2006{\natexlab{b}}){Heymans}, {White},
  {Heavens}, {Vale}, \& {van Waerbeke}}]{2006MNRAS.371..750H}
{Heymans} C., {White} M., {Heavens} A., {Vale} C., {van Waerbeke} L.,
  2006{\natexlab{b}}, \mnras, 371, 750

\bibitem[{{Hoekstra} {et~al}\mbox{.}(2002){Hoekstra}, {van Waerbeke},
  {Gladders}, {Mellier}, \& {Yee}}]{2002ApJ...577..604H}
{Hoekstra} H., {van Waerbeke} L., {Gladders} M.~D., {Mellier} Y., {Yee}
  H.~K.~C., 2002, \apj, 577, 604

\bibitem[{{Huff} {et~al}\mbox{.}(2011){Huff}, {Hirata}, {Mandelbaum},
  {Schlegel}, {Seljak}, \& {Lupton}}]{2011arXiv1111.6958H}
{Huff} E.~M., {Hirata} C.~M., {Mandelbaum} R., {Schlegel} D., {Seljak} U.,
  {Lupton} R.~H., 2011, ArXiv e-prints

\bibitem[{{Ilbert} {et~al}\mbox{.}(2006){Ilbert}, {Arnouts}, {McCracken},
  {Bolzonella}, {Bertin}, {Le F{\`e}vre}, {Mellier}, {Zamorani}, {Pell{\`o}},
  {Iovino}, {Tresse}, {Le Brun}, {Bottini}, {Garilli}, {Maccagni}, {Picat},
  {Scaramella}, {Scodeggio}, {Vettolani}, {Zanichelli}, {Adami}, {Bardelli},
  {Cappi}, {Charlot}, {Ciliegi}, {Contini}, {Cucciati}, {Foucaud}, {Franzetti},
  {Gavignaud}, {Guzzo}, {Marano}, {Marinoni}, {Mazure}, {Meneux}, {Merighi},
  {Paltani}, {Pollo}, {Pozzetti}, {Radovich}, {Zucca}, {Bondi}, {Bongiorno},
  {Busarello}, {de La Torre}, {Gregorini}, {Lamareille}, {Mathez}, {Merluzzi},
  {Ripepi}, {Rizzo}, \& {Vergani}}]{Ilbert_06}
{Ilbert} O. {et~al.}, 2006, \aap, 457, 841

\bibitem[{{Ilbert} {et~al}\mbox{.}(2009){Ilbert}, {Capak}, {Salvato}, {Aussel},
  {McCracken}, {Sanders}, {Scoville}, {Kartaltepe}, {Arnouts}, {Le Floc'h},
  {Mobasher}, {Taniguchi}, {Lamareille}, {Leauthaud}, {Sasaki}, {Thompson},
  {Zamojski}, {Zamorani}, {Bardelli}, {Bolzonella}, {Bongiorno}, {Brusa},
  {Caputi}, {Carollo}, {Contini}, {Cook}, {Coppa}, {Cucciati}, {de la Torre},
  {de Ravel}, {Franzetti}, {Garilli}, {Hasinger}, {Iovino}, {Kampczyk},
  {Kneib}, {Knobel}, {Kovac}, {Le Borgne}, {Le Brun}, {F{\`e}vre}, {Lilly},
  {Looper}, {Maier}, {Mainieri}, {Mellier}, {Mignoli}, {Murayama}, {Pell{\`o}},
  {Peng}, {P{\'e}rez-Montero}, {Renzini}, {Ricciardelli}, {Schiminovich},
  {Scodeggio}, {Shioya}, {Silverman}, {Surace}, {Tanaka}, {Tasca}, {Tresse},
  {Vergani}, \& {Zucca}}]{Ilbert_2009}
{Ilbert} O. {et~al.}, 2009, \apj, 690, 1236

\bibitem[{{Jones} {et~al}\mbox{.}(2009){Jones}, {Read}, {Saunders}, {Colless},
  {Jarrett}, {Parker}, {Fairall}, {Mauch}, {Sadler}, {Watson}, {Burton},
  {Campbell}, {Cass}, {Croom}, {Dawe}, {Fiegert}, {Frankcombe}, {Hartley},
  {Huchra}, {James}, {Kirby}, {Lahav}, {Lucey}, {Mamon}, {Moore}, {Peterson},
  {Prior}, {Proust}, {Russell}, {Safouris}, {Wakamatsu}, {Westra}, \&
  {Williams}}]{2009MNRAS.399..683J}
{Jones} D.~H. {et~al.}, 2009, \mnras, 399, 683

\bibitem[{{Jullo} {et~al}\mbox{.}(2012){Jullo}, {Rhodes}, {Kiessling},
  {Taylor}, {Massey}, {Berge}, {Schimd}, {Kneib}, \&
  {Scoville}}]{2012ApJ...750...37J}
{Jullo} E. {et~al.}, 2012, \apj, 750, 37

\bibitem[{{Kaiser}(1984)}]{1984ApJ...284L...9K}
{Kaiser} N., 1984, \apjl, 284, L9

\bibitem[{{Kaiser} {et~al}\mbox{.}(1995){Kaiser}, {Squires}, \&
  {Broadhurst}}]{1995ApJ...449..460K}
{Kaiser} N., {Squires} G., {Broadhurst} T., 1995, \apj, 449, 460

\bibitem[{{Kilbinger} {et~al}\mbox{.}(2011){Kilbinger}, {Benabed}, {Cappe},
  {Cardoso}, {Coupon}, {Fort}, {McCracken}, {Prunet}, {Robert}, \&
  {Wraith}}]{2011arXiv1101.0950K}
{Kilbinger} M. {et~al.}, 2011, ArXiv e-prints

\bibitem[{{Kilbinger} {et~al}\mbox{.}(2009){Kilbinger}, {Benabed}, {Guy},
  {Astier}, {Tereno}, {Fu}, {Wraith}, {Coupon}, {Mellier}, {Balland},
  {Bouchet}, {Hamana}, {Hardin}, {McCracken}, {Pain}, {Regnault}, {Schultheis},
  \& {Yahagi}}]{2009A&A...497..677K}
{Kilbinger} M. {et~al.}, 2009, \aap, 497, 677

\bibitem[{{Kilbinger} {et~al}\mbox{.}(2006){Kilbinger}, {Schneider}, \&
  {Eifler}}]{2006A&A...457...15K}
{Kilbinger} M., {Schneider} P., {Eifler} T., 2006, \aap, 457, 15

\bibitem[{{Kilbinger} {et~al}\mbox{.}(2010){Kilbinger}, {Wraith}, {Robert},
  {Benabed}, {Capp{\'e}}, {Cardoso}, {Fort}, {Prunet}, \&
  {Bouchet}}]{2010MNRAS.405.2381K}
{Kilbinger} M. {et~al.}, 2010, \mnras, 405, 2381

\bibitem[{{Kirkby} {et~al}\mbox{.}(2013){Kirkby}, {Margala}, {Slosar},
  {Bailey}, {Busca}, {Delubac}, {Rich}, {Bautista}, {Blomqvist}, {Brownstein},
  {Carithers}, {Croft}, {Dawson}, {Font-Ribera}, {Miralda-Escud{\'e}}, {Myers},
  {Nichol}, {Palanque-Delabrouille}, {P{\^a}ris}, {Petitjean}, {Rossi},
  {Schlegel}, {Schneider}, {Viel}, {Weinberg}, \&
  {Y{\`e}che}}]{2013JCAP...03..024K}
{Kirkby} D. {et~al.}, 2013, \jcap, 3, 24

\bibitem[{{Kitching} {et~al}\mbox{.}(2012){Kitching}, {Balan}, {Bridle},
  {Cantale}, {Courbin}, {Eifler}, {Gentile}, {Gill}, {Harmeling}, {Heymans},
  {Hirsch}, {Honscheid}, {Kacprzak}, {Kirkby}, {Margala}, {Massey}, {Melchior},
  {Nurbaeva}, {Patton}, {Rhodes}, {Rowe}, {Taylor}, {Tewes}, {Viola},
  {Witherick}, {Voigt}, {Young}, \& {Zuntz}}]{2012MNRAS.423.3163K}
{Kitching} T.~D. {et~al.}, 2012, \mnras, 423, 3163

\bibitem[{{Komatsu} {et~al}\mbox{.}(2011){Komatsu}, {Smith}, {Dunkley},
  {Bennett}, {Gold}, {Hinshaw}, {Jarosik}, {Larson}, {Nolta}, {Page},
  {Spergel}, {Halpern}, {Hill}, {Kogut}, {Limon}, {Meyer}, {Odegard}, {Tucker},
  {Weiland}, {Wollack}, \& {Wright}}]{Komatsu_2011}
{Komatsu} E. {et~al.}, 2011, \apjs, 192, 18

\bibitem[{{Kravtsov} {et~al}\mbox{.}(2004){Kravtsov}, {Berlind}, {Wechsler},
  {Klypin}, {Gottl{\"o}ber}, {Allgood}, \& {Primack}}]{2004ApJ...609...35K}
{Kravtsov} A.~V., {Berlind} A.~A., {Wechsler} R.~H., {Klypin} A.~A.,
  {Gottl{\"o}ber} S., {Allgood} B., {Primack} J.~R., 2004, \apj, 609, 35

\bibitem[{{Landy} \& {Szalay}(1993)}]{1993ApJ...412...64L}
{Landy} S.~D., {Szalay} A.~S., 1993, \apj, 412, 64

\bibitem[{{Laureijs} {et~al}\mbox{.}(2011){Laureijs}, {Amiaux}, {Arduini},
  {Augu{\`e}res}, {Brinchmann}, {Cole}, {Cropper}, {Dabin}, {Duvet}, {Ealet},
  \& et~al.}]{2011arXiv1110.3193L}
{Laureijs} R. {et~al.}, 2011, ArXiv e-prints

\bibitem[{{Leauthaud} {et~al}\mbox{.}(2010){Leauthaud}, {Finoguenov}, {Kneib},
  {Taylor}, {Massey}, {Rhodes}, {Ilbert}, {Bundy}, {Tinker}, {George}, {Capak},
  {Koekemoer}, {Johnston}, {Zhang}, {Cappelluti}, {Ellis}, {Elvis}, {Giodini},
  {Heymans}, {Le F{\`e}vre}, {Lilly}, {McCracken}, {Mellier},
  {R{\'e}fr{\'e}gier}, {Salvato}, {Scoville}, {Smoot}, {Tanaka}, {Van
  Waerbeke}, \& {Wolk}}]{2010ApJ...709...97L}
{Leauthaud} A. {et~al.}, 2010, \apj, 709, 97

\bibitem[{{Leauthaud} {et~al}\mbox{.}(2011){Leauthaud}, {Tinker}, {Behroozi},
  {Busha}, \& {Wechsler}}]{2011ApJ...738...45L}
{Leauthaud} A., {Tinker} J., {Behroozi} P.~S., {Busha} M.~T., {Wechsler} R.~H.,
  2011, \apj, 738, 45

\bibitem[{{Leauthaud} {et~al}\mbox{.}(2012){Leauthaud}, {Tinker}, {Bundy},
  {Behroozi}, {Massey}, {Rhodes}, {George}, {Kneib}, {Benson}, {Wechsler},
  {Busha}, {Capak}, {Cort{\^e}s}, {Ilbert}, {Koekemoer}, {Le F{\`e}vre},
  {Lilly}, {McCracken}, {Salvato}, {Schrabback}, {Scoville}, {Smith}, \&
  {Taylor}}]{2012ApJ...744..159L}
{Leauthaud} A. {et~al.}, 2012, \apj, 744, 159

\bibitem[{{Lee} {et~al}\mbox{.}(2013){Lee}, {Bailey}, {Bartsch}, {Carithers},
  {Dawson}, {Kirkby}, {Lundgren}, {Margala}, {Palanque-Delabrouille}, {Pieri},
  {Schlegel}, {Weinberg}, {Y{\`e}che}, {Aubourg}, {Bautista}, {Bizyaev},
  {Blomqvist}, {Bolton}, {Borde}, {Brewington}, {Busca}, {Croft}, {Delubac},
  {Ebelke}, {Eisenstein}, {Font-Ribera}, {Ge}, {Hamilton}, {Hennawi}, {Ho},
  {Honscheid}, {Le Goff}, {Malanushenko}, {Malanushenko}, {Miralda-Escud{\'e}},
  {Myers}, {Noterdaeme}, {Oravetz}, {Pan}, {P{\^a}ris}, {Petitjean}, {Rich},
  {Rollinde}, {Ross}, {Rossi}, {Schneider}, {Simmons}, {Snedden}, {Slosar},
  {Spergel}, {Suzuki}, {Viel}, \& {Weaver}}]{2013AJ....145...69L}
{Lee} K.-G. {et~al.}, 2013, \aj, 145, 69

\bibitem[{{Li} {et~al}\mbox{.}(2006){Li}, {Kauffmann}, {Jing}, {White},
  {B{\"o}rner}, \& {Cheng}}]{2006MNRAS.368...21L}
{Li} C., {Kauffmann} G., {Jing} Y.~P., {White} S.~D.~M., {B{\"o}rner} G.,
  {Cheng} F.~Z., 2006, \mnras, 368, 21

\bibitem[{{Limber}(1954)}]{1954ApJ...119..655L}
{Limber} D.~N., 1954, \apj, 119, 655

\bibitem[{{Luppino} \& {Kaiser}(1997)}]{1997ApJ...475...20L}
{Luppino} G.~A., {Kaiser} N., 1997, \apj, 475, 20

\bibitem[{{Ma} \& {Fry}(2000)}]{2000ApJ...543..503M}
{Ma} C.-P., {Fry} J.~N., 2000, \apj, 543, 503

\bibitem[{{Massey} {et~al}\mbox{.}(2007){Massey}, {Heymans}, {Berg{\'e}},
  {Bernstein}, {Bridle}, {Clowe}, {Dahle}, {Ellis}, {Erben}, {Hetterscheidt},
  {High}, {Hirata}, {Hoekstra}, {Hudelot}, {Jarvis}, {Johnston}, {Kuijken},
  {Margoniner}, {Mandelbaum}, {Mellier}, {Nakajima}, {Paulin-Henriksson},
  {Peeples}, {Roat}, {Refregier}, {Rhodes}, {Schrabback}, {Schirmer}, {Seljak},
  {Semboloni}, \& {van Waerbeke}}]{2007MNRAS.376...13M}
{Massey} R. {et~al.}, 2007, \mnras, 376, 13

\bibitem[{{Mehta} {et~al}\mbox{.}(2011){Mehta}, {Seo}, {Eckel}, {Eisenstein},
  {Metchnik}, {Pinto}, \& {Xu}}]{2011ApJ...734...94M}
{Mehta} K.~T., {Seo} H.-J., {Eckel} J., {Eisenstein} D.~J., {Metchnik} M.,
  {Pinto} P., {Xu} X., 2011, \apj, 734, 94

\bibitem[{{More} {et~al}\mbox{.}(2013){More}, {van den Bosch}, {Cacciato},
  {More}, {Mo}, \& {Yang}}]{2013MNRAS.430..747M}
{More} S., {van den Bosch} F.~C., {Cacciato} M., {More} A., {Mo} H., {Yang} X.,
  2013, \mnras, 430, 747

\bibitem[{{Mostek} {et~al}\mbox{.}(2013){Mostek}, {Coil}, {Cooper}, {Davis},
  {Newman}, \& {Weiner}}]{2013ApJ...767...89M}
{Mostek} N., {Coil} A.~L., {Cooper} M., {Davis} M., {Newman} J.~A., {Weiner}
  B.~J., 2013, \apj, 767, 89

\bibitem[{{Newman} \& {Davis}(2002)}]{2002ApJ...564..567N}
{Newman} J.~A., {Davis} M., 2002, \apj, 564, 567

\bibitem[{{Nuza} {et~al}\mbox{.}(2013){Nuza}, {S{\'a}nchez}, {Prada}, {Klypin},
  {Schlegel}, {Gottl{\"o}ber}, {Montero-Dorta}, {Manera}, {McBride}, {Ross},
  {Angulo}, {Blanton}, {Bolton}, {Favole}, {Samushia}, {Montesano}, {Percival},
  {Padmanabhan}, {Steinmetz}, {Tinker}, {Skibba}, {Schneider}, {Guo}, {Zehavi},
  {Zheng}, {Bizyaev}, {Malanushenko}, {Malanushenko}, {Oravetz}, {Oravetz}, \&
  {Shelden}}]{2013MNRAS.tmp.1188N}
{Nuza} S.~E. {et~al.}, 2013, \mnras

\bibitem[{{Oke} \& {Gunn}(1983)}]{1983ApJ...266..713O}
{Oke} J.~B., {Gunn} J.~E., 1983, \apj, 266, 713

\bibitem[{{P{\^a}ris} {et~al}\mbox{.}(2012){P{\^a}ris}, {Petitjean}, {Aubourg},
  {Bailey}, {Ross}, {Myers}, {Strauss}, {Anderson}, {Arnau}, {Bautista},
  {Bizyaev}, {Bolton}, {Bovy}, {Brandt}, {Brewington}, {Browstein}, {Busca},
  {Capellupo}, {Carithers}, {Croft}, {Dawson}, {Delubac}, {Ebelke},
  {Eisenstein}, {Engelke}, {Fan}, {Filiz Ak}, {Finley}, {Font-Ribera}, {Ge},
  {Gibson}, {Hall}, {Hamann}, {Hennawi}, {Ho}, {Hogg}, {Ivezi{\'c}}, {Jiang},
  {Kimball}, {Kirkby}, {Kirkpatrick}, {Lee}, {Le Goff}, {Lundgren}, {MacLeod},
  {Malanushenko}, {Malanushenko}, {Maraston}, {McGreer}, {McMahon},
  {Miralda-Escud{\'e}}, {Muna}, {Noterdaeme}, {Oravetz},
  {Palanque-Delabrouille}, {Pan}, {Perez-Fournon}, {Pieri}, {Richards},
  {Rollinde}, {Sheldon}, {Schlegel}, {Schneider}, {Slosar}, {Shelden}, {Shen},
  {Simmons}, {Snedden}, {Suzuki}, {Tinker}, {Viel}, {Weaver}, {Weinberg},
  {White}, {Wood-Vasey}, \& {Y{\`e}che}}]{2012A&A...548A..66P}
{P{\^a}ris} I. {et~al.}, 2012, \aap, 548, A66

\bibitem[{{Parkinson} {et~al}\mbox{.}(2012){Parkinson}, {Riemer-S{\o}rensen},
  {Blake}, {Poole}, {Davis}, {Brough}, {Colless}, {Contreras}, {Couch},
  {Croom}, {Croton}, {Drinkwater}, {Forster}, {Gilbank}, {Gladders},
  {Glazebrook}, {Jelliffe}, {Jurek}, {Li}, {Madore}, {Martin}, {Pimbblet},
  {Pracy}, {Sharp}, {Wisnioski}, {Woods}, {Wyder}, \&
  {Yee}}]{2012PhRvD..86j3518P}
{Parkinson} D. {et~al.}, 2012, \prd, 86, 103518

\bibitem[{{Planck Collaboration} {et~al}\mbox{.}(2013){Planck Collaboration},
  {Ade}, {Aghanim}, {Armitage-Caplan}, {Arnaud}, {Ashdown}, {Atrio-Barandela},
  {Aumont}, {Baccigalupi}, {Banday}, \& et~al.}]{2013arXiv1303.5076P}
{Planck Collaboration} {et~al.}, 2013, ArXiv e-prints

\bibitem[{{Prada} {et~al}\mbox{.}(2012){Prada}, {Klypin}, {Cuesta},
  {Betancort-Rijo}, \& {Primack}}]{2012MNRAS.423.3018P}
{Prada} F., {Klypin} A.~A., {Cuesta} A.~J., {Betancort-Rijo} J.~E., {Primack}
  J., 2012, \mnras, 423, 3018

\bibitem[{{Riebe} {et~al}\mbox{.}(2011){Riebe}, {Partl}, {Enke},
  {Forero-Romero}, {Gottloeber}, {Klypin}, {Lemson}, {Prada}, {Primack},
  {Steinmetz}, \& {Turchaninov}}]{2011arXiv1109.0003R}
{Riebe} K. {et~al.}, 2011, ArXiv e-prints

\bibitem[{{Schlegel} {et~al}\mbox{.}(2011){Schlegel}, {Abdalla}, {Abraham},
  {Ahn}, {Allende Prieto}, {Annis}, {Aubourg}, {Azzaro}, {Baltay}, {Baugh},
  {Bebek}, {Becerril}, {Blanton}, {Bolton}, {Bromley}, {Cahn}, {Carton},
  {Cervantes-Cota}, {Chu}, {Cortes}, {Dawson}, {Dey}, {Dickinson}, {Diehl},
  {Doel}, {Ealet}, {Edelstein}, {Eppelle}, {Escoffier}, {Evrard}, {Faccioli},
  {Frenk}, {Geha}, {Gerdes}, {Gondolo}, {Gonzalez-Arroyo}, {Grossan},
  {Heckman}, {Heetderks}, {Ho}, {Honscheid}, {Huterer}, {Ilbert}, {Ivans},
  {Jelinsky}, {Jing}, {Joyce}, {Kennedy}, {Kent}, {Kieda}, {Kim}, {Kim},
  {Kneib}, {Kong}, {Kosowsky}, {Krishnan}, {Lahav}, {Lampton}, {LeBohec}, {Le
  Brun}, {Levi}, {Li}, {Liang}, {Lim}, {Lin}, {Linder}, {Lorenzon}, {de la
  Macorra}, {Magneville}, {Malina}, {Marinoni}, {Martinez}, {Majewski},
  {Matheson}, {McCloskey}, {McDonald}, {McKay}, {McMahon}, {Menard},
  {Miralda-Escude}, {Modjaz}, {Montero-Dorta}, {Morales}, {Mostek}, {Newman},
  {Nichol}, {Nugent}, {Olsen}, {Padmanabhan}, {Palanque-Delabrouille}, {Park},
  {Peacock}, {Percival}, {Perlmutter}, {Peroux}, {Petitjean}, {Prada},
  {Prieto}, {Prochaska}, {Reil}, {Rockosi}, {Roe}, {Rollinde}, {Roodman},
  {Ross}, {Rudnick}, {Ruhlmann-Kleider}, {Sanchez}, {Sawyer}, {Schimd},
  {Schubnell}, {Scoccimaro}, {Seljak}, {Seo}, {Sheldon}, {Sholl},
  {Shulte-Ladbeck}, {Slosar}, {Smith}, {Smoot}, {Springer}, {Stril}, {Szalay},
  {Tao}, {Tarle}, {Taylor}, {Tilquin}, {Tinker}, {Valdes}, {Wang}, {Wang},
  {Weaver}, {Weinberg}, {White}, {Wood-Vasey}, {Yang}, {Yeche}, {Zakamska},
  {Zentner}, {Zhai}, \& {Zhang}}]{bigBOSS_2011}
{Schlegel} D. {et~al.}, 2011, ArXiv e-prints

\bibitem[{{Schneider} {et~al}\mbox{.}(1998){Schneider}, {van Waerbeke},
  {Mellier}, {Jain}, {Seitz}, \& {Fort}}]{1998A&A...333..767S}
{Schneider} P., {van Waerbeke} L., {Mellier} Y., {Jain} B., {Seitz} S., {Fort}
  B., 1998, \aap, 333, 767

\bibitem[{{Scoville} {et~al}\mbox{.}(2007){Scoville}, {Aussel}, {Brusa},
  {Capak}, {Carollo}, {Elvis}, {Giavalisco}, {Guzzo}, {Hasinger}, {Impey},
  {Kneib}, {LeFevre}, {Lilly}, {Mobasher}, {Renzini}, {Rich}, {Sanders},
  {Schinnerer}, {Schminovich}, {Shopbell}, {Taniguchi}, \&
  {Tyson}}]{2007ApJS..172....1S}
{Scoville} N. {et~al.}, 2007, \apjs, 172, 1

\bibitem[{{Seljak}(2000)}]{2000MNRAS.318..203S}
{Seljak} U., 2000, \mnras, 318, 203

\bibitem[{{Shan} {et~al}\mbox{.}(2012){Shan}, {Kneib}, {Tao}, {Fan}, {Jauzac},
  {Limousin}, {Massey}, {Rhodes}, {Thanjavur}, \&
  {McCracken}}]{2012ApJ...748...56S}
{Shan} H. {et~al.}, 2012, \apj, 748, 56

\bibitem[{{Sheth} {et~al}\mbox{.}(2001){Sheth}, {Mo}, \&
  {Tormen}}]{2001MNRAS.323....1S}
{Sheth} R.~K., {Mo} H.~J., {Tormen} G., 2001, \mnras, 323, 1

\bibitem[{{Sheth} \& {Tormen}(1999)}]{1999MNRAS.308..119S}
{Sheth} R.~K., {Tormen} G., 1999, \mnras, 308, 119

\bibitem[{{Simon}(2007)}]{2007A&A...473..711S}
{Simon} P., 2007, \aap, 473, 711

\bibitem[{{Simon} {et~al}\mbox{.}(2007){Simon}, {Hetterscheidt}, {Schirmer},
  {Erben}, {Schneider}, {Wolf}, \& {Meisenheimer}}]{2007A&A...461..861S}
{Simon} P., {Hetterscheidt} M., {Schirmer} M., {Erben} T., {Schneider} P.,
  {Wolf} C., {Meisenheimer} K., 2007, \aap, 461, 861

\bibitem[{{Skibba} \& {Sheth}(2009)}]{2009MNRAS.392.1080S}
{Skibba} R.~A., {Sheth} R.~K., 2009, \mnras, 392, 1080

\bibitem[{{Slosar} {et~al}\mbox{.}(2013){Slosar}, {Ir{\v s}i{\v c}}, {Kirkby},
  {Bailey}, {Busca}, {Delubac}, {Rich}, {Aubourg}, {Bautista}, {Bhardwaj},
  {Blomqvist}, {Bolton}, {Bovy}, {Brownstein}, {Carithers}, {Croft}, {Dawson},
  {Font-Ribera}, {Le Goff}, {Ho}, {Honscheid}, {Lee}, {Margala}, {McDonald},
  {Medolin}, {Miralda-Escud{\'e}}, {Myers}, {Nichol}, {Noterdaeme},
  {Palanque-Delabrouille}, {P{\^a}ris}, {Petitjean}, {Pieri}, {Pi{\v s}kur},
  {Roe}, {Ross}, {Rossi}, {Schlegel}, {Schneider}, {Suzuki}, {Sheldon},
  {Seljak}, {Viel}, {Weinberg}, \& {Y{\`e}che}}]{2013JCAP...04..026S}
{Slosar} A. {et~al.}, 2013, \jcap, 4, 26

\bibitem[{{Smith} {et~al}\mbox{.}(2003){Smith}, {Peacock}, {Jenkins}, {White},
  {Frenk}, {Pearce}, {Thomas}, {Efstathiou}, \&
  {Couchman}}]{2003MNRAS.341.1311S}
{Smith} R.~E. {et~al.}, 2003, \mnras, 341, 1311

\bibitem[{{Suzuki} {et~al}\mbox{.}(2012){Suzuki}, {Rubin}, {Lidman},
  {Aldering}, {Amanullah}, {Barbary}, {Barrientos}, {Botyanszki}, {Brodwin},
  {Connolly}, {Dawson}, {Dey}, {Doi}, {Donahue}, {Deustua}, {Eisenhardt},
  {Ellingson}, {Faccioli}, {Fadeyev}, {Fakhouri}, {Fruchter}, {Gilbank},
  {Gladders}, {Goldhaber}, {Gonzalez}, {Goobar}, {Gude}, {Hattori}, {Hoekstra},
  {Hsiao}, {Huang}, {Ihara}, {Jee}, {Johnston}, {Kashikawa}, {Koester},
  {Konishi}, {Kowalski}, {Linder}, {Lubin}, {Melbourne}, {Meyers}, {Morokuma},
  {Munshi}, {Mullis}, {Oda}, {Panagia}, {Perlmutter}, {Postman}, {Pritchard},
  {Rhodes}, {Ripoche}, {Rosati}, {Schlegel}, {Spadafora}, {Stanford},
  {Stanishev}, {Stern}, {Strovink}, {Takanashi}, {Tokita}, {Wagner}, {Wang},
  {Yasuda}, {Yee}, \& {Supernova Cosmology Project}}]{2012ApJ...746...85S}
{Suzuki} N. {et~al.}, 2012, \apj, 746, 85

\bibitem[{{Tegmark} \& {Bromley}(1999)}]{1999ApJ...518L..69T}
{Tegmark} M., {Bromley} B.~C., 1999, \apjl, 518, L69

\bibitem[{{Tegmark} \& {Peebles}(1998)}]{1998ApJ...500L..79T}
{Tegmark} M., {Peebles} P.~J.~E., 1998, \apjl, 500, L79

\bibitem[{{Tinker} {et~al}\mbox{.}(2005){Tinker}, {Weinberg}, {Zheng}, \&
  {Zehavi}}]{2005ApJ...631...41T}
{Tinker} J.~L., {Weinberg} D.~H., {Zheng} Z., {Zehavi} I., 2005, \apj, 631, 41

\bibitem[{{Trujillo-Gomez} {et~al}\mbox{.}(2011){Trujillo-Gomez}, {Klypin},
  {Primack}, \& {Romanowsky}}]{2011ApJ...742...16T}
{Trujillo-Gomez} S., {Klypin} A., {Primack} J., {Romanowsky} A.~J., 2011, \apj,
  742, 16

\bibitem[{{Vale} \& {Ostriker}(2004)}]{2004MNRAS.353..189V}
{Vale} A., {Ostriker} J.~P., 2004, \mnras, 353, 189

\bibitem[{{van den Bosch} {et~al}\mbox{.}(2013){van den Bosch}, {More},
  {Cacciato}, {Mo}, \& {Yang}}]{2013MNRAS.430..725V}
{van den Bosch} F.~C., {More} S., {Cacciato} M., {Mo} H., {Yang} X., 2013,
  \mnras, 430, 725

\bibitem[{{Wake} {et~al}\mbox{.}(2011){Wake}, {Whitaker}, {Labb{\'e}}, {van
  Dokkum}, {Franx}, {Quadri}, {Brammer}, {Kriek}, {Lundgren}, {Marchesini}, \&
  {Muzzin}}]{2011ApJ...728...46W}
{Wake} D.~A. {et~al.}, 2011, \apj, 728, 46

\bibitem[{{Wall} \& {Jenkins}(2012)}]{2012psa..book.....W}
{Wall} J.~V., {Jenkins} C.~R., 2012, {Practical Statistics for Astronomers}

\bibitem[{{White} {et~al}\mbox{.}(2011){White}, {Blanton}, {Bolton},
  {Schlegel}, {Tinker}, {Berlind}, {da Costa}, {Kazin}, {Lin}, {Maia},
  {McBride}, {Padmanabhan}, {Parejko}, {Percival}, {Prada}, {Ramos}, {Sheldon},
  {de Simoni}, {Skibba}, {Thomas}, {Wake}, {Zehavi}, {Zheng}, {Nichol},
  {Schneider}, {Strauss}, {Weaver}, \& {Weinberg}}]{2011ApJ...728..126W}
{White} M. {et~al.}, 2011, \apj, 728, 126

\bibitem[{{Wright} {et~al}\mbox{.}(2010){Wright}, {Eisenhardt}, {Mainzer},
  {Ressler}, {Cutri}, {Jarrett}, {Kirkpatrick}, {Padgett}, {McMillan},
  {Skrutskie}, {Stanford}, {Cohen}, {Walker}, {Mather}, {Leisawitz}, {Gautier},
  {McLean}, {Benford}, {Lonsdale}, {Blain}, {Mendez}, {Irace}, {Duval}, {Liu},
  {Royer}, {Heinrichsen}, {Howard}, {Shannon}, {Kendall}, {Walsh}, {Larsen},
  {Cardon}, {Schick}, {Schwalm}, {Abid}, {Fabinsky}, {Naes}, \&
  {Tsai}}]{2010AJ....140.1868W}
{Wright} E.~L. {et~al.}, 2010, \aj, 140, 1868

\bibitem[{{York} {et~al}\mbox{.}(2000){York}, {Adelman}, {Anderson},
  {Anderson}, {Annis}, {Bahcall}, {Bakken}, {Barkhouser}, {Bastian}, {Berman},
  {Boroski}, {Bracker}, {Briegel}, {Briggs}, {Brinkmann}, {Brunner}, {Burles},
  {Carey}, {Carr}, {Castander}, {Chen}, {Colestock}, {Connolly}, {Crocker},
  {Csabai}, {Czarapata}, {Davis}, {Doi}, {Dombeck}, {Eisenstein}, {Ellman},
  {Elms}, {Evans}, {Fan}, {Federwitz}, {Fiscelli}, {Friedman}, {Frieman},
  {Fukugita}, {Gillespie}, {Gunn}, {Gurbani}, {de Haas}, {Haldeman}, {Harris},
  {Hayes}, {Heckman}, {Hennessy}, {Hindsley}, {Holm}, {Holmgren}, {Huang},
  {Hull}, {Husby}, {Ichikawa}, {Ichikawa}, {Ivezi{\'c}}, {Kent}, {Kim},
  {Kinney}, {Klaene}, {Kleinman}, {Kleinman}, {Knapp}, {Korienek}, {Kron},
  {Kunszt}, {Lamb}, {Lee}, {Leger}, {Limmongkol}, {Lindenmeyer}, {Long},
  {Loomis}, {Loveday}, {Lucinio}, {Lupton}, {MacKinnon}, {Mannery}, {Mantsch},
  {Margon}, {McGehee}, {McKay}, {Meiksin}, {Merelli}, {Monet}, {Munn},
  {Narayanan}, {Nash}, {Neilsen}, {Neswold}, {Newberg}, {Nichol}, {Nicinski},
  {Nonino}, {Okada}, {Okamura}, {Ostriker}, {Owen}, {Pauls}, {Peoples},
  {Peterson}, {Petravick}, {Pier}, {Pope}, {Pordes}, {Prosapio},
  {Rechenmacher}, {Quinn}, {Richards}, {Richmond}, {Rivetta}, {Rockosi},
  {Ruthmansdorfer}, {Sandford}, {Schlegel}, {Schneider}, {Sekiguchi}, {Sergey},
  {Shimasaku}, {Siegmund}, {Smee}, {Smith}, {Snedden}, {Stone}, {Stoughton},
  {Strauss}, {Stubbs}, {SubbaRao}, {Szalay}, {Szapudi}, {Szokoly}, {Thakar},
  {Tremonti}, {Tucker}, {Uomoto}, {Vanden Berk}, {Vogeley}, {Waddell}, {Wang},
  {Watanabe}, {Weinberg}, {Yanny}, {Yasuda}, \& {SDSS
  Collaboration}}]{2000AJ....120.1579Y}
{York} D.~G. {et~al.}, 2000, \aj, 120, 1579

\bibitem[{{Zehavi} {et~al}\mbox{.}(2011){Zehavi}, {Zheng}, {Weinberg},
  {Blanton}, {Bahcall}, {Berlind}, {Brinkmann}, {Frieman}, {Gunn}, {Lupton},
  {Nichol}, {Percival}, {Schneider}, {Skibba}, {Strauss}, {Tegmark}, \&
  {York}}]{2011ApJ...736...59Z}
{Zehavi} I. {et~al.}, 2011, \apj, 736, 59

\end{thebibliography}

\end{document}